\documentclass[a4paper]{aa}
\usepackage{amssymb, longtable}
\usepackage{graphicx}
\usepackage{babel}
\usepackage{hyperref}
\usepackage{soul}
\usepackage{color}
\usepackage{enumerate}
\usepackage{natbib}
\bibpunct{(}{)}{;}{a}{}{,} 
\usepackage{floatrow}
\newfloatcommand{capbtabbox}{table}[][\FBwidth]

\begin{document}


\def\rxy{$r_{xy}$~}
\def\pr{Proxima Cen}
\def\ga{G~224-58~A~}
\def\Tef{$T_{\rm eff}$~}
\def\Teff{$T_{\rm eff}$}
\def\Vr{$V_{\rm r}$~}
\def\Vsini{$v$ sin $i$~}
\def\Vm{$V_{\rm m}$~}
\def\EE{$E^{\prime\prime}$~}
\def\kmps{kms$^{\rm -1}$~}
\def\um{$\mu$m~}
\def\logg{$\log~g$~}
\def\loggg{$\log~g$}
\def\hev{He$_{5876}$}
\def\heiv{He$_{4026}$}
\def\Ha{$H_{\alpha}$}
\def\Haa{$H_{\alpha}$}
\def\DO{D$_1$}
\def\DT{D$_2$}
\def\pew{$pEW$}
\def\vlt{VLT/X-shooter}
\def\Po{P$_o$}
\def\Heps{H$_{\epsilon}$}

\title{Temporal changes of the flare activity of Proxima Cen}

\author{
Ya.~V.\ Pavlenko \inst{1,2,3}
A. Su\'arez Mascare\~no \inst{4}, 
M.~R.\ Zapatero Osorio\inst{5},
R.\ Rebolo \inst{2,6,7},
N.\ Lodieu \inst{2,6},\\
V. J. S. B\'ejar \inst{2,6},
J.~I. Gonz\'alez Hern\'andez\inst{2,6},
M.\ Mohorian\inst{8} 
}
\institute{
Main Astronomical Observatory of the National Academy of Sciences of Ukraine, 27 Zanolotnoho, Kyiv, 03143, Ukraine;
         \email{yp@mao.kiev.ua}
         \and
Instituto de Astrof\'isica de Canarias (IAC), Calle V\'ia L\'actea s/n, E-38200 La Laguna, Tenerife, Spain,          
         \and
          Center for Astrophysics Research, University of Hertfordshire, College Lane, Hatfield, Hertfordshire AL10 9AB, UK
          \and
          Observatoire Astronomique de l'Universit\'e de Gen\`eve, E-1290 Versoix, Gen\`eve, Switzerland,
          \and
          Centro de Astrobiolog\'ia (CSIC-INTA), Carretera de Ajalvir km 4, 28850 Torrej\'on de Ardoz, Madrid, Spapin,
         \and
         Departamento de Astrof\'isica, Universidad de La Laguna (ULL), E-38205 La Laguna, Tenerife, Spain,
         \and
          Consejo Superior de Investigaciones Cient\'ificas, CSIC, Spain, 
          \and
          Taras Shevchenko National University of Kyiv, Kyiv, Ukraine
}

\offprints{Yakiv Pavlenko}
\mail{yp@mao.kiev.ua}

\date{\today{}}

\authorrunning{Pavlenko et al.}
\titlerunning{Temporal changes of the flare activity of Proxima Cen}

\abstract{
We study temporal variations of the emission lines of \Ha{}, \Heps{}, H and K Ca{\small{I}}, 
\DO{} and \DT{} Na{\small{I}},  \heiv{}, and \hev{} in the HARPS spectra of Proxima Centauri 
across an extended time of 13.2 years, from May 27, 2004, to September 30, 2017\@.
}{
We analyse the common behaviour and differences in the intensities and profiles of different emission lines
in flare and quiet modes of Proxima activity. 
}{
We compare the pseudo-equivalent widths ($pEW$) and profiles of the emission lines in the 
HARPS high-resolution (R\,$\sim$\,115,000) spectra observed at the same epochs.
}{
All emission lines show variability with a timescale of at least 10 min.
The strength of all lines except \heiv{} correlate with \Ha{}.
During strong flares the `red asymmetry' appears in the \Ha{} emission line indicating the 
infall of hot condensed matter into the chromosphere with velocities greater than 100 km/s disturbing chromospheric layers.
As a result, the strength of the Ca II lines anti-correlates with \Ha{} during strong flares.
The He{\small{I}} lines at 4026 and 5876 \AA{} appear in the strong flares. 
The cores of \DO{} and \DT{} Na{\small{I}} lines are also seen in emission.
During the minimum activity of Proxima Centauri, Ca{\small{II}} lines and \Heps{} almost 
disappear while the blue part of the Na{\small{I}} emission lines is 
affected by the absorption in the extending and condensing flows. 
}{
We see different behaviour of emission lines formed in the flare regions and chromosphere.  
Chromosphere layers of Proxima Cen are likely heated by the flare 
events; these layers are  cooled in the `non-flare' mode. 
The self-absorption structures in
cores of our emission lines vary with time due to the presence of a 
complicated system of inward and outward matter flows in the absorbing
layers.
}

\maketitle 

%
%
\section{Introduction}

In this paper we define `flare activity' as a strong and time-variable phenomenon observed  
in stellar spectra and caused by the dissipation of magnetic field originated by dynamo 
processes in the stellar photosphere, in the chromosphere, or in the corona \citep[e.g.][]{park55}. 
These processes differ comparatively from the case of slow phenomena of `stellar activity' responsible 
for variations of observed fluxes due to comparatively slow variations of spots and 
filament structures on the stellar surface. Observationally, the dynamo mechanism can be probed 
through the relationship between rotation and magnetic activity, and the evolution 
of these properties \citep{suar15,newt17}.

 Magnetic activity in late-type stars is often measured in terms of \Ha{} emission and
increases with decreasing stellar mass \citep[e.g.][]{hawl96}.
Magnetic activity increases with decreasing stellar mass; for example, \cite{hawl96}.
In addition, magnetically active stars are known to flare more frequently than inactive stars. 
Magnetic activity, stellar rotation and magnetic field strength configuration 
are intrinsically linked. We believe that stellar flares are caused by the
reconnection of surface magnetic field loops (see, e.g. \cite{yoko98} and reference therein).  
The correlation between magnetic field in active regions is very clear in the case of solar
flares \citep{flet11} meaning that stellar flares may be used as a proxy for magnetic activity.

The fraction of stars showing \Ha{} in their emission \citep{delf98} or frequent flares 
\citep{hawl91,schm14,hawl14} increases with decreasing mass, mainly due to their 
longer rotational braking times \citep{delf98,barn03,delo11}. 
The braking rates strongly depend on the spectral type (or mass) of the
star \citep{barn03,barn10}, which results in partially convective low-mass stars spinning 
down slower than their higher-mass counterparts. 
Fully convective stars have even longer rotational braking timescales \citep{rein08,brow10}. 
Low-mass M dwarfs exhibit stronger chromospheric and coronae emissions for a given 
rotation period than more massive G--K dwarfs \citep[see fig.\ 6 in][]{kira07}.
On the other hand, rotation and magnetic activity of stars decrease with time 
as a consequence of magnetic braking, that is\ the angular momentum is extracted from the 
convective envelope and lost through a magnetized wind \citep{skum72,mest84,mest87}.

M dwarfs make up 70\% of the stars in the solar neighbourhood. They are
small, cool main sequence stars with temperatures in the range 2400--3800\,K
and radii between 0.10 and 0.63\,R$\odot$. Radii of M dwarfs are smaller than 
solar-like stars, implying that the flare activity phenomena are more pronounced 
at the background of their cool photospheres. M dwarfs with spectral types later 
than M4 have fully convective atmospheres with no boundary zone between the radiative 
and convective zones \citep{hawl14}, implying that we do not expect significant magnetic 
fields such as those generated in the Sun \citep{doyl18}.
The magnetic fields present at the surface of low-mass M dwarfs (M5--M8) originate
either from strong, axisymmetric dipole fields or from weak, higher-order multipole 
fields \citep{mori10, mori08}.

Some of the first detailed optical observations of stellar flares on M dwarfs were reported 
by \cite{bopp73} and \cite{gers83}; see also \cite{gers05}. X-ray observations with the
EXOSAT satellite detected the first flare in M dwarfs in the atmosphere of YZ CMi \citep{heis75}.
Since then, the physics of stellar flares has been studied by different teams over a wide
range of energy, from $\gamma$-ray to radio frequencies. 
The characterization of M dwarf flares has become possible thanks to the advent of large-scale
surveys such as the Sloan Digital Sky Survey \citep[SDSS;][]{york00}.
\cite{kowa09} identified flares in the SDSS Stripe 82, where the authors
found that active (\Ha{} in emission) M dwarfs flared at approximately 30 times the
rate of inactive M dwarfs. They also found that the flaring fraction increases for
cooler stars, which they attribute in part to a higher fraction of magnetically
active stars among late-M dwarfs. This is compatible with lower-mass stars rotating
more rapidly over a long time-span (and therefore magnetically active) than their more massive 
counterparts.
The presence of higher rates of flares in late-type M dwarfs impacts 
the exoplanet community because flares release large amounts of energy in the UV,
which could potentially affect the atmospheres and habitability of orbiting exoplanets.

Recent large-scale radial velocity (RV) searches for extra-solar planets 
\citep[][and references therein]{jenk17}
provide a new framework and motivation to study stellar and flare activity phenomena.
Indeed, they provide extensive time series of high-resolution spectra obtained 
by the best spectrographs on the biggest telescopes for large samples of stars. 
The study of activity in exoplanet science is of special interest because spots, 
plages, and other inhomogeneities of the stellar surface affect the shape of spectral lines, 
shift their centroids, and consequently bias the measured RV 
\citep[e.g.][and reference therein]{barn11}. Stellar and flare activity produce additional 
signals usually handled as random noise (the so-called RV jitter), which should be added 
to other known systematics such as photon noise and instrumental instabilities 
\citep{astu15, suar17}.

Today, M dwarfs represent important targets for searches of exoplanets.
Due to their low mass and luminosity, M dwarfs are ideal targets to find 
low-mass, Earth-size planets with the RV technique
\citep[see][and references therein]{silv12} and look for possible transits.
The first rocky planets around M dwarfs were detected by RV and transits 
\citep{rive15, char09}. We note that the first rocky planet Corot-7b orbiting a G9V star
was announced by the COROT team \citep{lege09,quel09}.

Proxima Cen was recently highlighted as a planet-host mid-M dwarf \citep{anglada_escude16}. 
Proxima Cen b orbits its host star with a period of 11.2 days, corresponding to a semi-major 
axis distance of 0.05 AU, and has a mass close to that of the Earth (from 1.10 
to 1.46  Earth masses) and orbits in the temperate zone \citep{anglada_escude16}. 
 
Proxima Cen is a known flare star that exhibits random but significant increases in brightness 
due to magnetic activity \citep{christian04a}. The
spectrum of Proxima Cen contains numerous emission lines \citep{fuhr11,pavl17}.
These features most likely originate from plages, spots, or a combination of both.
\citet{pavl17} found the presence of hot evaporated matter flow originating from the outer envelope.
The atmosphere 
of Proxima Cen shows a rather complicated structure consisting of the cool photosphere,
chromosphere, and hot outer layers where even lines of He{\small{I}} form. 
In general, the flare rate of Proxima Cen is lower than that of other flare stars 
of similar spectral type, but is unusually high given its slow rotation period \citep{dave16}.
The star has an estimated rotation period of $\sim$83 days and a magnetic cycle of 
about 7 years \citep{bene98,suar16,warg17}.
 The X-ray coronal and chromospheric activity have been studied in detail by \citet{fuhr11} 
and \citet{warg17}. Recently, \citet{thom17} claimed a detection of 
rotational modulation of emission lines in the spectrum of Proxima Cen. 

 \cite{west15} suggested that the activity-rotation relation for the slow rotating stars
may be more complicated than predicted by a simple spin-down model. One possibility could
be that while persistent magnetic activity in late-type M dwarfs is rare (or non-existent) 
for slow rotators, the presence of magnetic cycles may produce
observed activity in a small number of stars at any given time. As stated by
\cite{newt16}, many slowly rotating mid-to-late M dwarfs show variability
amplitudes of half a per cent or more, implying that they have maintained
strong-enough magnetic fields to produce the requisite spot contrasts. The lack
of correlation between rotation period and amplitude for these stars indicates
that the spot contrast is not changing significantly, even when they undergo
substantial spin-down.

To date, most of the rocky planets in the habitable zone have been found around these 
very low-mass stars \citep{udry07,bonf11,quin14,torr15,wrig16,anglada_escude16}.
However, the flare activity of the hosting stars may affect the structure and temperature 
regime of the exoplanetary atmospheres, thereby affecting the size of the habitable zone. 
In March 2016, the Evryscope observed the first  superflare visible to the naked eye detected from 
Proxima Cen \citep{howa18}. The brightness of Proxima Cen increased by a factor of 68 
during this superflare and released a bolometric energy of 10$^{33.5}$ erg, 
i.e. approximately ten times larger than any flare detected before. Modelling photochemical 
effects of NOx atmospheric species generated by the severe particle events in the 
atmosphere of Proxima Cen b 
shows that the high level of the similar extreme strong flare activity in the long term
is sufficient to reduce the ozone of an Earth-like atmosphere by 90\% within five years. 
In other words, a conception of habitable zone should be reconsidered taking into account 
the impact of flare activity on exoplanet atmospheres \citep{howa18}. 

In this paper, we focus on the temporal changes of emission lines in the optical spectra
of Proxima Cen. We describe the observations in Sect. \ref{_observ}. We detail our 
procedure in Sect. \ref{_proc}. Section \ref{_results} presents our results, which we discuss 
and put into context in Sect. \ref{_diss}.

%
%
\section{ Optical spectra of Proxima obtained with 3.6-m/HARPS}
\label{_observ}

We analysed 386 optical spectra obtained with HARPS, a fibre-fed, echelle, high-resolution 
spectrograph installed on the 3.6-m European Southern Observatory (ESO) telescope in the 
La Silla Observatory, Chile. The instrument offers a resolving power of R$\sim$115,000 over 
a spectral range from 3780 to 6810\AA{}. The dates cover a time interval of over 13 years,
from Barycentric Julian dates (BJD) between [245]3152.600, that is\ the night of May 27, 2004, 
and [245]6418.643, the night of September 30, 2017. In the remainder of the paper, we omit the prefix [245]. Most observations include one or two exposures per night, with some extended sets of 
exposures taken every $\sim$10 min at dates BJD\,=\,6417 (May 4, 2013; number of exposures on that
night is N$_t$\,=\,55), 6418 (May 5, 2013; N$_t$\,=\,64), 6420 (May 7, 2013; N$_t$\,=\,22), and 6426 
(May 13, 2013; N$_t$\,=\,9).
The HARPS observations cover an extended time interval providing a unique set of data 
to analyse the behaviour of stellar flares with time over a long period. 

%
\section{Procedure}
\label{_proc}
We discuss the temporal changes of a few emission lines listed in Table \ref{_t1}. 
These lines result from different excitation potentials. Their formation requires different 
physical conditions, which occur in distant parts of the active atmosphere of Proxima Cen.
Changes in the \pew{} and line profiles of these emission lines reflect the direct or indirect
impact of the flare activity on the whole stellar atmosphere, as well as their changes in time.

%
%
\begin{table*}
\caption{Emission lines studied in our paper.}
\label{_t1}
\begin{tabular}{ccccc}
\hline\hline 
Line        & $\lambda_{air}$ (\AA) & Transition & $E_l$ (eV)  &  Formation regions   \\
\hline
\Ha{}         & 6562.81        &     2P$^0$ - 3S,D                        &    10.199  &  Flares, facules     \\
Ca{\small{II}} (K)    & 3933.66      &   4$^2$S$_{1/2}$ - 4$^2$P$^{0}_{1/2}$)   & 0.0         & Chromosphere \\    
Ca{\small{II}} (H)    & 3968.47      &  4$^2$S$_{1/2}$ - 4$^2$P$^{0}_{3/2}$    & 0.0          &  Chromosphere \\
He$_{5876}$ &  5875.64   &    2$^3$P$_0$-3$^3$D  & 20.964                         &  Flares, transition zone \\
He$_{4026}$ &  4026.19       &   2$^3$P$_0$-5$^3$D                 &  20.964     &  Flares, transition zone \\
\Heps{}       & 3970.72       & 2P$^0$ - 7S,D                        &       10.199      &  Flares          \\
Na{\small{I}} (D$_1$) & 5889.95       & 3$^2$S$_{1/2}$ -3$^2$P$^{0}_{3/2}$   &        0.0    &  Chromospheres \\
Na{\small{I}} (D$_2$) & 5895/92       & 3$^2$S$_{1/2}$ -3$^2$P$^{0}_{3/2}$   &        0.0    &  Chromospheres \\
\hline
\end{tabular}
\end{table*}

We re-normalised the final 1D observed spectra to get the same flux in the background of 
these emission lines. This allows us to investigate the relative changes of \pew{} of emission 
lines with time. As shown in \cite{pavl17}, the changes of the background of emission lines 
appear rather marginal. In other words, activity processes do not affect photospheric levels, 
where the most absorption features and continuum form. 

We measured \pew{} of all emission lines listed in Table \ref{_t1}. Here we note that all these 
emission lines in the spectrum of Proxima Cen were analysed on the same spectra and, 
thus, at the same times. To investigate the temporal correlation between different emission lines formed in the flare events, we computed the Pearson $k_p$ and Stockman $k_s$ 
rank correlation coefficients to evaluate the changes in \pew{} of our emission lines versus \Ha{}:

$${\displaystyle k_p={\frac {n\sum x_{i}y_{i}-\sum x_{i}\sum y_{i}}
{{\sqrt {n\sum x_{i}^{2}-(\sum x_{i})^{2}}}~
{\sqrt {n\sum y_{i}^{2}-(\sum y_{i})^{2}}}}},} $$

\begin{equation}
\begin{aligned}
k_s = {\frac{\sum(R(x_i) - (n+1)/2)\times(R(y_i ) - (n+1)/2)} 
 {(\sum(R(x_i ) -  (n+1)/2))^2 \times \sum(R(y_i) - (n+1)/2))^2 )^{0.5}}} 
\end{aligned}
.\end{equation}

These coefficients change in the range  $-$1\,$\leq$\,$r_{xy}$\,$\leq$\,1\@.
By definition, the cases $k_{p}$, $k_{s}$\,=\,1, 0, $-$1 correspond to the strong
correlation, no correlation, and strong anti-correlation, respectively.

%
%
\begin{figure*}
  \centering
  \includegraphics[width=0.3\linewidth, angle=0]{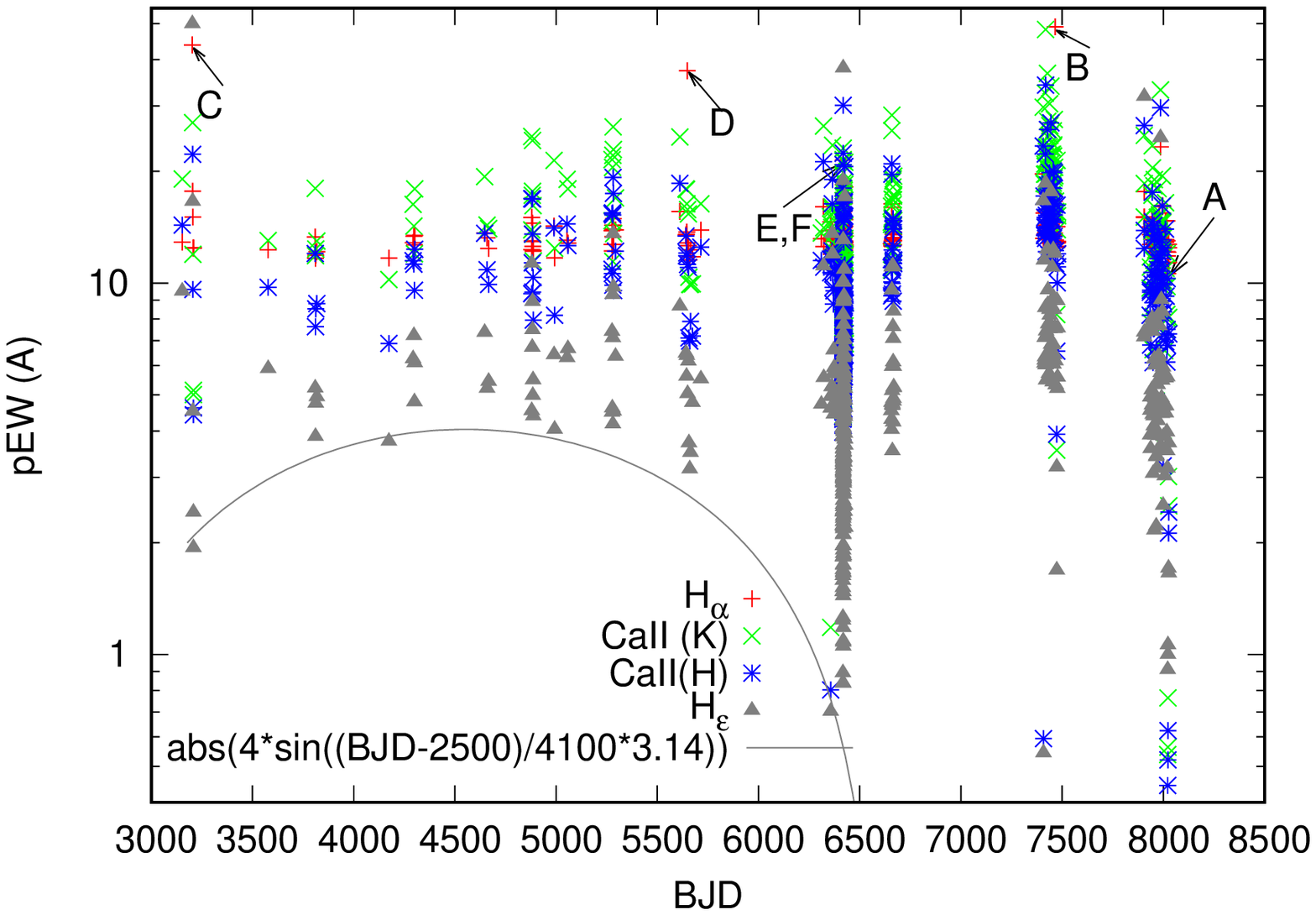}  
  \includegraphics[width=0.3\linewidth, angle=0]{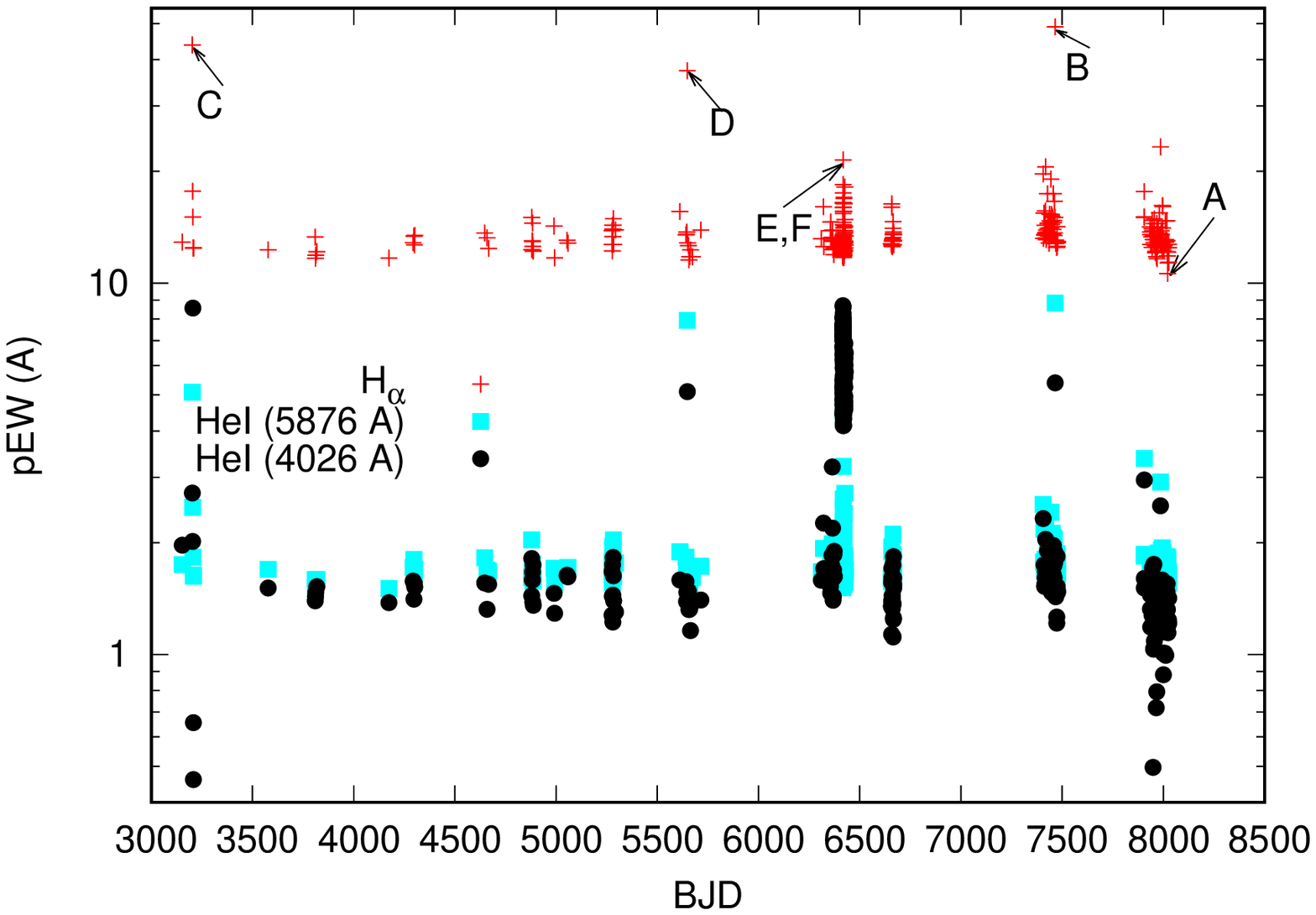}  
  \includegraphics[width=0.3\linewidth, angle=0]{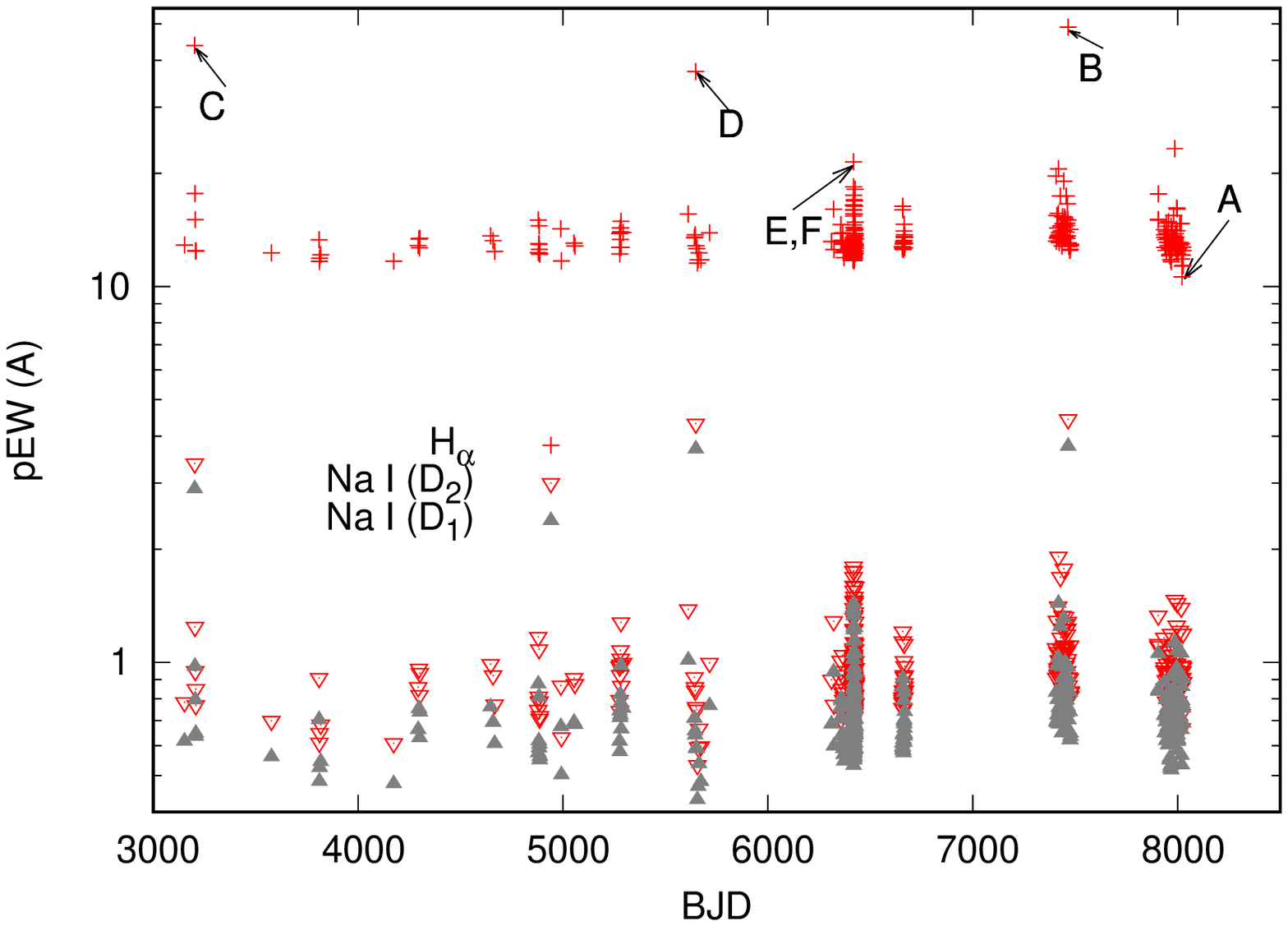}  
\caption{Variations of \pew{} of emission lines at all dates of observations. 
The letter `A' marks the absolute minimum of the \pew{} of \Ha{} (Section \ref{_sminn}), 
letters `B', `C' and `D' label the flares occasionally observed at BJD\,=\,7465.8823, 
3202.5647, and 5647.7300, respectively (Sect. \ref{_sminn}) and letters `E' and `F' label 
the \pew{} of two \Ha{} flares observed across BJD\,=\,6417 (Sect. \ref{_sf}). Grey line
shows the approximation of the set of minimum values of \pew of \Heps by a sinusoid; see Sect.
\ref{_minHeps}.}
   \label{_allpew}
\end{figure*}
%

%
%
\section{Results}
\label{_results}
The variability of the \pew{} of the selected emission lines on different dates is depicted in 
Fig.\ \ref{_allpew}. These \pew{} were determined for one dataset of the observed spectra. 
We draw our attention to a few expositions marked with the letters A, B, C, D, E, and F:

\begin{itemize}
\item The time mark `A' locates the minimum level of activity of Proxima Cen, determined 
by the lowest \pew{} of \Ha{} (Sect. \ref{_sminn}).
\item The letters B, C, and D mark three flare events during observations on BJD\,=\,6417 
and 6418 (Sect. \ref{_smaxx}).
\item The letters F and E mark two peaks of the `strong' flare seen in \Ha{} at BJD\,=\,6414
(Sect. \ref{_ext}). 
\end{itemize}

\subsection{Long-term variatons of min \Heps}
\label{_minHeps}

The temporal changes of the strength of the \Heps{} line provide some 
evidence about the presence of long-term variability in Proxima Cen. 
The emission in \Heps{} is weaker in comparison with \Ha{}, still  its relative intensity 
is more sensitive to the level of activity  in comparison with other lines.
We derive a quasi-periodic variability of $\sim$\,8200 days by approximating the minimum 
values of the observed \pew{} of \Heps \ at the dates BJD \,=\,3000--8000 by the relation
$4*sin(2*\pi*(BJD-2500)/8200)$.

\subsection{Temporal changes of the emission lines}
\label{_sf}

We draw our attention to the temporal variability of emission lines in the spectrum of Proxima 
Cen. We argue that if all lines are formed in a single flare region of high temperature, we 
should see a strong correlation in their behaviours.

For simplicity, we adopt that the level of activity of Proxima Cen corresponds to the
 \pew{} of \Ha{} measured at a given date, that is, the minimum activity level of Proxima Cen
 corresponds to the minimum of the \pew{} of \Ha{} and vice versa. 
Indeed, the formation of \Ha{} seen in emission is possible only in high-temperature 
regions heated by the flare events. 

To analyse the relative temporal behaviour of \Ha{} and other lines we computed the Pearson 
and Stockman correlation coefficients for each \pew{} (Table \ref{_corr1}).
We note that the spectra observed between BJD\,=\,6417 and 6429 were obtained with the 
Fabry-Perot wavelength calibration which generates some flux distortion at the exact position 
of the He{\small{I}} at 4206 \AA{}. Therefore, we excluded this line from our analysis for those dates.

In Table \ref{_corr1}, we show the result of computations of the Pearson and Stockman 
correlation coefficients of the emission line strength with \Ha{} for different time intervals. This was carried out (1) for all dates covered our interval of observations, from 3152.6001 to 6418.6431, 
except for BJD\, =\, 6417 and 6429 in the case of the \heiv{} line, and (2)
for two extended  continuous sub-sets of observations of our lines except \heiv{} on 
BJD\,=\,6417 and 6418 (Sect.\ \ref{_ext}).

In all cases we derive high positive values for the two correlation coefficients. 
In other words, we see that flare processes modulate the strength of all lines. Indeed, 
all lines follow the changes in intensity of \Ha{}. The He$_{4026}$ line shows the weakest
correlation with \Ha{}. We interpret this correlation as evidence that all lines 
are formed in or are affected by the same flare events which covered the broad range of 
heights in the stellar atmosphere. 

In Fig.\ \ref{_2mr}, we show the response of other lines on changes in the \pew{} of \Ha{}. 
In fact, we can see three groups of lines with different correlation with \Ha{}, which 
can be roughly characterised by the gradient $b\,=\,d$(\pew{})/$d$(\pew{}(\Ha{})).
Ca{\small{II}} H, K, and \Heps{} show larger $b$, while both helium lines show 
the weakest dependence; the relative response of Na{\small{I}} 
is stronger than \heiv{}, but weaker than Ca{\small{II}} H and K.

For convenience we implement a rather phenomenological classification of the strength 
of flares using quantitative criteria based on the values of the \pew{} of \Heps{}
(Table \ref{_Hepsi}; Fig.\ \ref{_2mr}).
\begin{table}
\label{_Hepsi}
\caption{Classification of the strength of flares adopted in this paper}
\begin{tabular}{cc}
\hline\hline
Classification & \pew(\Heps) \\
\hline
0$_\epsilon$ & 0 \pew(\Heps) $<$ 2 \AA{} \\
1$_\epsilon$ & 2 $<$ \pew(\Heps) $<$ 4 \AA{} \\
2$_\epsilon$ & 4 $<$ \pew(\Heps) $<$ 16 \AA{} \\
3$_\epsilon$ &    \pew(\Heps) $>$ 16 \AA{}  \\
\hline
\end{tabular}
\end{table}

We favour the use of \Heps{} to classify flares because its \pew{} is strongly sensitive
to the strength of flares. At the minimum of the flare activity, this line almost disappears
(see Sect.\ \ref{_sminn}). At maximum activity, the \pew{} of \Heps{} might even
be higher than the \pew(\Ha), see
Fig. \ref{_2mr}. However, this is only true for \pew{} computed 
with respect to the local pseudo-continuum, which can itself change during the flares
(Sect. \ref{_fa}).
Nonetheless, the equivalent widths of \Ha{} always exceed the equivalent widths of \Heps{}
due to the exponential drop of the flux towards the blue part of the spectrum.

%
%
\begin{table*}
\caption{Pearson and Stockman correlation coefficients of the computed \pew{} of the 
emission lines considered in this work vs. the \pew{} of \Ha{}. Dates between 
BJD\,=\,6417 and 6429 were not accounted for \heiv{} line (see Sect. \ref{_sf}).
}
\label{_corr1}
\begin{tabular}{ccccccc}
\hline\hline
 & \multicolumn{2}{c}{all, N=386} & \multicolumn{2}{c}{6417$^d$, N=55} & \multicolumn{2}{c}{6418$^d$, N=23} \\
\hline
\Ha{} vs.\ line  & $k_p$ & $k_s$  &  $k_p$ & $k_s$  &  $k_p$  & $k_s$   \\
\hline
Ca{\small{II}} (K)       &  0.8045 & 0.7229 &  0.9531 & 0.7832 & 0.8042 & 0.7955 \\
Ca{\small{II}} (H)       &  0.8897 & 0.7847 &  0.9753 & 0.8188 & 0.8068 & 0.8098 \\
He$_{5876}$             &  0.9288 & 0.8389 &  0.9377 & 0.7286 & 0.6407 & 0.6055 \\
He$_{4026}$             &  0.5265 & 0.4420 &              &             &             &           \\
\Heps{}                      &  0.9068 & 0.7506 &  0.9210 & 0.8353 & 0.5907 & 0.6118 \\
Na{\small{I}}(D$_2$) &  0.9397 & 0.8824 &  0.9456 & 0.7877 & 0.9673 & 0.9555 \\
Na{\small{I}}(D$_1$) &  0.9496 & 0.8832 &  0.9489 & 0.8108 & 0.9628 & 0.9617 \\
\hline
\end{tabular}\\
$^*$ computed without dates  between BJD\,=\,6417 and 6429\@.
\end{table*}
%

%
%
\begin{figure*}
  \centering
  \includegraphics[width=0.33\linewidth, angle=0]{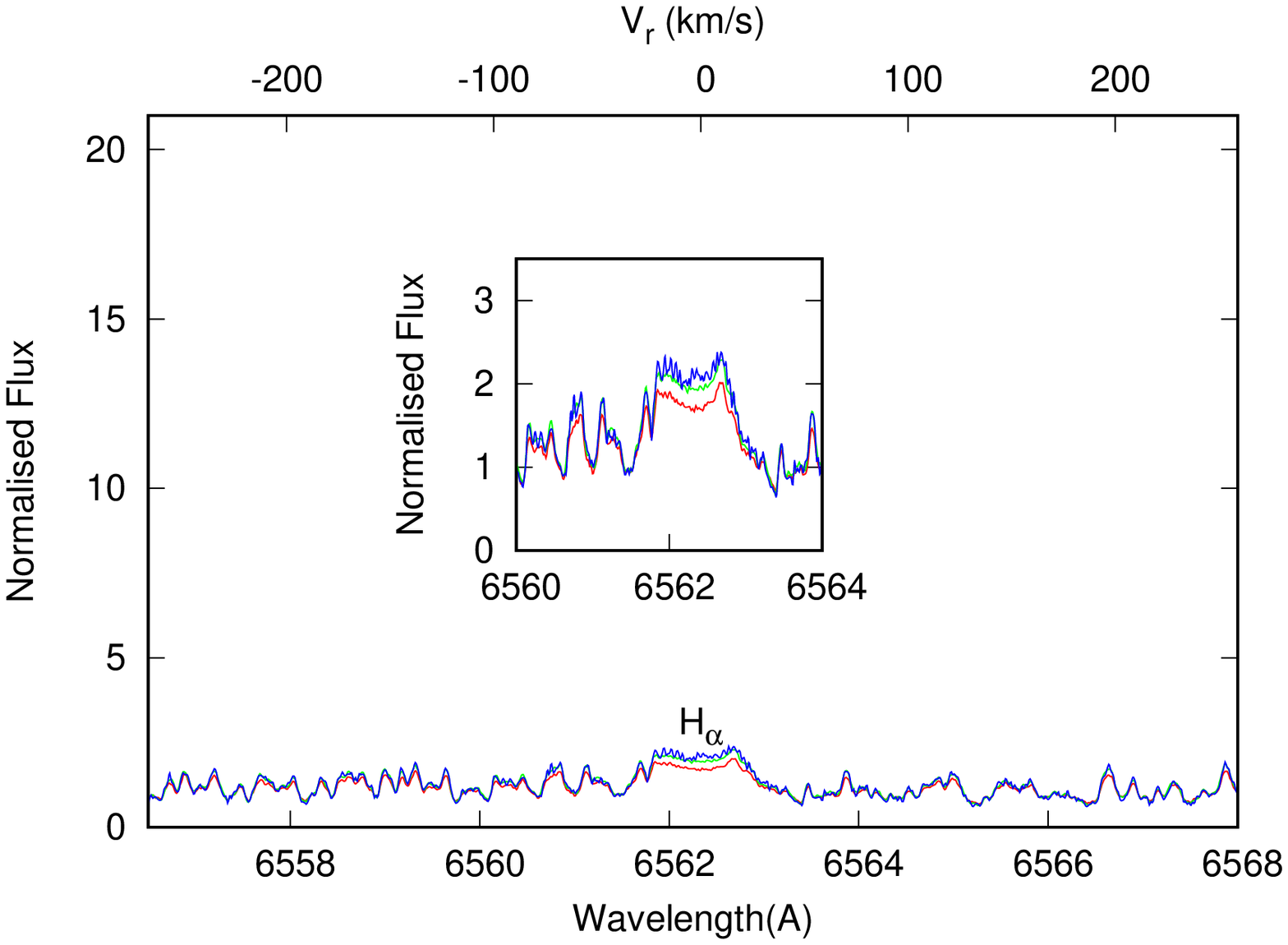}  
  \includegraphics[width=0.33\linewidth, angle=0]{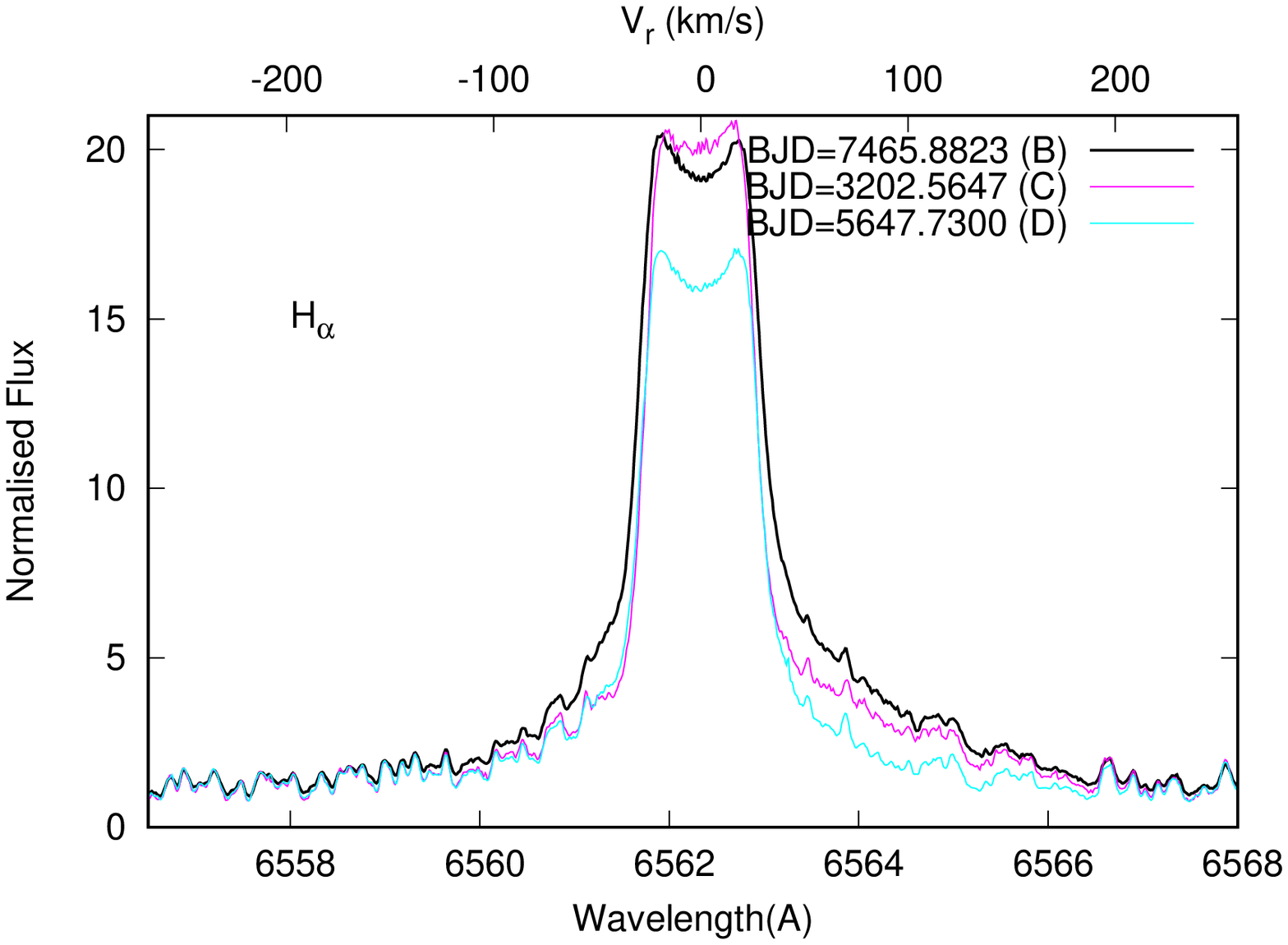}  \\
  \includegraphics[width=0.33\linewidth, angle=0]{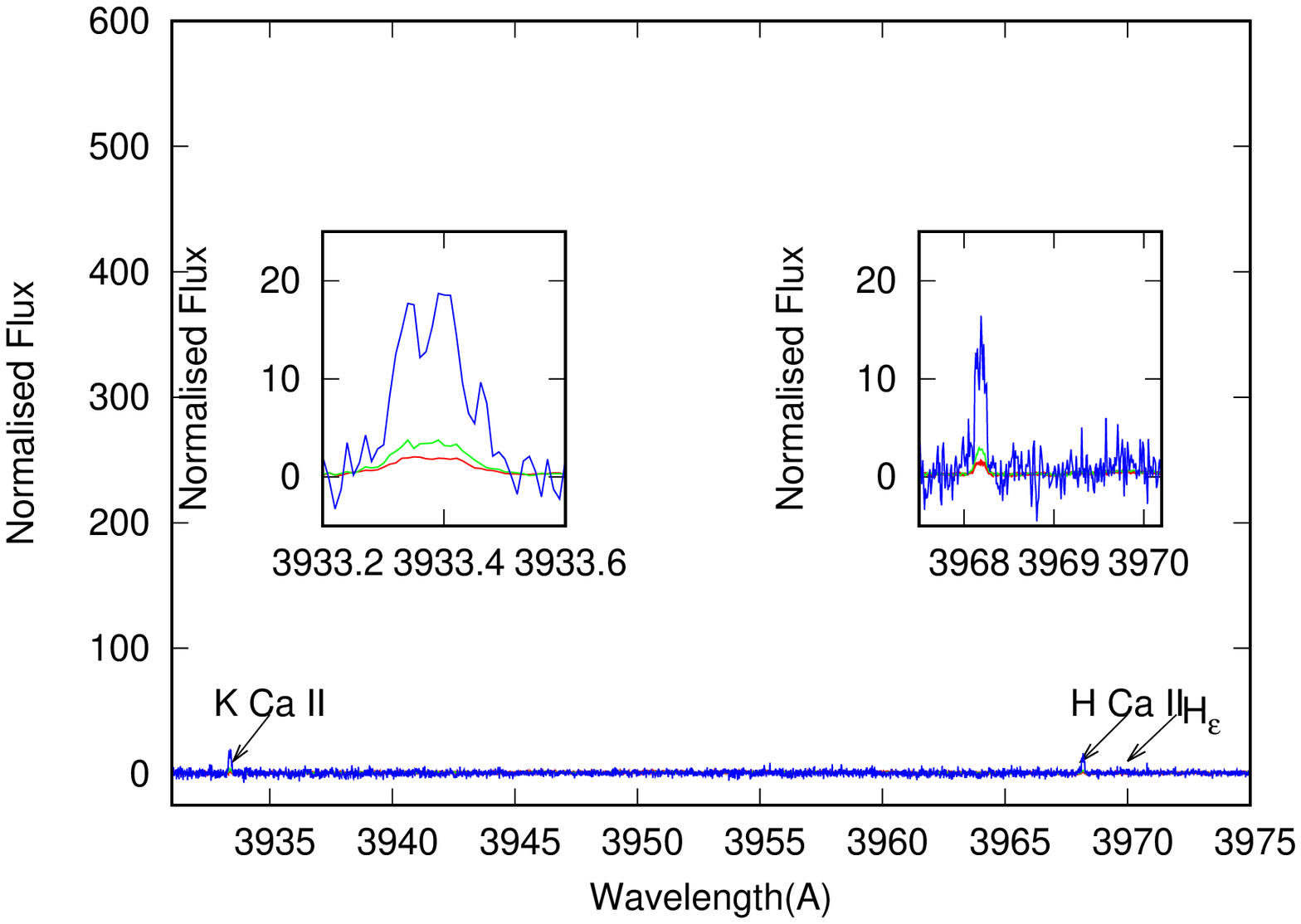}    
  \includegraphics[width=0.33\linewidth, angle=0]{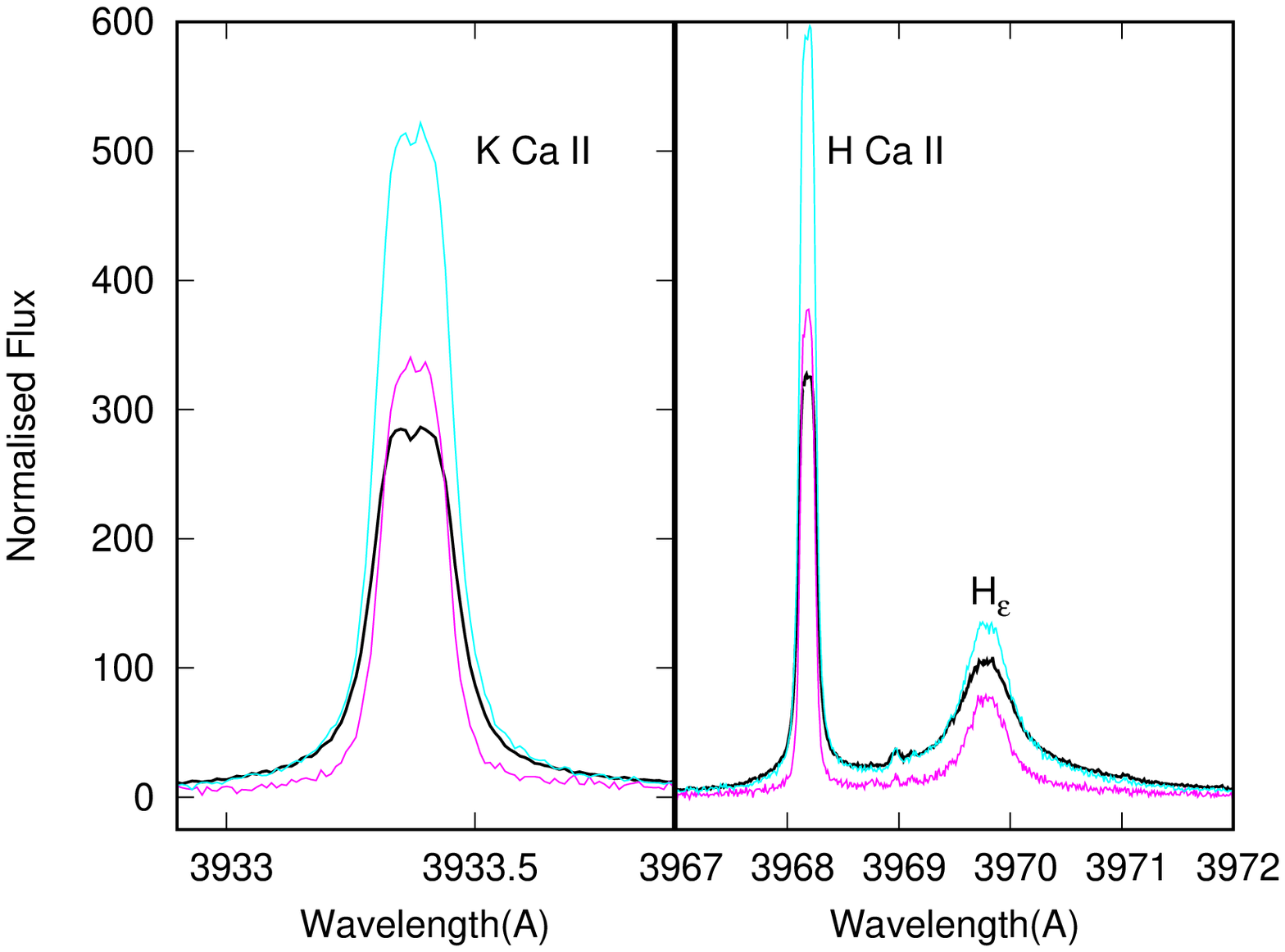} \\
  \includegraphics[width=0.33\linewidth, angle=0]{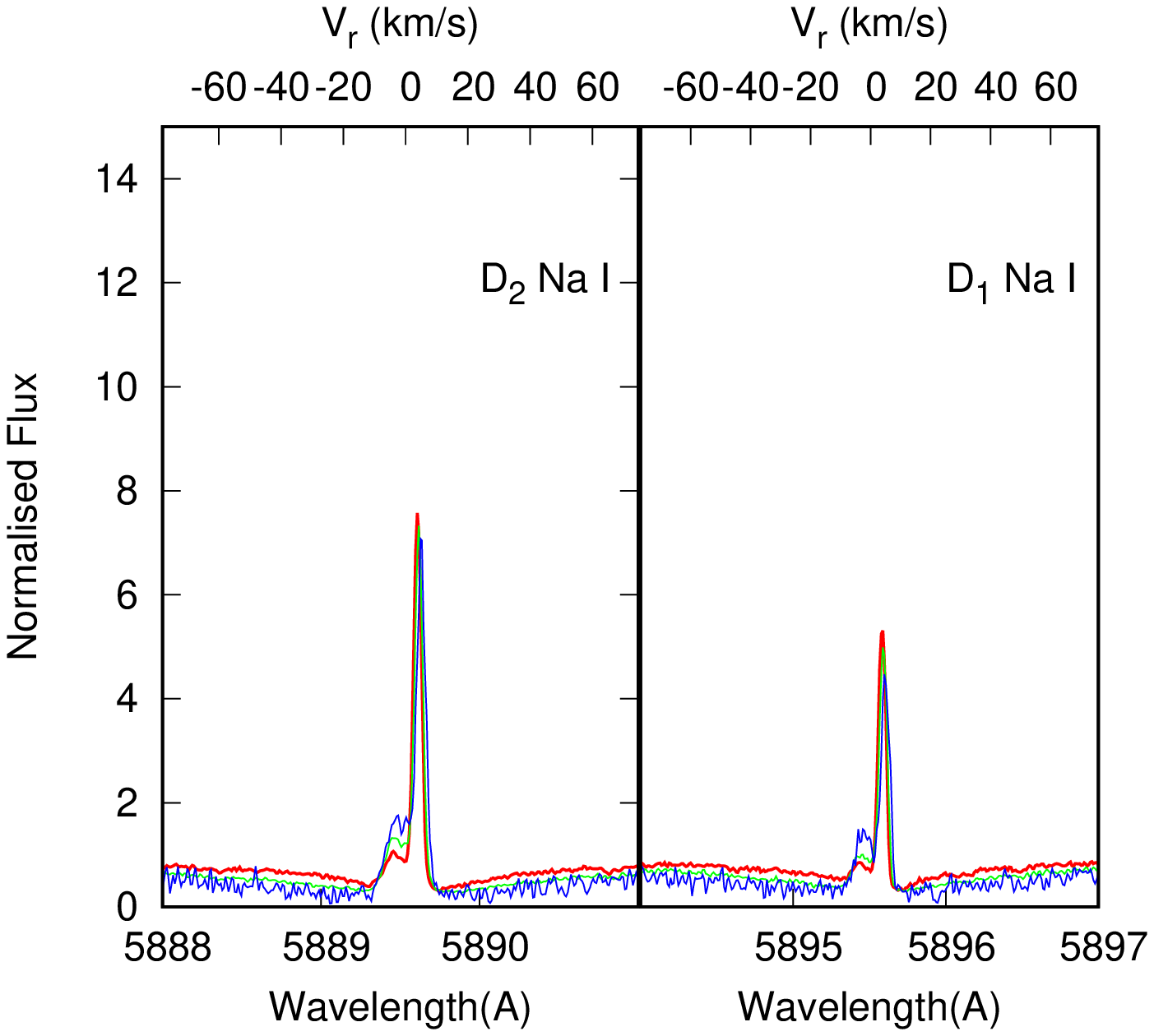}  
  \includegraphics[width=0.33\linewidth, angle=0]{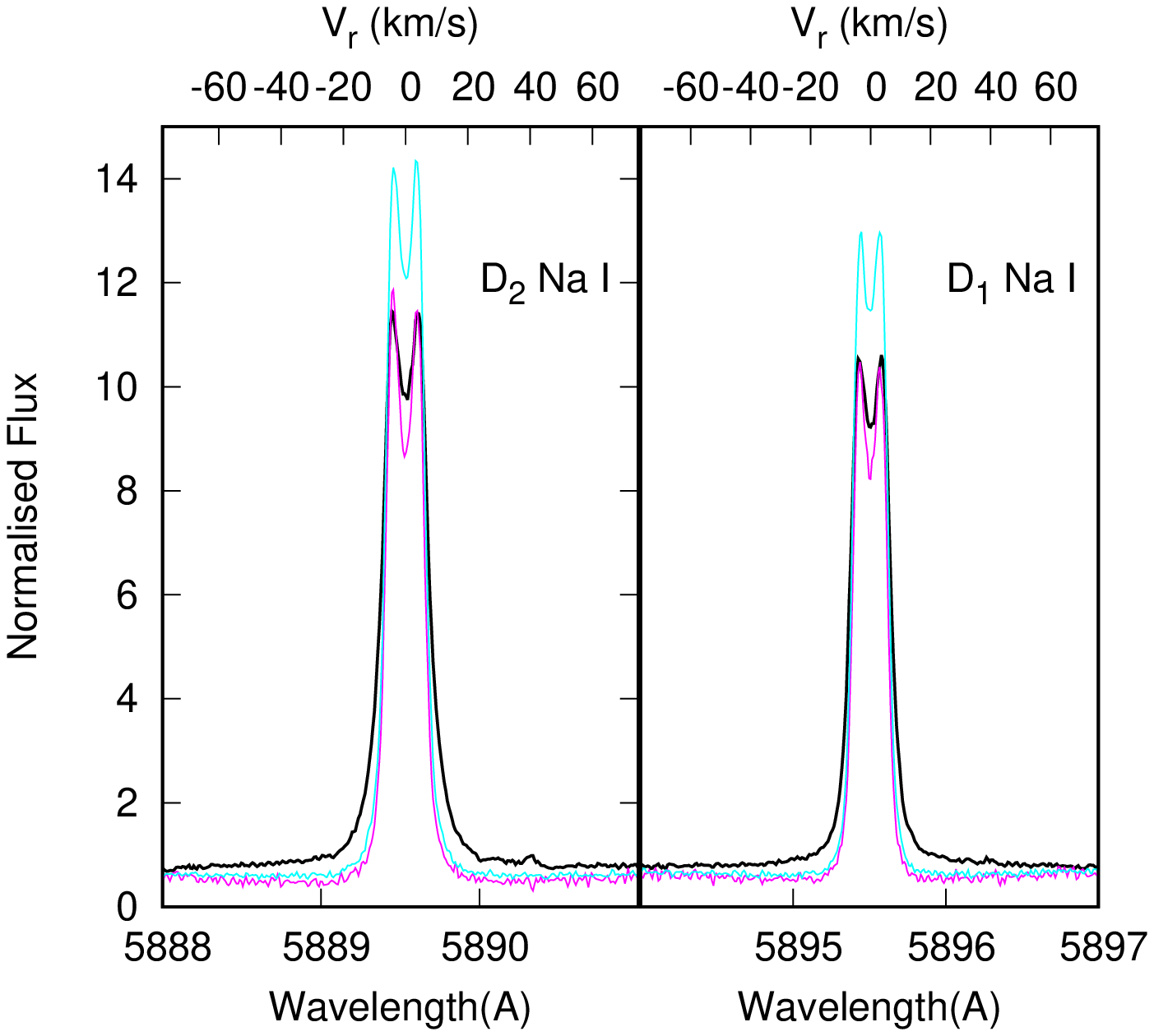}  \\
 \includegraphics[width=0.33\linewidth, angle=0]{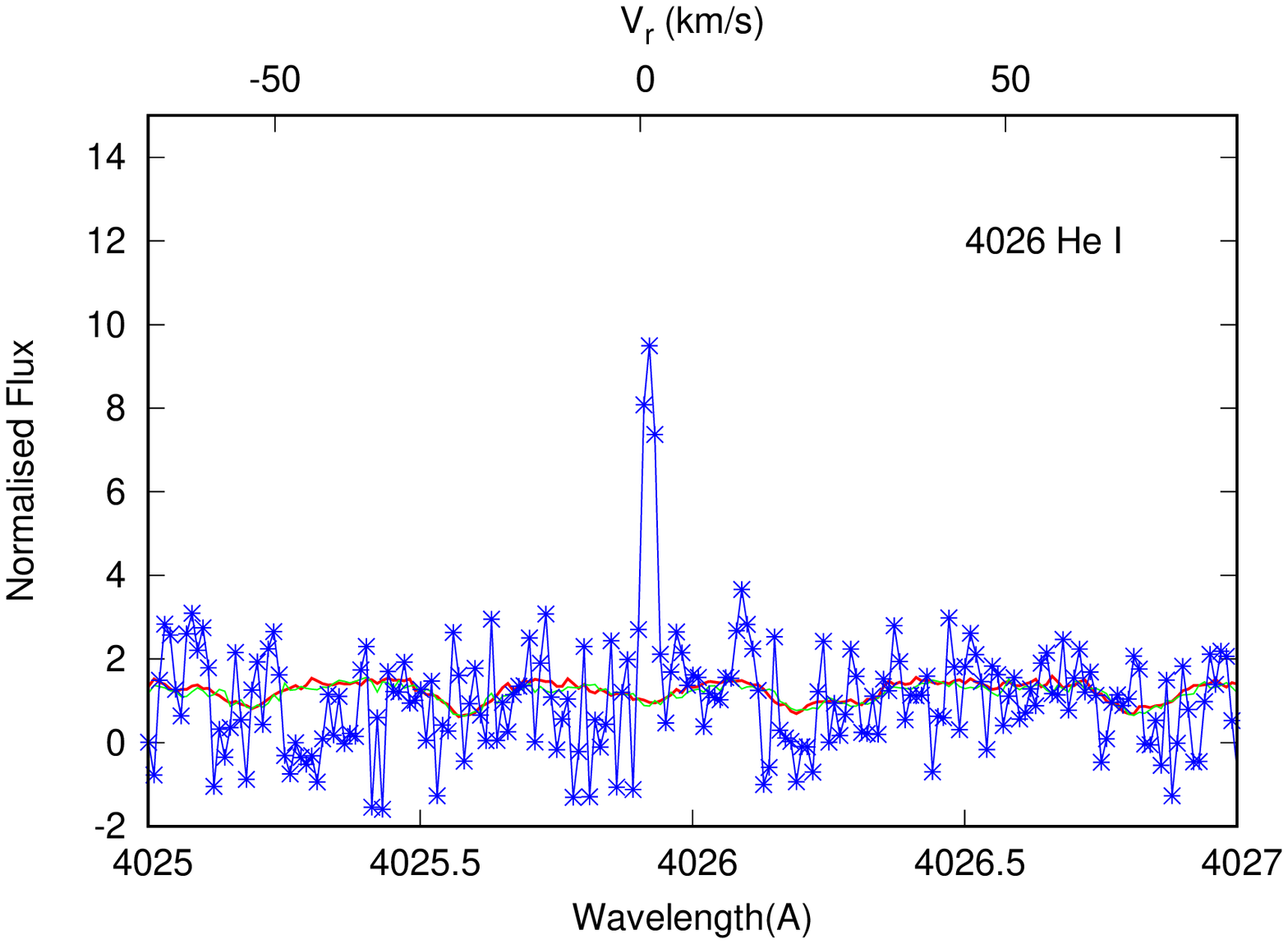}  
  \includegraphics[width=0.33\linewidth, angle=0]{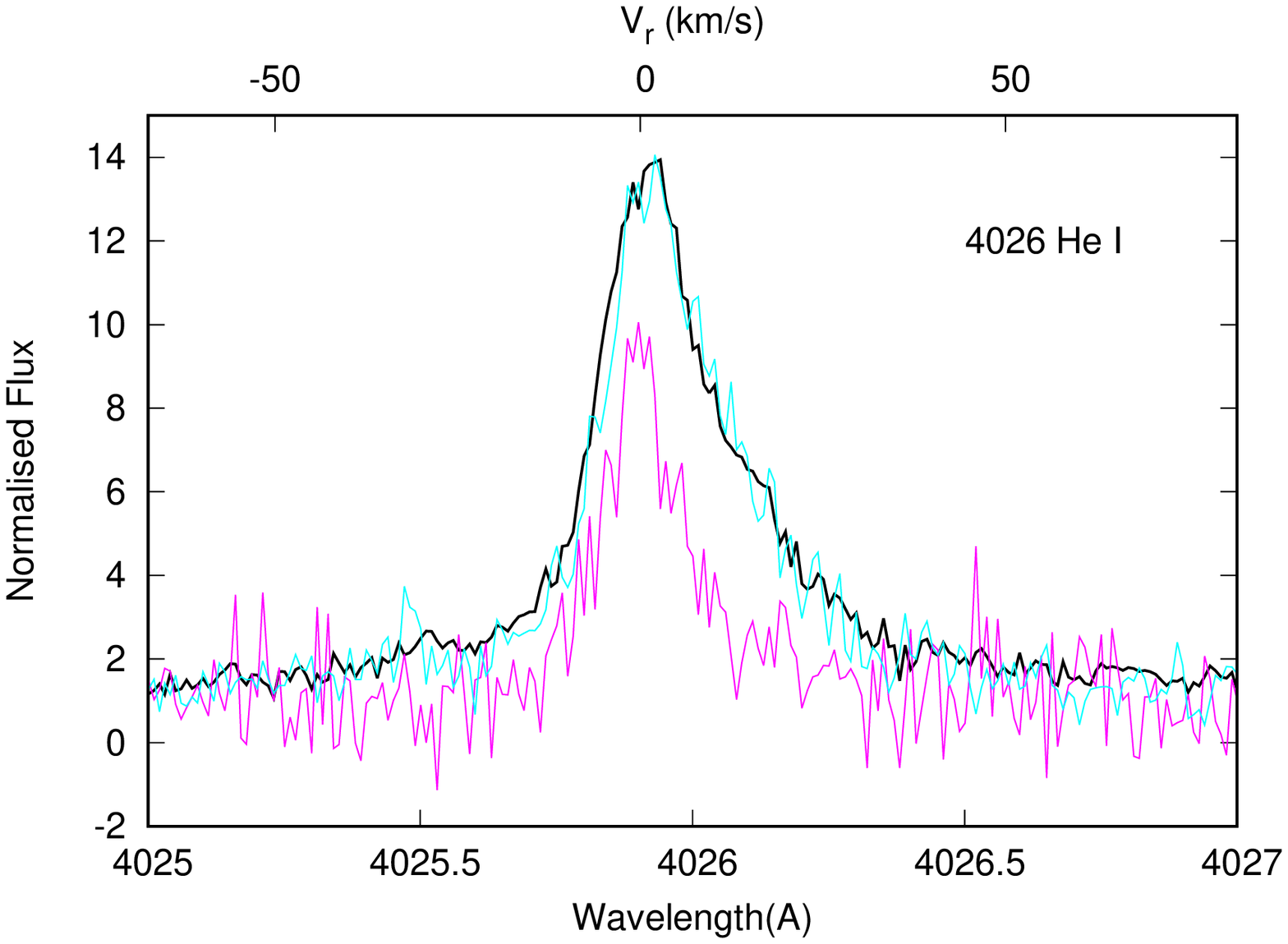} \\
  \includegraphics[width=0.33\linewidth, angle=0]{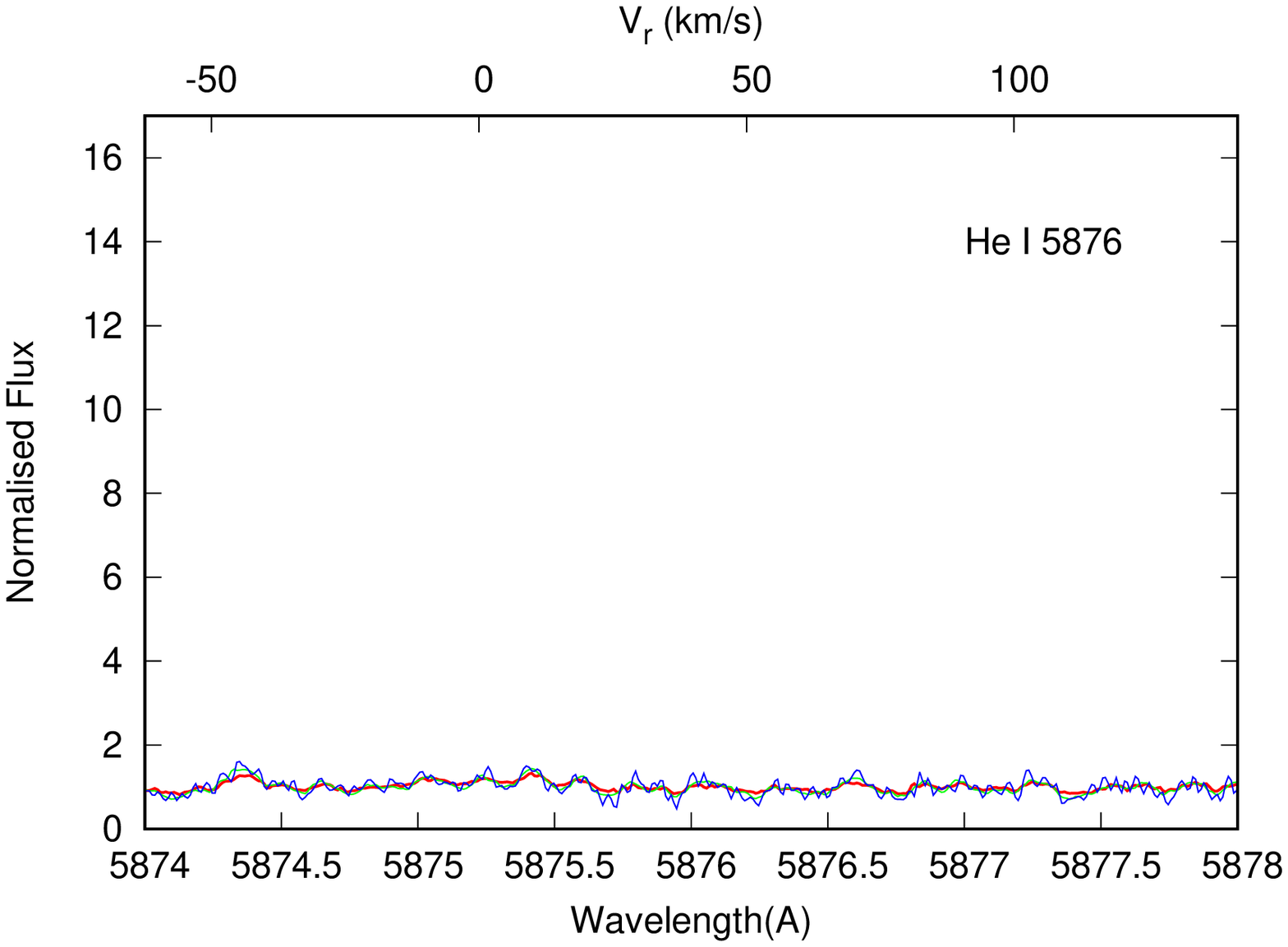}  
  \includegraphics[width=0.33\linewidth, angle=0]{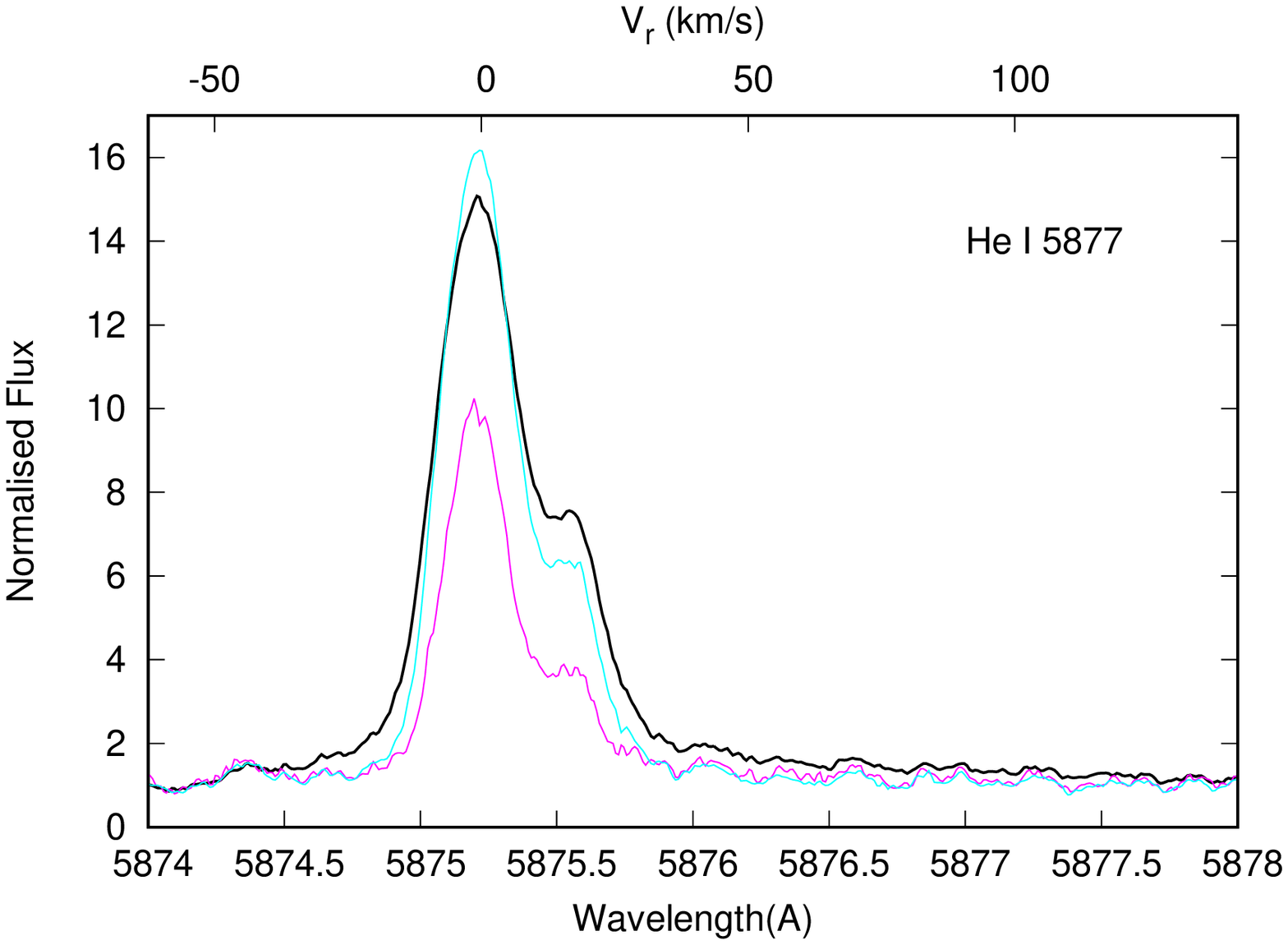}  
  \caption{{\it Left panel:} Profiles of several emission lines, including \Ha{}, 
  Ca{\small{II}}  K and H,  Na{\small{I}} D$_2$ and D$_1$, \hev{}, and \heiv{} at the three dates of 
  minimum activity of Proxima Cen: BJD\,=\,8021.4600 (red), 8023.4639 (green), and 
  8027.4785 (blue). These dates are marked with letter `A' in Fig.\ \ref{_allpew}. 
{\it Right panel:} `Same lines at the dates of maximum of activity of Proxima Cen, 
i.e. BDJ\,=\,7465.8823 (B; black), 3202.5647 (C; cyan), 5647.7300 (D; magenta)
displayed in Fig.\ \ref{_allpew}}.
   \label{_minmax}
\end{figure*}

\subsection{Photosphere, chromosphere, and flare regions}
 In this paper, we discuss different aspects of the flare activity of Proxima Cen.
For simplicity, we adopt that the atmosphere of Proxima Cen consists of three parts:

-- The photosphere is the deepest part of the atmosphere,
in which molecular and atomic absorption spectrum of Proxima. 
The photosphere of Proxima seems to be stable; flares do not affect its structure, 
because molecular lines are practically not affected by the flare events.

-- The chromosphere lies above the photosphere and the temperature increases
outwards. Here the formation of most emission lines occurs in NLTE\@. 
The chromosphere is a quasi-stable structure 
heated by the dissipation of non-thermal flows from the convective photosphere.
Chromospheric lines are comparatively narrow due to absence of large-scale moves.

-- According to modern notions, the flares are located in the outer part of a chromosphere and/or in the corona. 
Again, for simplicity we adopt flare conception described in \cite{yoko98}. 
Just after an avalanche-like process of the release of energy due to the re-connection of magnetic field 
loops a small volume of the atmosphere of Proxima Cen (up to a few percent of 
the surface in the case of strongest flares) is heated up to 10$^6$\,K or more. Afterwards, 
the upward flows of hot matter from chromosphere (evaporation) and downward flows 
of cooling matter (condensation) toward photosphere are formed. 
Due to the higher temperatures in the flare regions, thermal and non-thermal motions are much higher
and emission lines are broader than chromospheric lines.
We follow this simplified model to avoid the uncertainty of interpretation of our observed 
data due to different conceptions based on the numerous theoretical models. In reality,
a flare is a multi-dimensional magnetohydrodynamic process, which is variable on short timescales. 
Current models of flares are much more complicated \citep[][and references therein]{zhan16};
the analysis of which is beyond the scope of this paper.

\subsection{Emission lines at the minimum activity of Proxima Cen}
\label{_sminn}

The photospheric spectrum at least in the optical and near-infrared spectral 
region shows rather marginal response to the flares \citep{pavl17}.
On the contrary, both profiles and intensities of our emission lines 
respond to the changes of activity; see Fig. \ref{_allpew}.

Firstly, we find the three lowest \pew{} of \Ha{} in the spectrum of Proxima Cen, with
\pew{} of 10.62\AA{}, 11.34\AA{}, and 11.39\AA{} at dates BJD\,=\,8021.4600, 8023.4639, 
and 8027.4785, respectively. The exposure taken on BJD\,=\,8021.4600 is marked 
by `A' in Fig.\ \ref{_allpew}. These three dates differ by only a few days and show a 
minimum activity level of Proxima Cen.

The left-hand side panel of Fig.\ \ref{_minmax} compares the level of activity of the selected
lines on those three dates of minimum activity of Proxima Cen.
The right-hand side panel of Fig.\ \ref{_minmax} shows the relative change of these lines 
between the times of maximum and minimum activity. 

At times of minimum activity, we find the following.
\begin{itemize}
\item We still see an emission component in the blue wing of \Ha{} with $V$\,=\,$-$30 km/s 
\citep{pavl17}, formed by the rapidly moving evaporating plasma, see \citep{abbe99}. 
In the core of the \Ha{} emission, we see a self-absorption feature formed by the flows of 
neutral hydrogen moving towards and away from the observer; see the top-left panel of Fig.\ \ref{_minmax}.
\item The interesting phenomenon is the almost complete disappearance of emission
Ca{\small{II}} H and K  as well as emission of \Heps{} (second plot from the 
top in the left panel of Fig.\ \ref{_minmax}).
Then, at the moment of the relatively stronger \Ha{} we see the weaker 
emission of both Ca{\small{II}} H and K lines. 
\item  A notable difference in the H$_{\epsilon}$ and Ca{\small{II}} H lines in 
Fig.\ \ref{_minmax} (right panel) is the broadening. \citet{paul06} discussed the broadening
of Balmer lines formed during flare events in the Barnard star and noted some problems in the 
physical explanation of their observed widths. However, in our case the Ca{\small{II}} lines 
and \Heps{} form in different parts of the atmosphere of Proxima Cen. Ca{\small{II}} H and K lines 
are collisionally controlled lines that form in the quasi-stable chromosphere whereas Balmer 
lines are lines controlled by the photoinisation; they form in the high-temperature flare region(s) 
where dispersion of thermal and/or 
non-thermal velocities is larger, generating broader emission lines \citep{thom57}.
\item The emission in the cores of the Na{\small{I}} $D_1$ and $D_2$ doublet correlates 
with the \Ha{} line (third left panel from the top of Fig.\ \ref{_minmax}). 
The emission cores of Na{\small{I}} D$_1$ and D$_2$ lines are still strong at minimum 
activity although two times weaker than at maximum activity (panels in third row from the top
in Fig. \ref{_minmax}).
Furthermore, these emission lines at the minimum of flare activity are affected by the 
absorption of cool condensed matter moving outwards from the star. Indeed, blue parts of 
the emission lines are more affected by absorption. In other words, moving outwards cool 
 condensed matter `eats' the blue part of the Na{\small{I}} emission lines.  
\item At the position of the He{\small{I}} line at 4026 \AA{}, we observe a narrow feature 
in the spectrum taken on BJD\,=\,8027.4785 (fourth plot on the left panel  
of Fig.\ \ref{_minmax}). However, the spectrum in the blue is noisier compared to other days. 
On the other hand, this `nebular-like' line is at least four pixels wide, and therefore it may be 
a real feature formed somewhere in the outer atmosphere of the star in specific circumstances. 
Unfortunately, we cannot investigate the temporal evolution of this feature to verify its origin.
\item \hev{}-like emission is absent on the days of minimum activity (left bottom panel of 
Fig.\ \ref{_minmax}).  \citet{paul06} noted that the \hev{} line only appears during 
strong-enough stellar flares. In other words, the lack of \hev{} at the minimum of activity 
corresponds to a low level of activity. We note that this line is located in the optical 
spectral region where observed {\it photosphetic} fluxes show little variation from quiet to flare modes; 
see Sect. \ref{_fa}.  
\end{itemize}

The time interval of the minimum activity of Proxima Cen likely extends for a few days to a
few weeks. In other words, at the times of the last observations made in 2017, we 
observe the star in the state of absolute minimum activity.

\subsection{Emission lines at the maximum activity of Proxima Cen}
\label{_smaxx}

During the maximum activity of Proxima Cen, we observe all our lines in emission
(right-hand panel of Fig. \ref{_minmax}). The three strongest flares are 
highlighted in Fig.\ \ref{_allpew} with the letters B, C, 
and D\@. All belong to the 3$_{\epsilon}$ flares:
\begin{itemize}
\item 
During these strong flares,  we see the fine structure of the self-absorption in the core of 
the strong \Ha{} emission line, which can be interpreted as evidence of small-scale neutral hydrogen flows in the outer part of the  
flares.
The strongest \Ha{} profile shows a well-defined red asymmetry 
due to inward movement of hot matter (top-right panel of Fig.\ \ref{_minmax}). 
We measure a velocity \Vr{} of 150 km/s for the B
flare, where \Vr{} is the motion of the hot plasma from the flare regions.
\item The strength of the Ca{\small{II}} K and H emission lines anti-correlates with \Ha{}. 
The strongest emission of H and K lines is observed at the same time as the `D' flare, 
the third strongest flare by intensity (second plot on the right panel of Fig.\ \ref{_minmax}).  
Interestingly, we find the self-absorption in the cores of Ca{\small{II}} H and K lines in all 
B, C, and D flares. The self-absorption is most likely formed in the outer, colder condensed flows, 
containing a number of Ca{\small{II}} ions.
\item The strength of the emission of the \DO{} and \DT{} Na{\small{I}} lines also 
anti-correlates with \Ha{} (third plot on the right panel of Fig.\ \ref{_minmax}). 
However, the strongest flare (B) provides the extended wings 
with \Vr{} up to $\pm$80 km/s. In this case the formation region of 
the H and K line wings is likely affected by the strong turbulence coming from the flare region. 
\item The strong flares generate strong helium emission lines of \heiv{} and \hev{}
(fourth and bottom plots on the right-hand side of Fig. \ref{_minmax}). 
We see that both lines are stronger when \Ha{} is stronger, although this correlation 
remains rather qualitative. The second strongest \Ha{} flare provides the weakest
He lines (lower left-hand panels of Fig. \ref{_minmax}). 
Again, in the case of \hev{} line, we witness an extended red wing up to \Vr{}\,=\,100 km/s, likely 
due to the infalling hot matter on the star.
\end{itemize}

\subsection{S/N and accuracy of the \pew measurements.}
\label{_fa}

We analyse spectra of Proxima Cen obtained by different authors on different dates.
One of the key questions is whether or not our measurements are reliable, and  therefore whether or not we can compare the
\pew{} obtained on different dates during different levels of activity. To answer this question, 
we compared spectra observed in the blue ($\lambda < 4200$ \AA)
and optical (4200~\AA $<\lambda<$~8000~\AA)
spectral regions at different levels of Proxima activity. 
The flux of M dwarfs drops towards the blue spectral region, and therefore the quality of the 
observed spectra in the blue and optical regions may differ simply due to `natural' reasons.
Furthermore, the blue-end of the spectrum is more sensitive to changes in the physical properties 
of emitted matter. 

\begin{figure*}
  \centering
 \includegraphics[width=0.48\linewidth, angle=0]{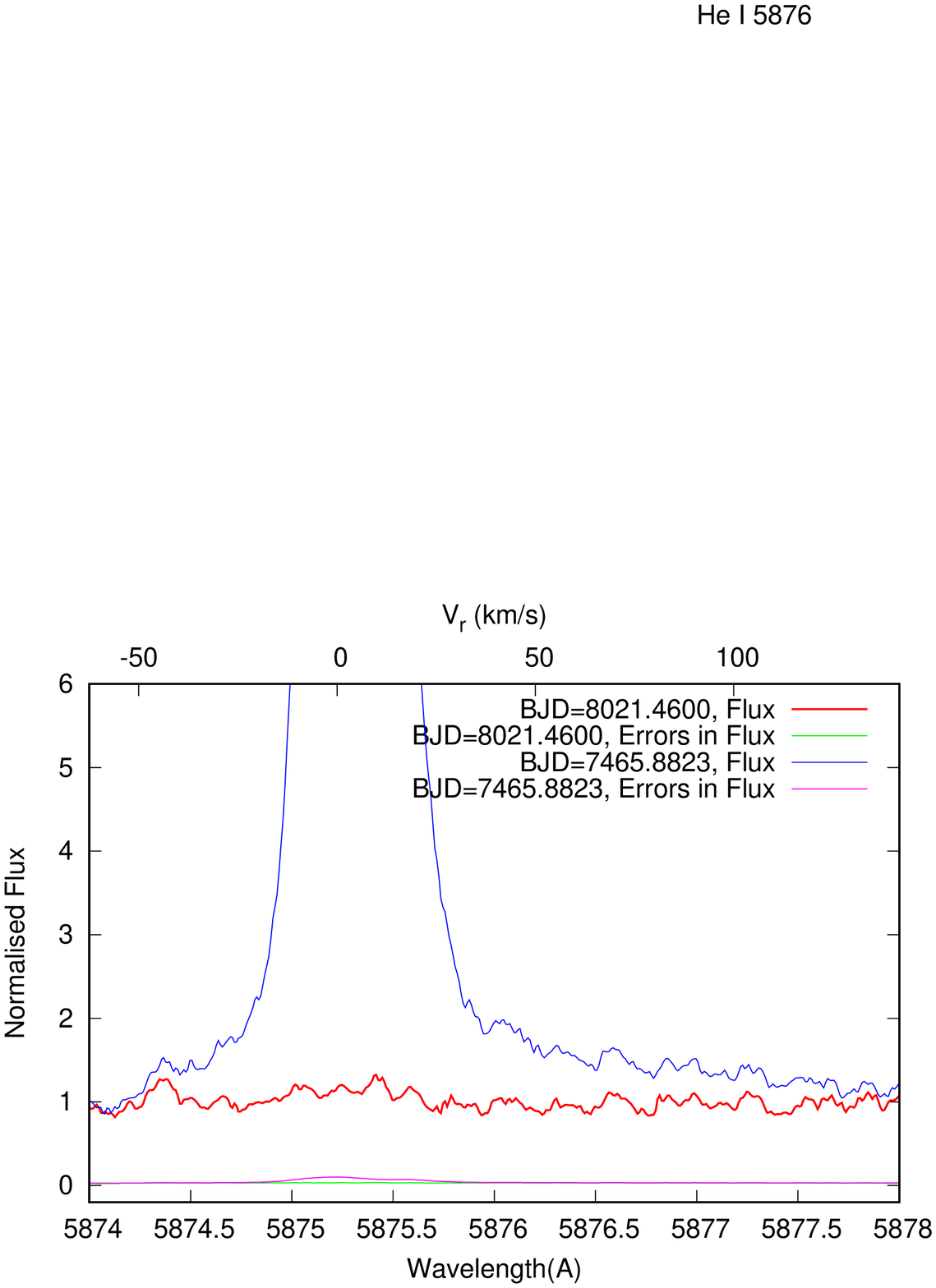}  
  \includegraphics[width=0.48\linewidth, angle=0]{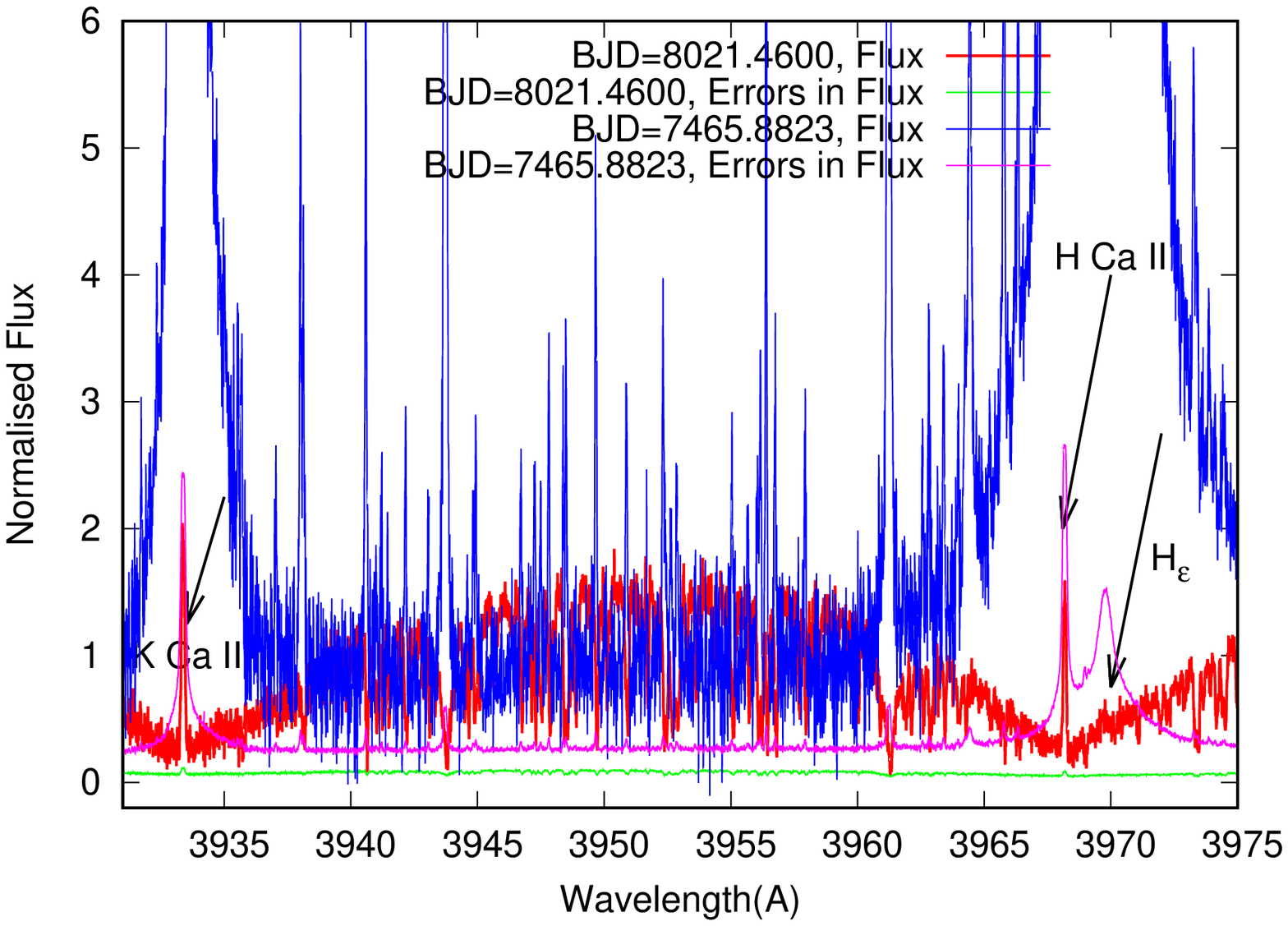} 
  \includegraphics[width=0.48\linewidth, angle=0]{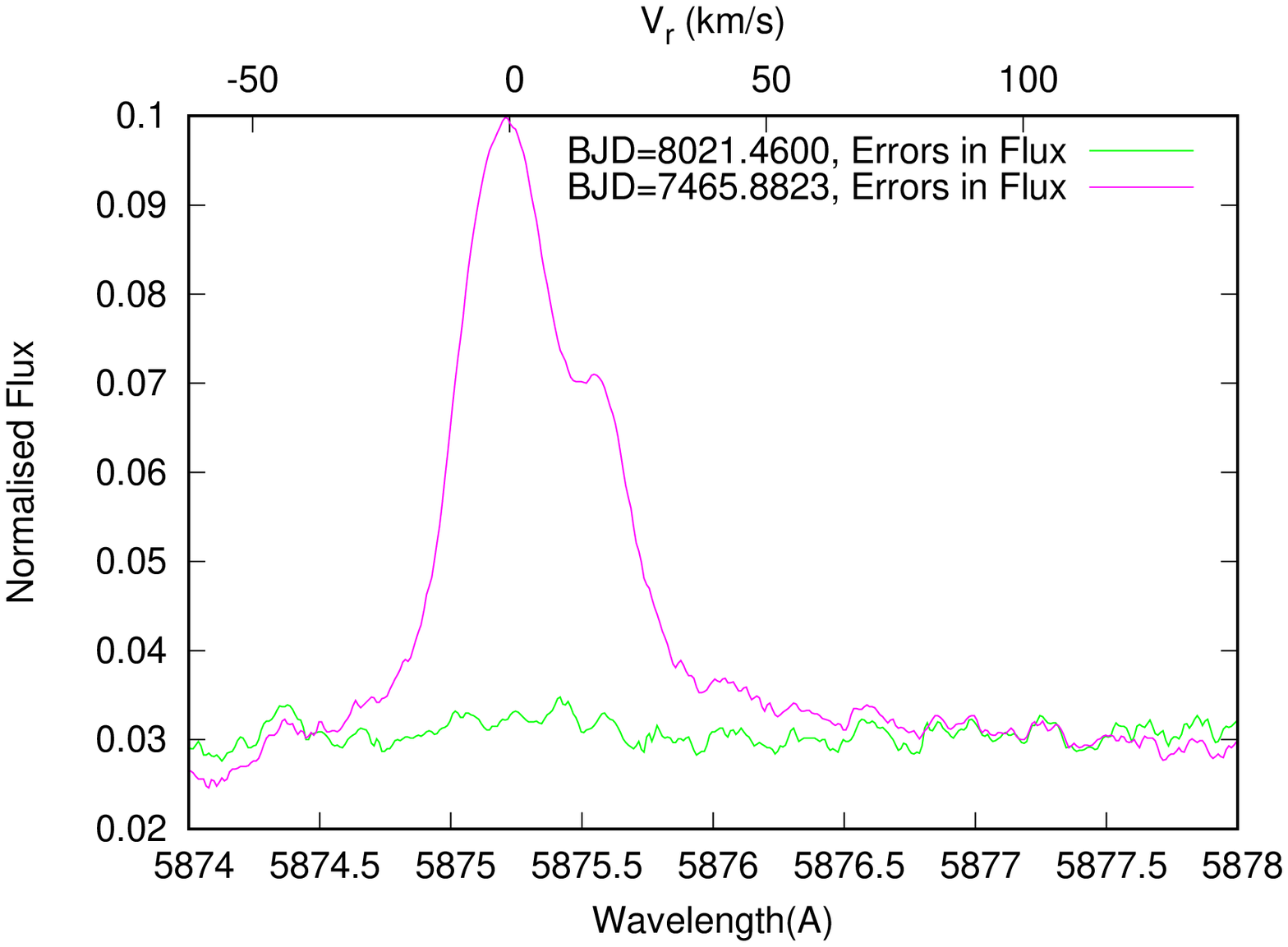}
   \includegraphics[width=0.48\linewidth, angle=0]{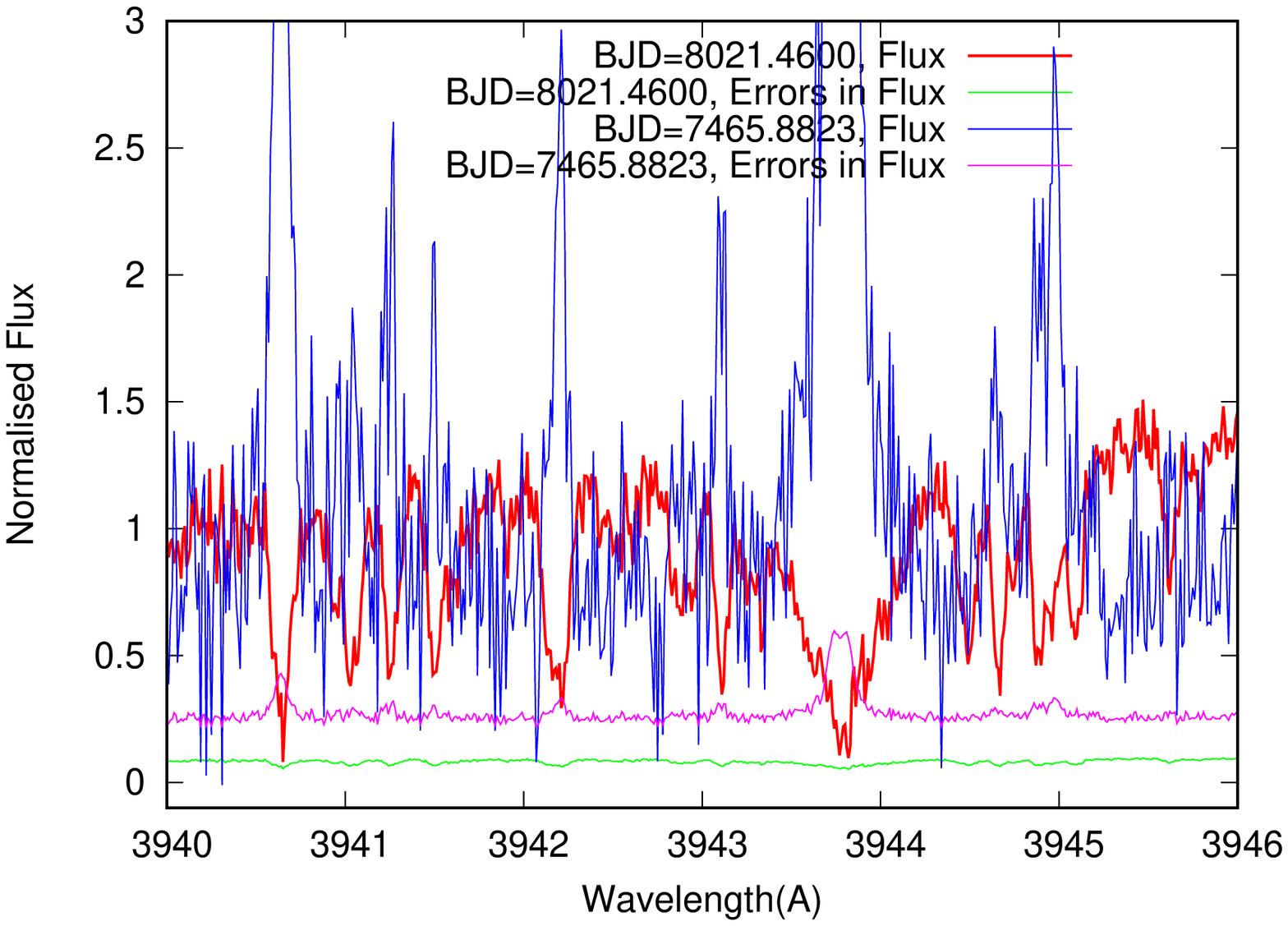}  
\caption{{\it Top-left panel}: Observed fluxes of emission lines across \hev  spectral region at the dates of minimum 
and maximum activity, shown by red and blue lines, respectively. Uncertainties of the flux determination are 
shown by the green and pink colours, respectively. {\it Bottom-left panel}: Uncertainties of the flux determinations at the
phases of minimum and maximum activity shown in larger scale.
{\it Top-right panel:} Observed fluxes of emission lines across Ca{\small{II}} H and K and \Heps{} spectral regions at the same 
dates of minimum  and maximum activity, shown by red and blue lines, respectively. 
{\it Bottom-right panel:} Part of the observed spectra shown in larger scale. Flux measurement uncertainties are also shown.}
   \label{_fminn}
\end{figure*}

In Fig.\ \ref{_fminn} we show optical (left panel) and blue (right panel) wavelength ranges of 
the spectra of Proxima obtained on the dates of minimum (BJD\,=\,8021.46) and maximum activity
(BJD\,=\,7465.8823). The optical and blue spectral ranges contain \hev{} and Ca{\small{II}} H and K 
emission lines, respectively. We emphasise that we work with the residual fluxes, that is\ fluxes
reduced to the level of the local pseudocontinuum by 1.0\@. Due the presence of molecular absorptions
in the optical spectral range, the true continuum is buried under the absorption features formed 
by the numerous strong molecular bands. To the contrary, the continuum is still seen in the blue 
spectral region, which is more or less free from molecular absorption \citep{pavl17}. 
The local pseudocontinuum in this part of the spectrum is formed by the wings of strong atomic lines 
seen in absorption and/or in emission.    

In Fig.\ \ref{_fminn} we show the estimations of the  accuracy of flux measurements, $F_a$, 
provided by the HARPS pipeline. In the first approach we can adopt $F_a/F$ $\approx$ S/N, where $F$ is the monochromatic flux. 

Accuracy of the flux measurements in the optical spectral region seems to be similar in the flare and quiet modes; 
see left panels of Fig. \ref{_fminn}. Furthermore, in the quiet mode the S/N is similar for the 
observed blue and optical spectra.

On the other hand, in the case of the Ca{\small{II}} H, K, and \Heps{} lines, we see different 
S/N for the spectra obtained in the flare and quiet modes. 
In the flare mode we observe a decrease in 
S/N from $\sim$ 10 to $\sim$ 3 (see the top-right panel of Fig.\ \ref{_fminn}), which cannot be explained by instrumental 
effects, such as different seeing or different
exposures, because observations in the optical spectral regions provide the same quality, that is S/Ns 
in the flare and quiet modes are similar.
Numerous emission lines appear in the flare mode. At the corresponding wavelengths we see absorption lines in the quiet mode, 
as shown in the
bottom-right panel of Fig.\ \ref{_fminn}. Nevertheless, we still clearly see the drop in S/N for the blue spectra obtained 
in the flare mode.

Likely, something happens with absolute fluxes in the blue spectral region. Namely, a decrease in S/N may be explained 
in the case of the drop of these fluxes in the blue spectral region during the flares. 
In turn, any decrease in flux may be caused by an increase in opacity in the photospheric layers due to the
non-thermal ionisation of metals (and/or hydrogen?), which are the main donors of free electrons in photospheres 
of cool stars. Over-ionisation of metals (and/or hydrogen?) increases the number density of H$^-$, which is 
a very important source of opacity in stellar atmospheres.

On the other hand, we do not see any significant differences in S/N in the optical spectra obtained in 
quiet and flare modes; this could be interpreted as a very weak dependence of the observed fluxes here on the flare activity.

Detailed analysis of these phenomena is beyond the scope of this paper. 
We simply note that despite the possible drop in flux in the blue spectral region, the level of the 
pseudocontinuum can be determined with sufficient accuracy during the quiet phase, but lower accuracy 
during flares.
However, in the last case the strong emission cores are the main contributors to the \pew{}. The emission cores 
can be measured with higher accuracy due to larger fluxes and higher S/N\@. We observe the opposite effect 
for weak emission lines in the quiet mode: both the pseudocontinuum and \pew{} are well determined when S/N is 
greater than 10\@.

Overall, we estimate an error budget of less than 10\% of the \pew{} 
after taking into account all effects
that may affect the quality of the spectra in the quiet and emission phases. We reiterate that \pew{}
measurements are done with respect to the local pseudocontinuum, which itself can change depending on physical 
processes in and/or over its formation levels.
  
\cite{fuhr11} reported activity in Proxima Cen on BJD\,=\,4901, 4902, and 4903, that is\  March 10--12, 2009\@.
On the third night, these latter authors observed a strong flare from X-ray to the optical, and UV excess (see their Fig.\ 13).
They also observed many strong lines in the blue spectral range, including Fe{\small{II}}, Si{\small{II}},
and Ti{\small{II}} in the 3200--4200\AA{} region. Likely,
\cite{fuhr11} observed an even stronger flare, known as a
white flare, a phenomenon known in solar and stellar astrophysics \citep[][and references therein]{walk11,dave14}.
In fig.\ 11 of \citet{fuhr11} we see that the S/N in their UVES spectra is similar during the quiet and flare modes,
suggesting that the increase of the flux in the blue spectral region does not affect the S/N in the observed
spectra. This effect is contrary to our case, where the S/N increases during flare events. Unfortunately
the dates of \cite{fuhr11} do not overlap with the dates of the HARPS database so we cannot carry out
a direct comparison.
\cite{howa18} also reported a white flare detected with the Euroscope telescope.

\subsection{Flare during the night of BJD\,=\,6417}
\label{_flare}

On the night of BJD\,=\,6417 we find a flare of strength 3$_\epsilon$ that was fully covered 
by uniform observations with time steps of 10 min until it recovered its `normal' state.
Although 10 min might seem long for flare processes, we can investigate the main 
stages of this event. Moreover, we are able to simultaneously cover a significant wavelength 
range containing several emission lines discussed in this work.

On the night of BJD\,=\,6417, we see a flare which started at 6417.6299 (labelled as 
`E$_{1}$' and corresponding to $t$\,=\,$-$34.6 min) and reached its maximum at 6417.65 
(i.e.\ $t$\,=\,0.0 min, marked as `E'  in the top-left panel of Fig.\ \ref{_4pew}). We observe
a temporal fading of that flare (BJD\,=\,6417.6841 or $t$\,=\,$+$43.02 min; F$_1$) and a secondary maximum (secondary flare?) of a strength of 2$_\epsilon$ at 6417.69 
($t$\,=\,$+$60.0 min; `F') that later restored to normal state at 6417.7642 ($t$\,=\,160.00 min, `F$_2$').
In Fig.\ \ref{_4pew}, we show the time mark E$_2$ (BJD\,=\,6417.6445 $t$\,=\,$-$12.6 min); at this time we see transit from  a comparatively slow increase to  an avalanche-like increase in flare intensity. The total duration of this flare is about 200 min, i.e.\  more than 3h. 

\begin{figure*}
  \centering
\includegraphics[width=0.48\linewidth, angle=0]{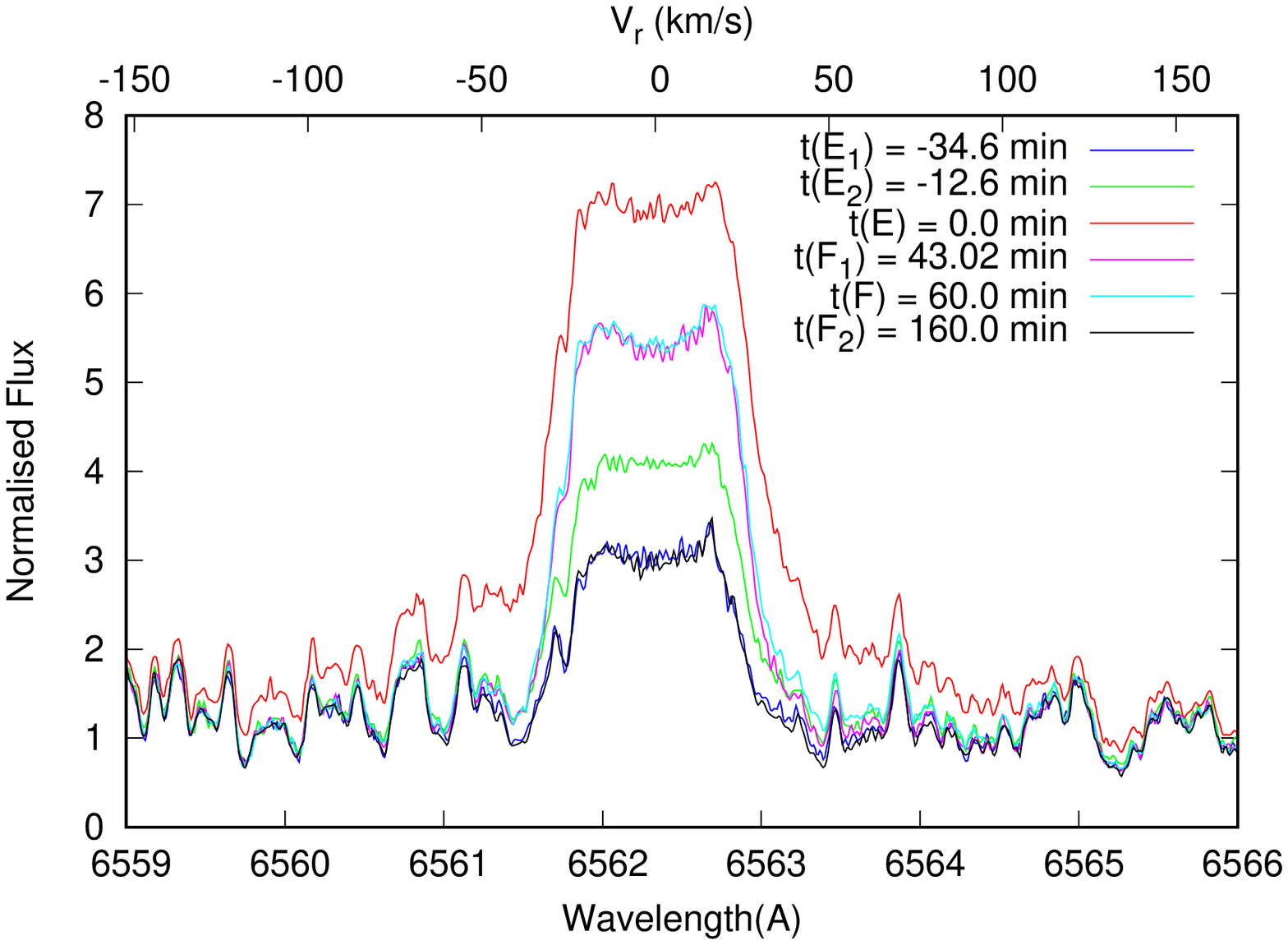}  
 \includegraphics[width=0.48\linewidth, angle=0]{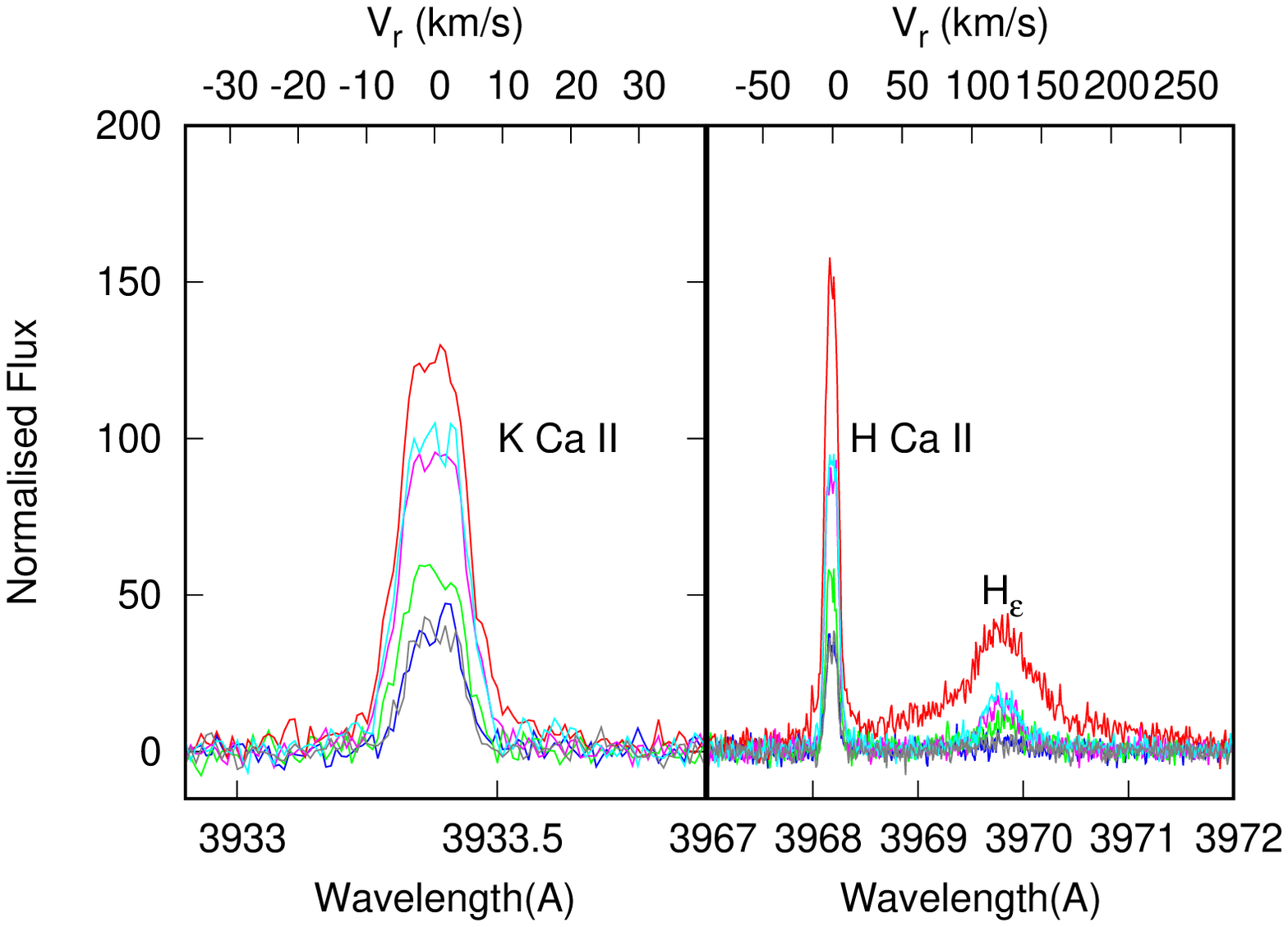}
  \includegraphics[width=0.48\linewidth, angle=0]{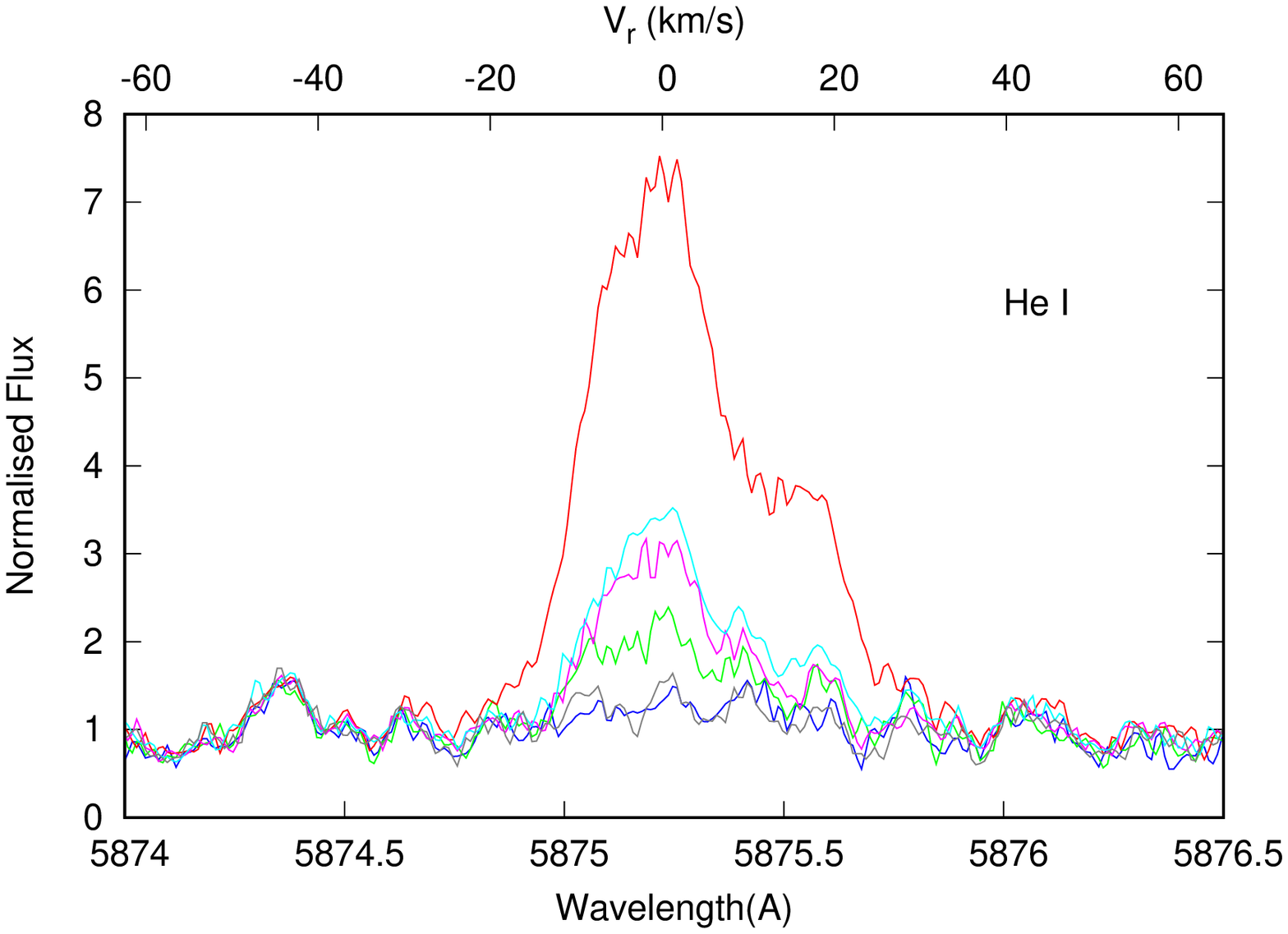}  
  \includegraphics[width=0.48\linewidth, angle=0]{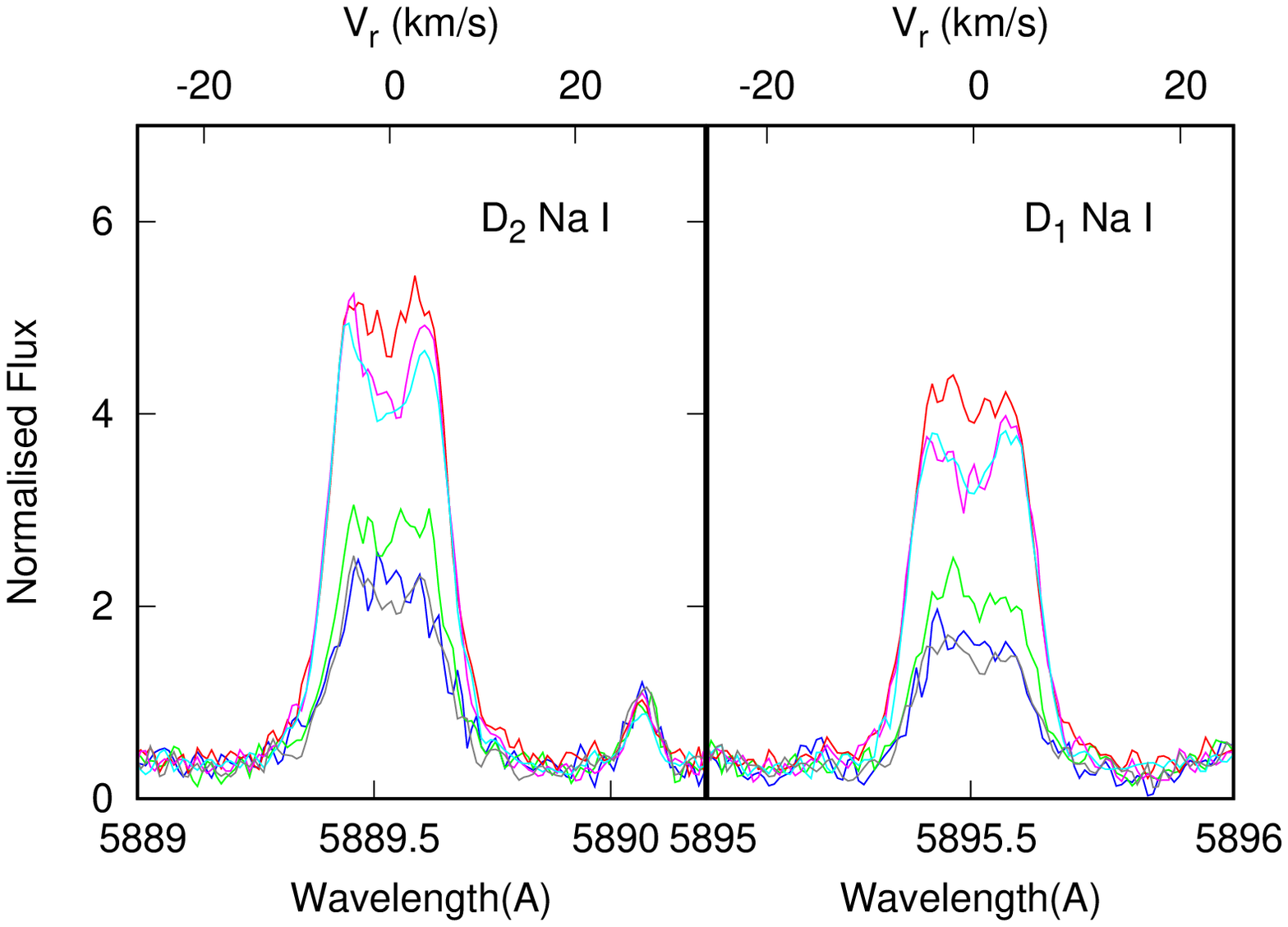}  
\caption{Temporal variations of the \pew{} of several emission lines during the flare on the 
night of BJD\,=\,6417\@. The times depicting the different phases are marked at the top 
of the top-left panel showing changes in the strength of \Ha{}.}
   \label{_4pp}
\end{figure*}

This flare is not the strongest in our observational set (Fig.\ \ref{_allpew}). Although this 
phenomenon may differ from the other strong flares marked with letters `B', `C', and `D' in 
Fig.\ \ref{_allpew}, we note a few interesting facts:

\begin{enumerate}
\item The \Ha{} line shows a symmetrical profile (see top-left panel of Fig. \ref{_4pp}), as in the case of flares C and D shown in Fig. \ref{_minmax}. 
\item We see a good correlation between the strength of all lines and the strength of \Ha{}. 
We interpret this as all lines most likely forming in one flare space. 
\item The cores of all lines are affected by the the self-absorption, suggesting that we see 
multi-component structures of the absorbing condensed matter. 
\item The fine structure of the outer layers affects the cores of all lines shown in Fig.\ \ref{_4pp}. 
This structure of the 
outer layers changes with time on short timescales. Indeed, we see differences in the cores 
of the lines observed at $t$\,=\,$+$43.02 and $+$60.0 mins displayed with pink and cyan lines, respectively,
in Fig.\ \ref{_4pp}. Even in the case of the Ca{\small{II}} K line, we note a self-absorption component shifted by different \Vr{} at different flare stages. 
\item We note the self-absorption components in the cores of the   Ca{\small{II} and \hev{}  
 lines which are} moving bluewards and redwards reflecting outward
and inward flows of hot matter in the outer atmosphere, respectively 
(top-right and bottom-left panels of Fig. \ref{_4pp}). 
\item The multi-component structure of the active regions of the atmosphere 
is also reflected in the shape of \DO{} and \DT{} Na{\small{I}} lines. In the `quiet' state, we 
recognise the absorption components of the cores of emission lines. In the `active' phases 
the emission line intensity increases and self-absorption becomes notable; see 
the bottom-right panel of Fig. \ref{_4pp}. However, this second  
emission core forms with another self-absorption component, which we interpret as at least
two regions of line formation in the cases of \DO{} and \DT{}.
\end{enumerate}
\subsection{Flare activity on BJD\,=\,6718, 6420, and 6426}
\label{_ext}

Other extended and well sampled sets of spectroscopic observations of the Proxima Cen 
were collected during the nights BJD\,= 6718, 6420, and 6426\@. From the comparison 
of spectral lines in emission obtained at the same times, we may study the physics of the 
flare variability over short timescales thanks to exposures taken every $\sim$10\,min.
Hence, we can investigate the relative responses of different lines during flare events.

\begin{figure*}
  \centering
 \includegraphics[width=0.48\linewidth, angle=0]{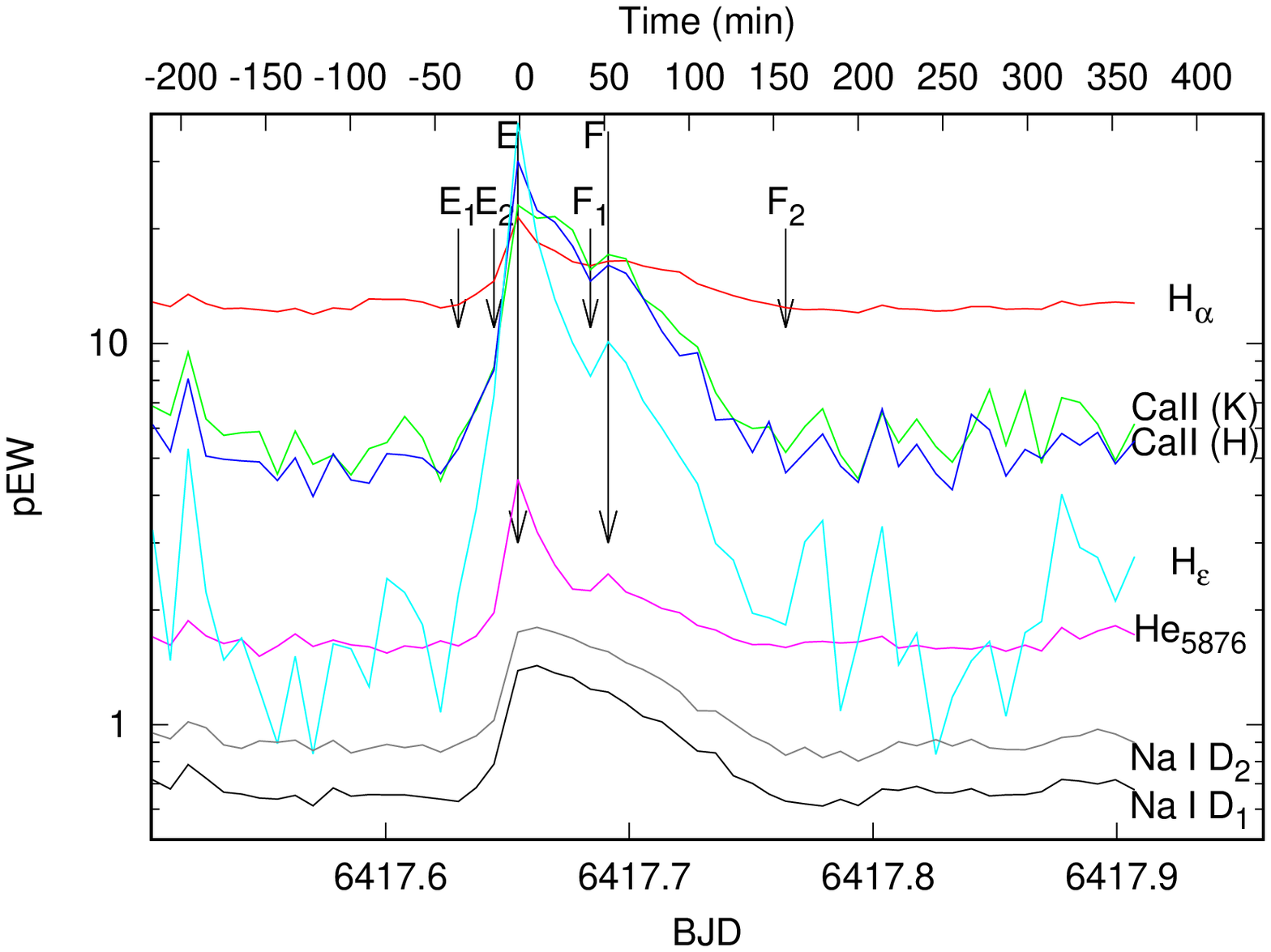}  
  \includegraphics[width=0.48\linewidth, angle=0]{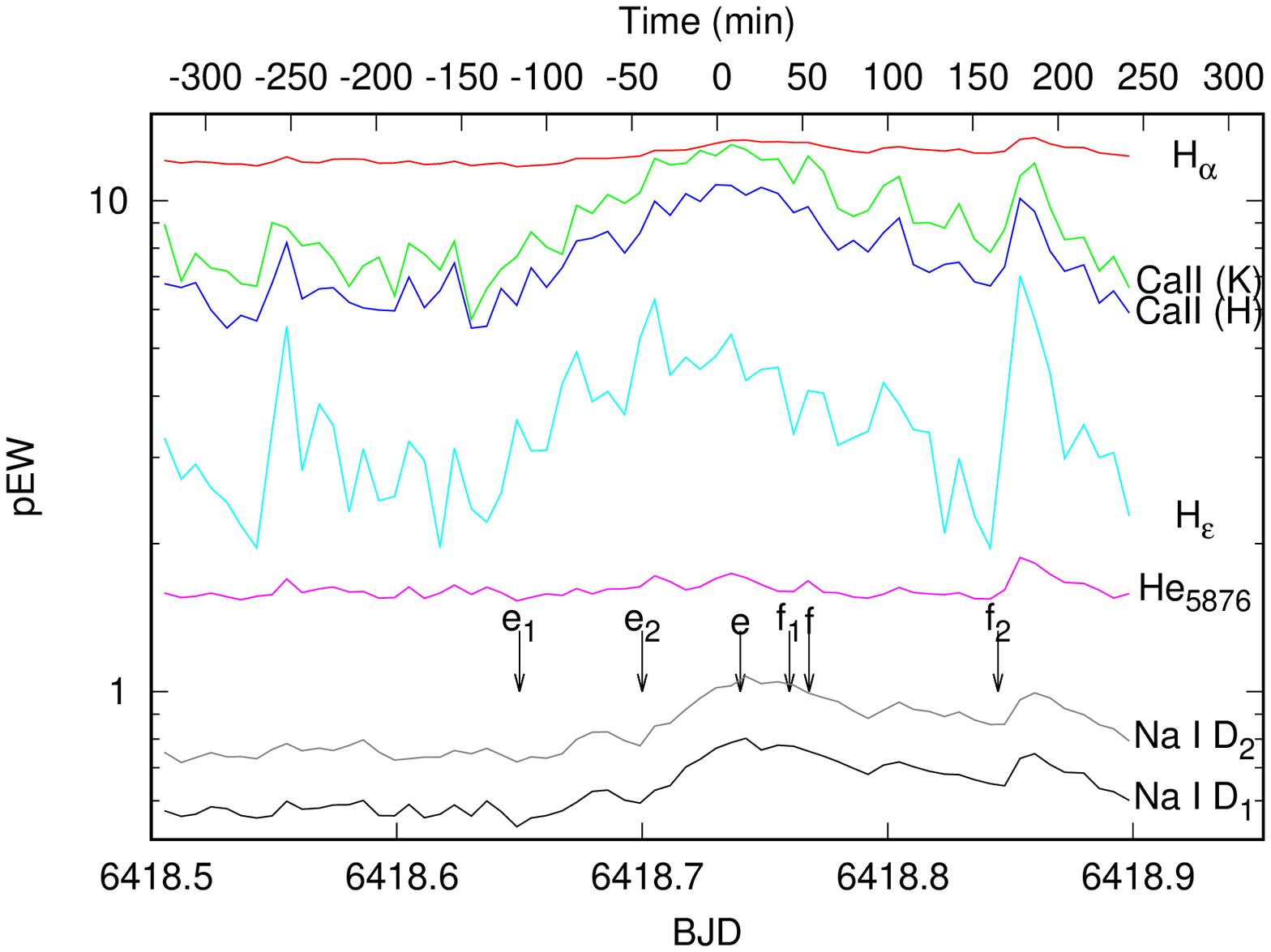}
  \includegraphics[width=0.48\linewidth, angle=0]{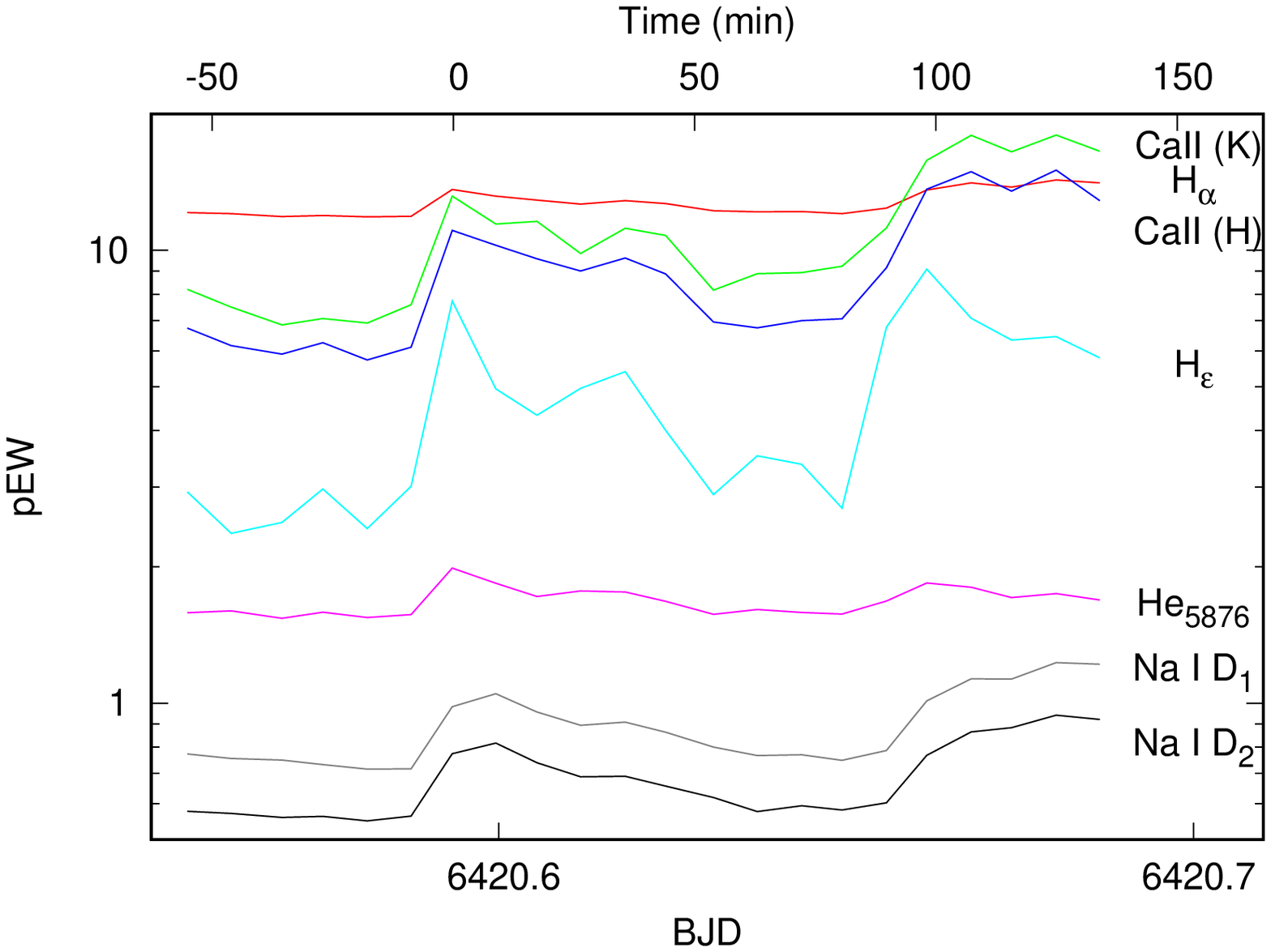}  
  \includegraphics[width=0.48\linewidth, angle=0]{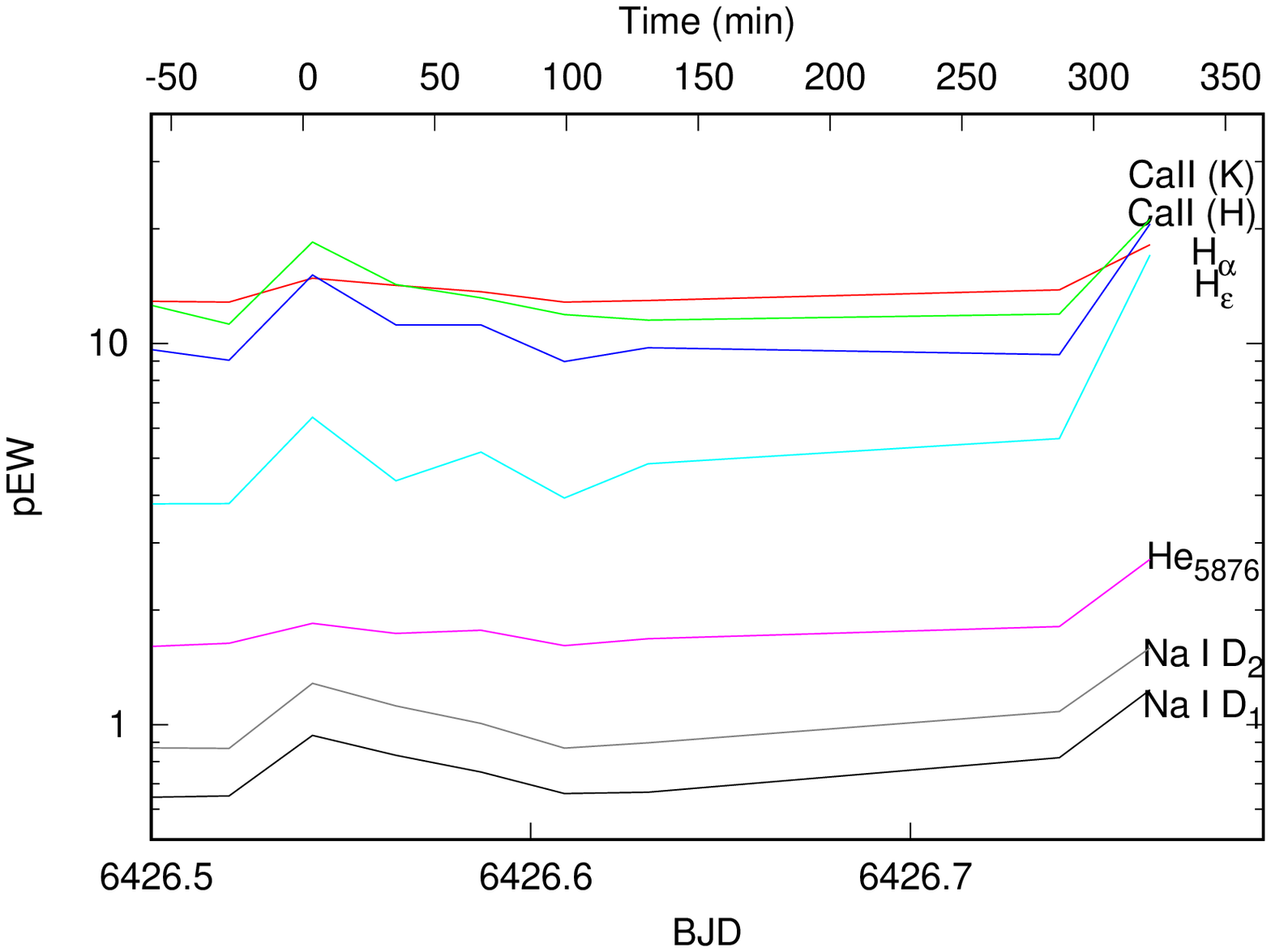}  
\caption{Temporal variations of the \pew{} of several emission lines during four nights on 
BJD\,=\,6417, 6418, 6420, and 6426. Time marks with small letters depict 
the phases of the flare on 6418 corresponding to the same phases as on BJD\,=\,6417\@.}
   \label{_4pew}
\end{figure*}

During those nights, we detect a few flare events covered by exposures made at
varying time intervals.  
We show the changes of \pew{} of several lines during those nights in Fig.\ \ref{_4pew}. 
For convenience, the upper x-axis is given in minutes. Time $t$\,=\,0 min corresponds 
to the maximum of the flares. 

On BJD\,=\,6418, we detect a flare in some lines with a duration of 250 min (top-right panel 
of Fig.\ \ref{_4pew}). The strength of the flare is of order 2$\epsilon$. We most likely capture 
part of the flare, which we explain from the weak changes in amplitude of the emission lines 
and the absence of reaction of the \hev{} line.  On the other hand, we see the same flare 
phases, marked with small letters, as on the night of 6417\@.

 On BJD\,=\,6420 and 6426, we detect a flare event of 100 min in duration with
a strength of 2$_\epsilon$ (bottom-left and bottom-right panels of Fig. \ref{_4pew}). 
Again, we see the secondary flares at $t$\,=\,$+$40 or $+$50 min.
Interestingly, on BJD\,=\,6426 we see a flare `extended in time', where the strength of 
\Heps{} corresponds to the 2$_\epsilon$ flare.
  
It is worth noting again that the behaviour of \Heps{} is more sensitive to the flare events 
compared with other lines. We suggest that this line is formed mostly in the active region of high temperature. We observe strong changes of the \pew{} of \Heps{} ; even in the case of the 
extended flare on BJD\,=\,6418, where changes of \Ha{} and other lines are weaker 
than on the previous night (top-left panel of Fig.\ \ref{_4pew}).

%
%

%
%
\subsection{Rotational modulation of the chromosphere and the photosphere}
\begin{figure}
  \centering
  \vskip1cm
\includegraphics[width=0.95\linewidth, angle=0]{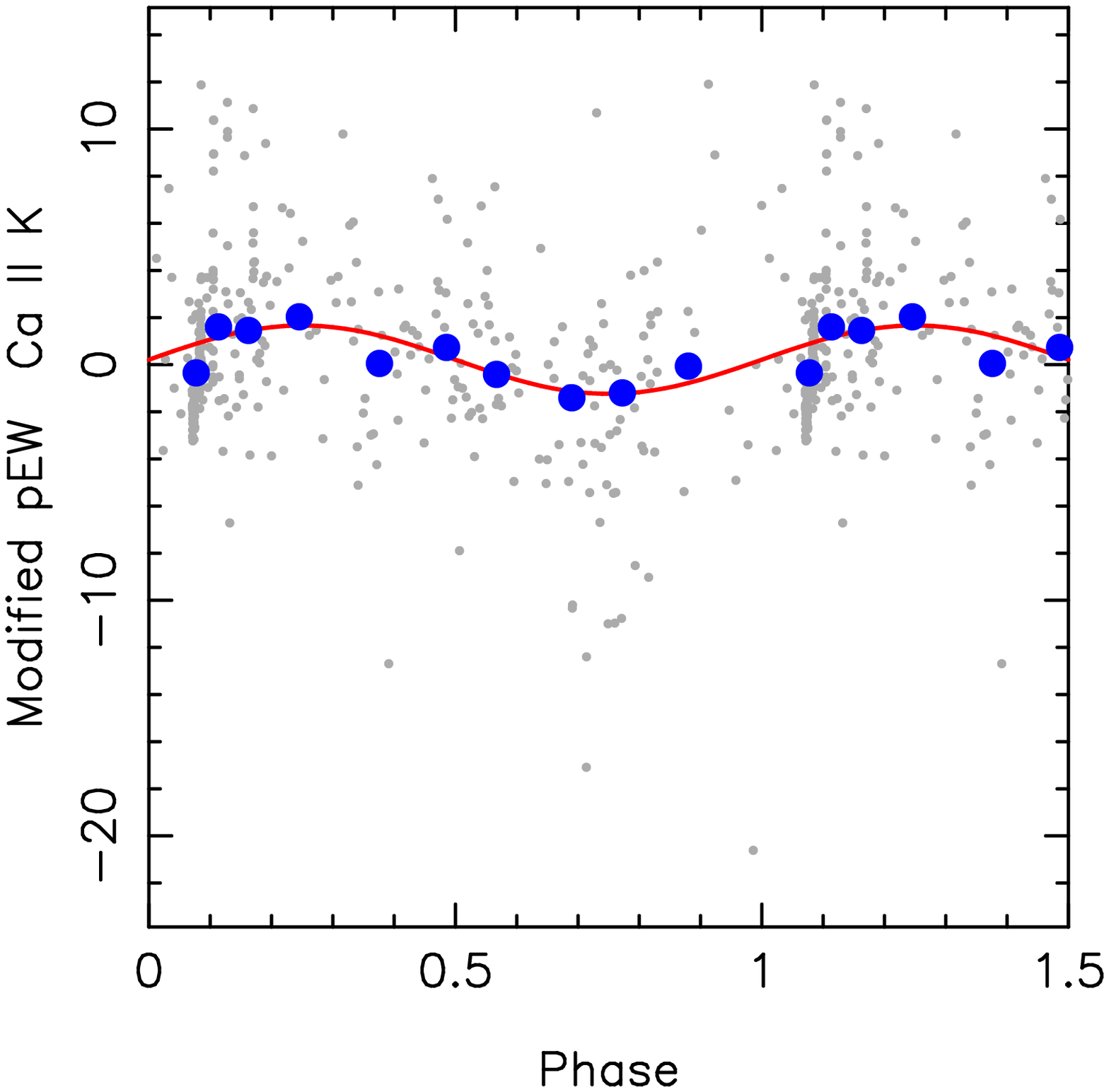} 
\caption{Median-subtracted pEWs of the Ca\,{\sc ii} K line (grey dots) folded in phase with the period of 90.72 days. The red line depicts the sinusoidal fit to the folded data (amplitude of 1.5 \AA), and the blue dots correspond to the data points averaged with their 21 neighbours. Flares and strongly deviant data points are not included in the diagram.
}
   \label{_mr_cak_phase90.73}
\end{figure}

 Here, we aim at detecting the rotation of the  chromosphere of  Proxima Cen thanks to the extended sets 
of HARPS high-resolution spectra. We removed the two flare events and the deviant data points from the 
\pew{} dataset of the Ca{\small{II}} K line, and subtracted the median values per observing epoch. Otherwise, 
it is not possible to see the modulation, if it exists, because of the different amplitudes of the line 
intensity at the different epochs and activity states. The Lomb-Scargle periodogram analysis 
\citep{lomb76,scargle82} yields no significant strong peak. This is probably because of the artificial noise 
introduced by the removal of the median values from the original data. However, in the interval 30--1100 days, 
the highest peak lies at 90.72 days (with an uncertainty of $\pm$1.5 days as derived from the FWHM of the peak). The second-highest peak of the periodogram lies at 81.68\,$\pm$\,1.5 days. 
We note however that the periodogram is rather sensitive to the removal of deviant data. 
Figure\ \ref{_mr_cak_phase90.73} displays the median-subtracted Ca\,{\sc ii} K \pew{} folded in phase with 
the period of 90.72 days. The data show a sinusoidal modulation with a small amplitude of 1.5\AA{}. 
A similar sinusoidal pattern is found when using the 81.68-day period, thus making the two periodicities 
indistinguishable with current HARPS spectra. The two values differ from the rotational period of 
116.6\,$\pm$\,0.7 days determined from a subset of 222 HARPS spectra included here by \citet{suar15}, 
but are closer to the period of $\sim$83 days reported by \citet{bene98}, \citet{kiraga07}, and \citet{warg17}. 
These latter authors used photometric observations obtained with the {\sl Hubble Space Telescope} (HST) 
and the automated sky survey ASAS \citep{pojmanski97,pojmanski04}. A periodicity of 82.6\,$\pm$\,0.1 days
is also reported by \citet{coll17} after combining UVES, HARPS, and ASAS photometry. These authors discussed 
that the 116.6$^d$ measurement published by \citet{suar15} is likely an alias induced by the specific HARPS 
observation times, and concluded that the most significant rotation period of Proxima Cen is 82.6\,$\pm$\,0.1 days.

\begin{figure}
  \centering
\includegraphics[width=0.95\linewidth, angle=0]{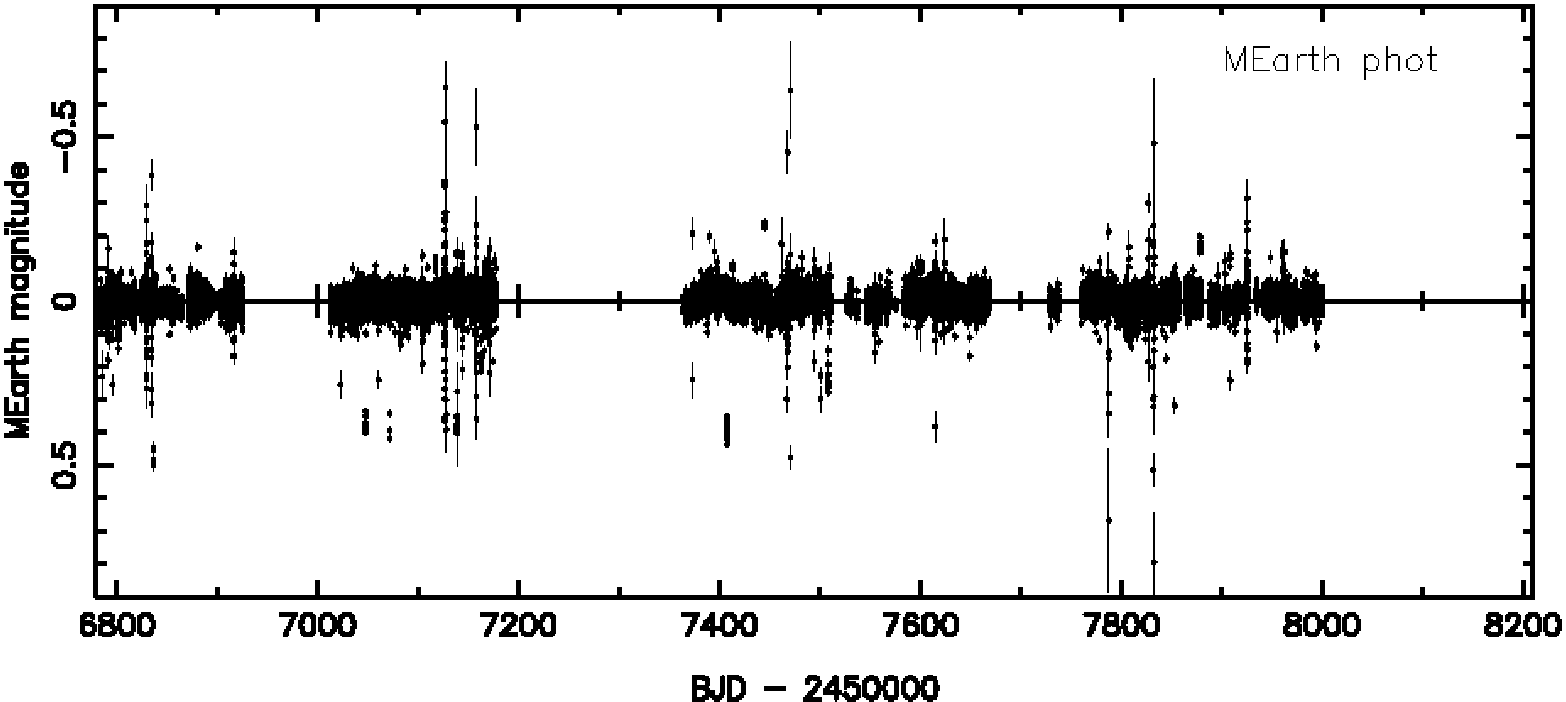} \\ \vskip 6mm
\includegraphics[width=0.95\linewidth, angle=0]{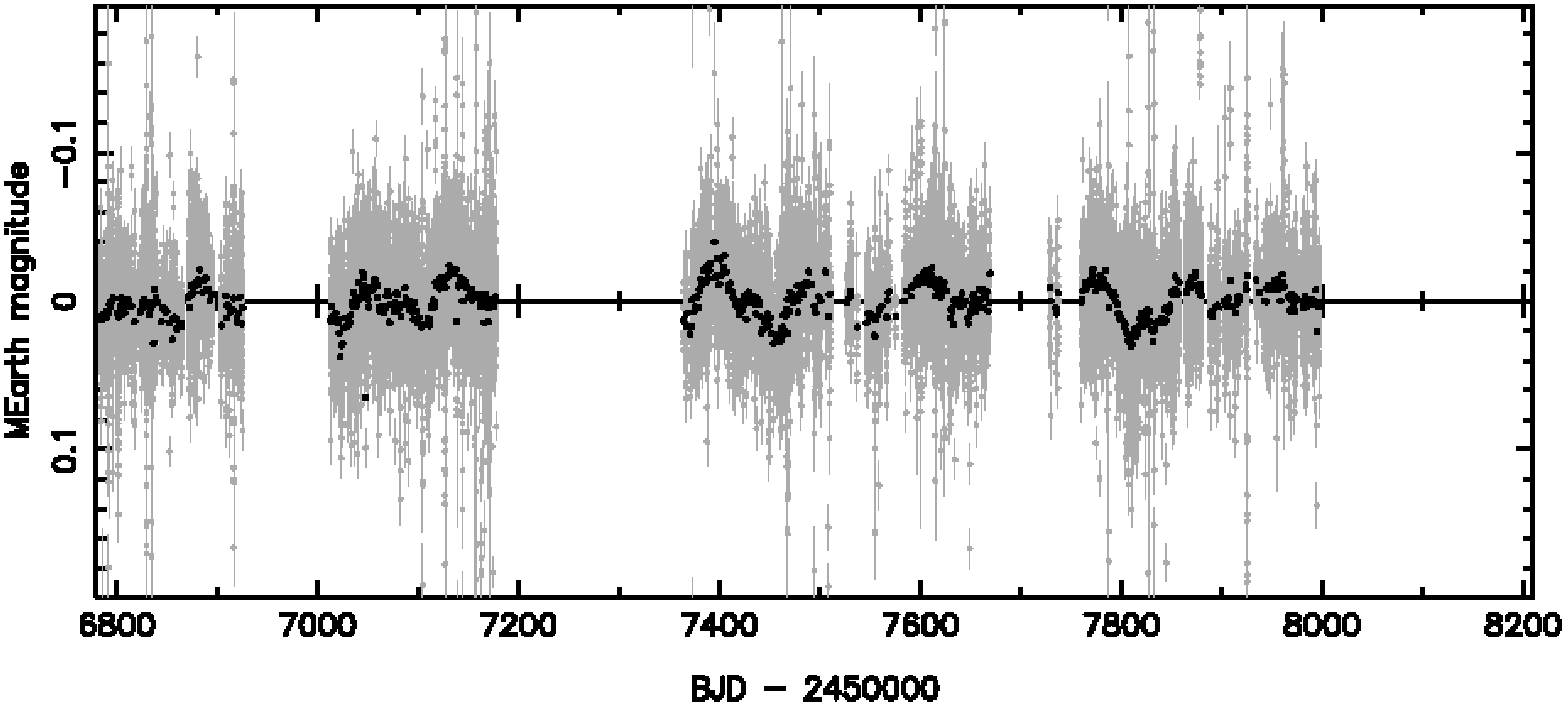} \\ \vskip 6mm
\includegraphics[width=0.95\linewidth, angle=0]{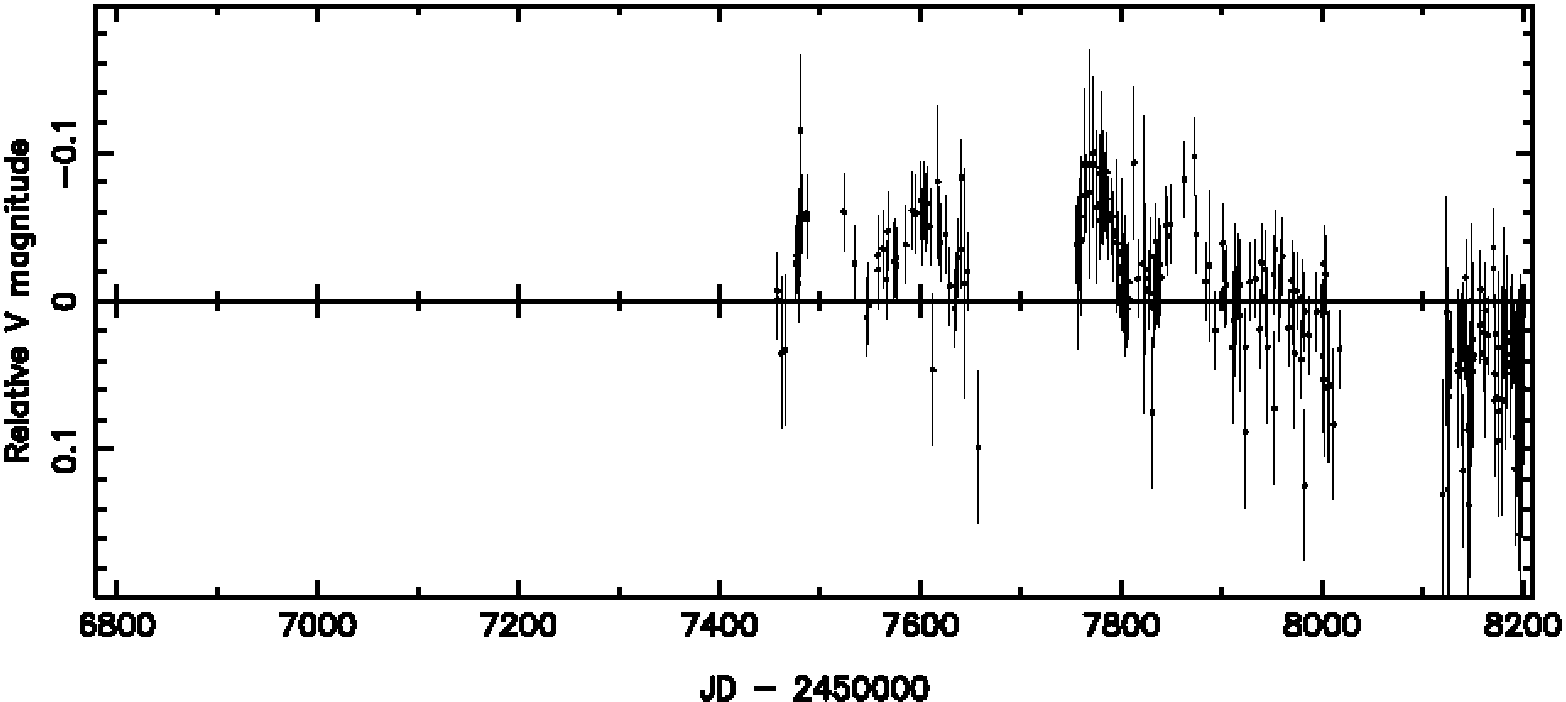} \\   \vskip 6mm

\caption{{\it Top:} Proxima Cen photometric light curve (filter RG715) delivered by MEarth DR7 (black dots). The data span a period of 3.2 yrs.
Negative magnitudes imply brighter states of the star. {\it Middle:} Enlargement of the MEarth DR7 light curve. Original data are plotted as grey dots. The day-averaged photometry is shown with black dots. {\it Bottom:} ASAS-SN light curve (black dots) spanning a time period of 2 yrs. The mean magnitude of 11.227 mag has been subtracted from the original data for a proper comparison with the MEarth DR7 light curve.}
   \label{_mr_mearth}
\end{figure}

\begin{figure}
  \centering
\includegraphics[height=0.8\linewidth]{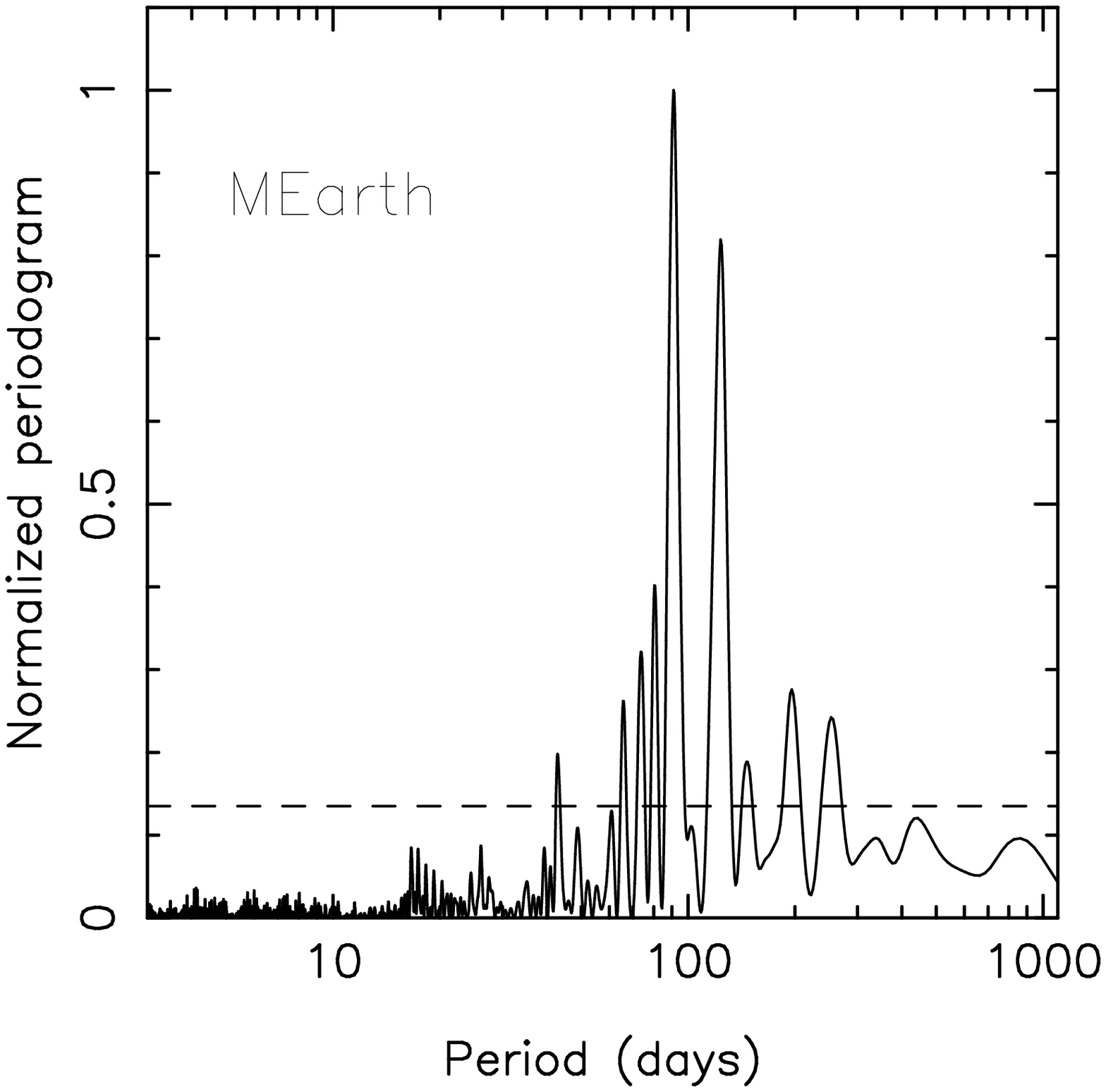} \\
\includegraphics[height=0.8\linewidth]{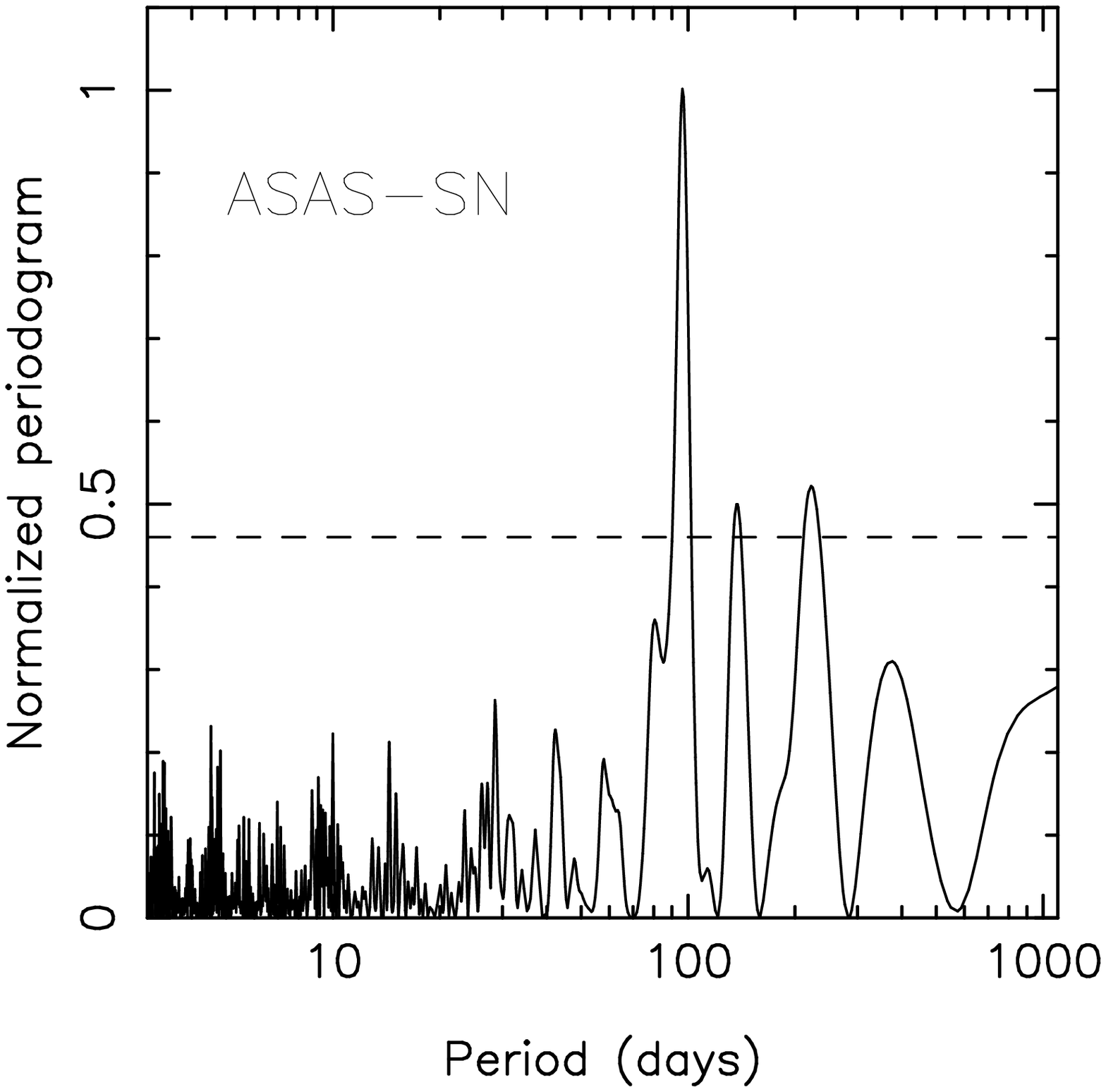} 
\caption{{\it Top:} Lomb-Scargle periodogram of the MEarth DR7 light curve for Proxima Cen. 
{\it Bottom:} Lomb-Scargle periodogram of the ASAS-SN light curve for  Proxima Cen after subtraction of the  
decreasing flux trend. The 0.1\%~FAP is indicated by the horizontal dashed lines.}
   \label{_mr_periodogram}
\end{figure}

\begin{figure}
  \centering
\includegraphics[width=0.95\linewidth, angle=0]{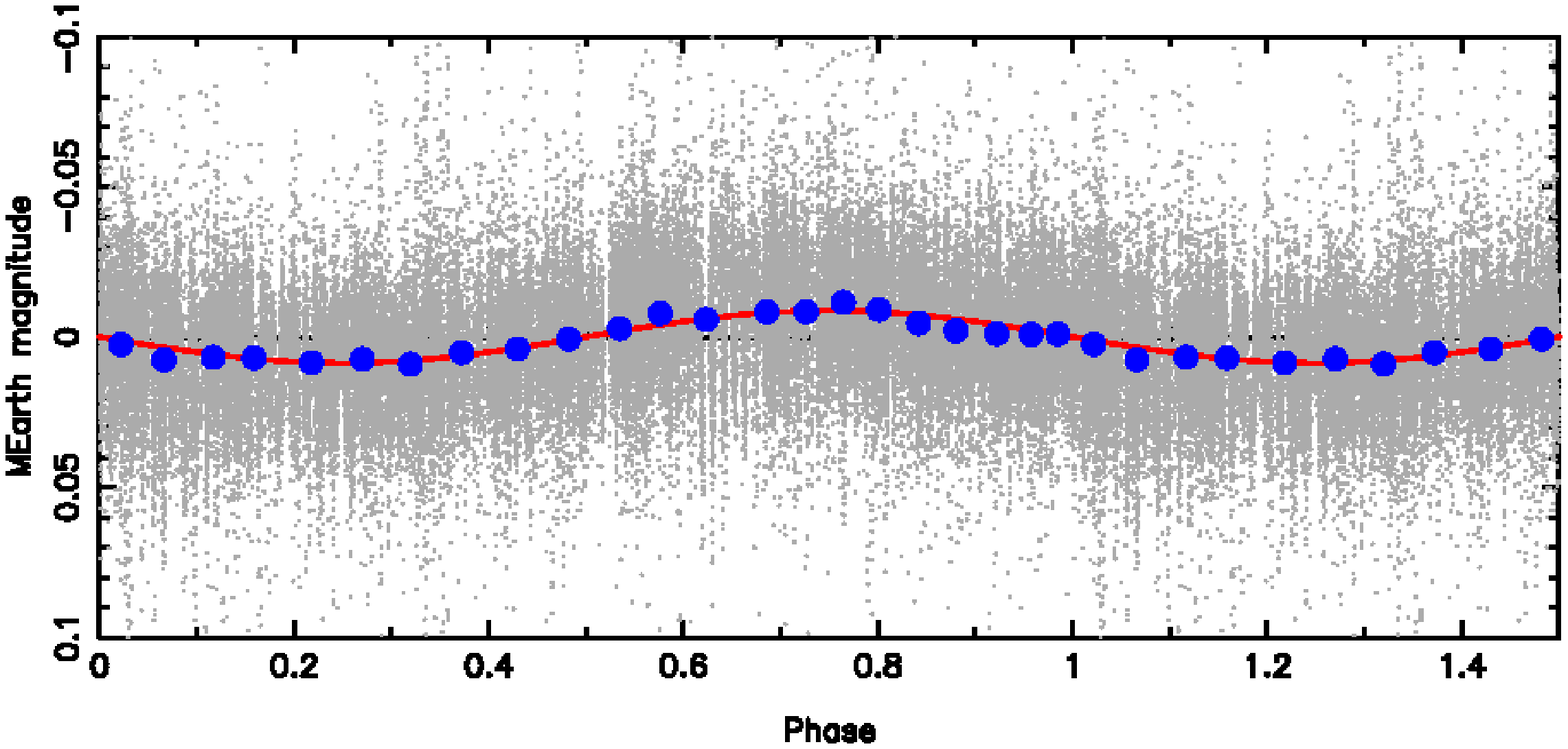}\\ \vskip 5mm
\includegraphics[width=0.95\linewidth, angle=0]{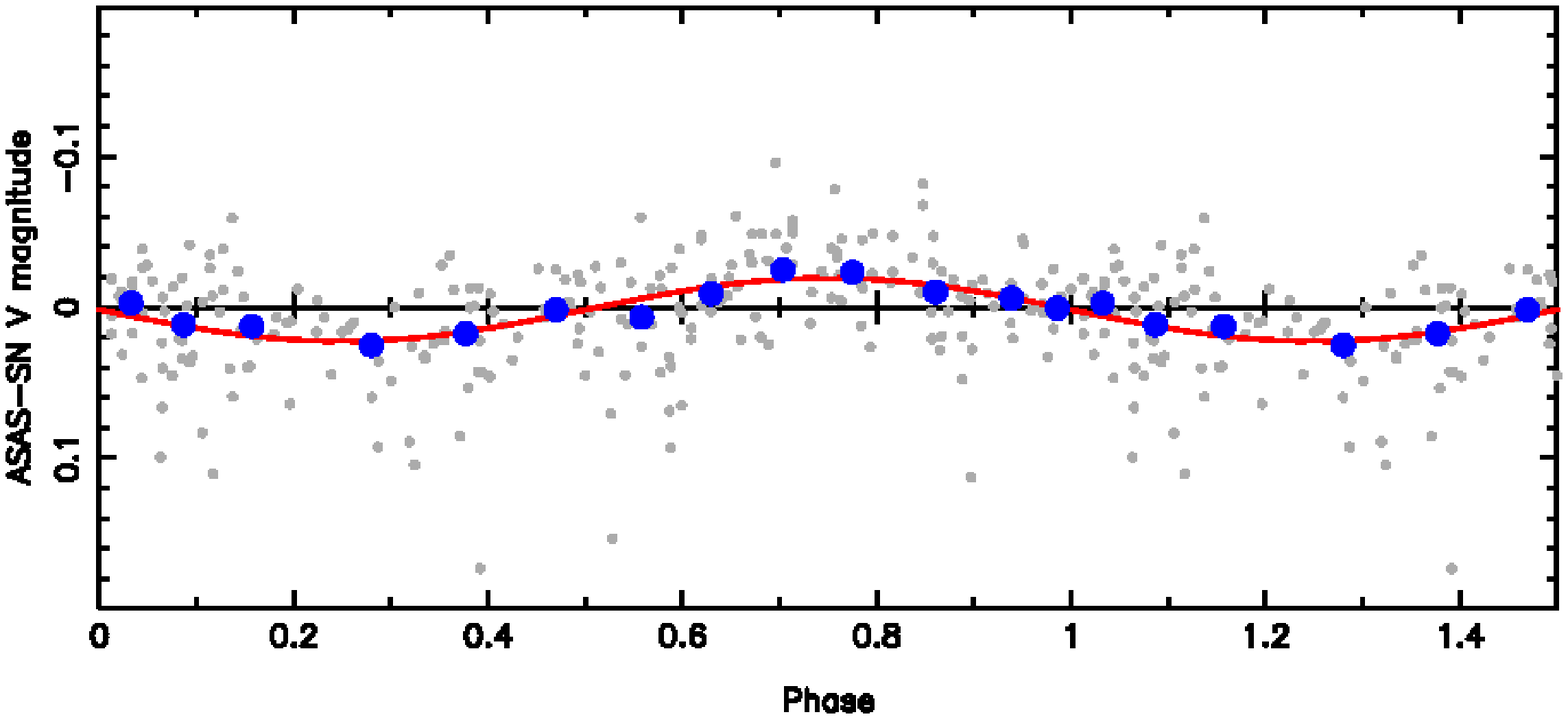} \vskip 1cm
\caption{MEarth DR7 ({\it top}) and ASAS-SN ({\it bottom}) light curves (grey dots) folded with a period of 91.0 and 96.5 d, respectively. The sinusoid fit to the data is shown with the red curve (amplitudes of 8.7 and 21.0 mmag). Each blue dot corresponds to the average of 3000 MEarth and 19 ASAS-SN individual data points. We note that the vertical scale is different by a factor of two different between the two panels.}
   \label{_mr_folded}
\end{figure}

 In the recent seventh data release (DR7) of the MEarth project \citep{bert12,irwi11}, Proxima Cen 
is included with more than 64200 photometric magnitudes obtained with the $RG715$ filter between 2014 
May and 2017 August. \citet{irwi11} provided a detailed review on the observing cadence, exposure times, and 
processing of the enormous amount of data of the MEarth monitoring campaigns. Concerning Proxima Cen, the 
resulting MEarth photometric light curve is illustrated in the top panel of Fig.\ \ref{_mr_mearth}, where 
the data present a dispersion of $\pm$30 mmag, which is three times larger than the typical photometric 
error bar. A quick inspection of the light curve reveals some sinusoidal patterns with small amplitudes 
at various epochs (e.g.\ BJD\,=\,2457400). After applying a 2-$\sigma$ clipping algorithm to remove the 
most deviant data points, we obtained the Lomb-Scargle periodogram of the MEarth DR7 light curve shown in 
the top panel of Fig.\ \ref{_mr_periodogram}. A strong peak is detected at 91.0\,$\pm$\,3.0 days (a second 
strong peak is seen at $\approx$124 days, but it has a non-negligible contribution from an alias of the 
91.0-day peak. This is most likely an alias of the annual window, i.e.\ $P_{alias}=1/(1/91-1/365)=121^d$). 
Interestingly, the MEarth DR7 periodogram shows quite a number of peaks well above the false-alarm probability 
(FAP) of 0.1\%, suggesting that they may be significant frequencies of the data. We removed the strong 
91.0-day periodicity from the original photometry, and a forest of significant peaks still remain in the 
periodogram. These peaks lie at 39.4, 42.9, 56.9, 83.2, 116.6, 129.2, and 167.3 days (with estimated 
uncertainty of 1--3 days), and all have associated amplitudes between 3 and 7 mmag (smaller than the quoted 
amplitude of the 91.0-day peak; see below). We note that two of these frequencies, 83.2 and 116.6 days, 
are consistent at the 1-$\sigma$ level with those previously reported in the literature using very different 
datasets \citep{bene98,kiraga07,suar15,warg17,coll17}. Therefore, there is little probability they are 
caused by the way the various data have been acquired. Rather, they may be true signals of the  
periodic activity of Proxima Cen. The MEarth DR7 data folded in phase with the 91.0-day period are shown in the bottom panel 
of Fig.\ \ref{_mr_folded}. A sinusoid curve with an amplitude of 8.7 mmag is plotted on top of the observations.

 Proxima Cen has also been part of the All-Sky Automated Survey for Supernovae (ASAS-SN) long-term 
photometric monitoring \citep{shappee14}. The ASAS-SN $V$-band light curve is shown in the bottom panel 
of Fig.\ \ref{_mr_mearth}. It overlaps with the MEarth DR7 data and extends over an additional period of 
several months of observations. Both the MEarth DR7 and ASAS-SN light curves cover a total of 3.9 years of 
continuous photometric monitoring. There are two obvious features highlighted by the comparison of both light 
curves: one is that the amplitude of the variations of the ASAS-SN data is larger than that of the 
MEarth DR7 photometry. This is very likely explained by the fact that the variability of  Proxima
Cen is more 
notorious at blue wavelengths. The second feature is that the ASAS-SN photometry shows a decreasing trend 
towards fainter $V$ magnitudes, which contrasts with the flat nature of the MEarth $RG715$ data. We note 
that the state of minimum activity labelled with the letter `A' in previous sections corresponds to epochs 
with the faintest $V$ magnitudes. Whether this trend is real or an artifact of the ASAS-SN data is not clear 
to us since the two curves were acquired with different filters. The decreasing trend, which is mostly 
dominated by the most recent epoch of observations, induces a strong signal/power at small frequencies in 
the Lomb-Scargle periodogram of the ASAS-SN data. To identify potential peaks at longer frequencies, we 
removed the trend by subtracting a fourth-order polynomial from the original photometry. The resulting 
Lomb-Scargle periodogram is shown in the bottom panel of Fig.\ \ref{_mr_periodogram}, where one strong peak 
is visible at 96.3\,$\pm$\,5.0 days. After removing this signal from the light curve, there remains only 
one significant peak at 82.8\,$\pm$\,4.0 days with a FAP above 0.1\%. 

\begin{figure}
  \centering
  \vskip1cm
\includegraphics[width=0.95\linewidth, angle=0]{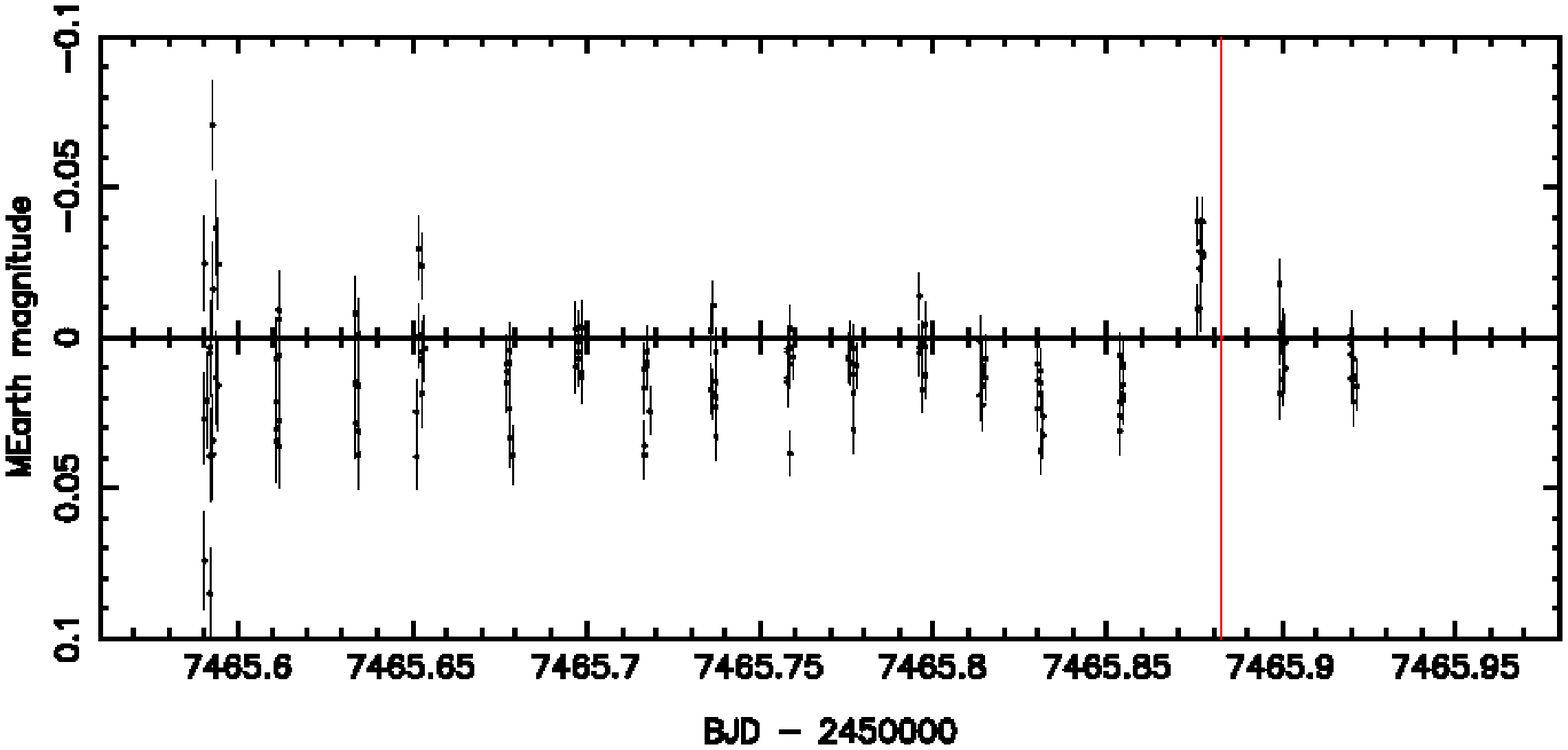} 
\vskip 1cm
\caption{
MEarth DR7 light curve of  March 18, 2016. The red vertical line indicates the location of the spectroscopic flare B.
}
   \label{_mr_flare}
\end{figure}

 Within 1-$\sigma$ of the quoted uncertainties, we found the  peaks at $\sim$91 and $\sim$83 days to be common 
to all three data sets (HARPS, MEarth, and ASAS-SN), with the peak at $\sim$91 days being the strongest signal. 
These two periodicities are likely related to the rotation period of Proxima Cen. The changes in the 
variability periods could correspond to changes and evolution (appearance and disappearance) of active 
regions located at different latitudes on the stellar surface. Proxima Cen shows multiple significant 
periods, which we interpret as differential rotation. On the other hand, we cannot exclude the possibility 
that the peaks at $\sim$91 or $\sim$83 days are another alias of the annual window.

The strong flare, event B, reported in previous sections is reasonably well covered by the MEarth data: 
photometric observations are available before, during, and after the flare. This is the only flare reported here for which there is such temporal coverage in the MEarth DR7 database. Figure\ \ref{_mr_flare} depicts the MEarth 
photometry corresponding to March 18,  2016; the location of the spectroscopic flare is indicated. The 
flare event induced a photometric increase of $\ge$45 mmag in the brightness of  Proxima Cen at red optical 
wavelengths, which is about nine times greater than the amplitude of the rotation curve.

\section{Discussion}
\label{_diss}

In this paper we analyse the flare activity of Proxima Cen using extended sets of 
HARPS spectra. We inspect the intensity of the \Ha{} line to study the "flare activity".
\citet{pavl17} showed that the molecular background of the emission lines in the spectrum
of Proxima Cen shows rather marginal response to the variations of the activity level. Using this discovery, we computed the \pew{} of multiple emission lines and present a detailed analysis of 
the profiles of several emission lines of interest to study dynamic processes occurring during
different flare phases. We find some similarities and differences in the temporal changes of 
\Ha{} and other emission lines. 

The hydrogen and helium emission lines are of special interest. Indeed, these lines form 
in the flare region, in layers of high temperature (T\,$>$\,10000 K) in the inner and/or outer chromosphere,
where we observe a large dispersion of velocities in the largest flares. He{\small{I} lines 
form in the regions of higher temperatures and their intensities are rather marginal during 
the minimum activity of Proxima Cen, contrary to \Ha{}.

 The minima in \pew{} of \Ha{} (`A') are just barely fainter than the mean \pew{}, yet 
the minima in \Heps{} are about ten times lower than the mean \pew{}. This raises the 
possibility that the blue end of the spectra is prone to very low S/N on certain 
dates, not that the blue emission lines are actually lower on the star.

The  H{\small I} emission line \Heps{} disappears almost completely during the minimum activity of Proxima Cen, an effect that cannot 
be explained by the low S/N on those dates (Sect. \ref{_fa}). 
Indeed, our analysis shows that the S/N in the blue part of the spectra containing 
Ca{\small{II}} H and K and \Heps{} is higher in quiet mode. 
On the contrary, we see a drop in
S/N during phases with strong flares. However, at these dates \Heps{} increases its intensity by
several orders of magnitude. Even if the increase of \Heps{} originates from the drop of the background flux formed
in the stellar atmospheres the formally computed \pew should increase during flares.

We suggest that differences in the physical conditions responsible for the formation of the emission 
lines in quiet and flare modes may provide a more realistic explanation of these phenomena. Indeed, the Balmer 
decrement should change in flare mode due to variable physical conditions in the regions where hydrogen forms.
A more detailed consideration of this problem is beyond the scope of this paper, but this last possibility 
cannot be discarded.

Moreover, the formation of the strong emission in the core of the Na{\small{I}} line appears very 
unusual, as already noted by \citet{pavl17}. Special conditions are needed to create such strong 
Na{\small{I}} emission. We see that the half-width of the Na{\small{I}} line shows rather 
marginal changes during flares. At the same time, the intensity of the core of this line
reduces in the strong flare modes while some emission appears in the wings.

Our study is based on the analysis of the \pew{} and emission line profiles of H{\small{I}},
He{\small{I}}, Ca{\small{II}}, and Na{\small{I}}. Firstly, we looked at their correlation with
the strength of \Ha{}. We selected \Ha{} as the reference line because it forms in 
high-temperature (flare) regions. All lines show a good correlation with the strength of \Ha{}. 
The \heiv{} line shows the weakest correlation with \Ha{}, indicating that the processes
and regions of the formation of this line differ from other lines.
Interestingly, we note that the other He{\small{I}} line \hev{} better follows \Ha{}.

From our dataset we selected three dates of minimum and maximum activity of Proxima Cen. 
During the minimum activity, the spectrum of Proxima Cen does not exhibit any \hev{} or 
\heiv{} lines. Moreover, the intensity of \Ha{} is reduced by a factor of eight with respect to its maximum values, while the intensity of the Ca{\small{II}} lines are 150 times lower.
We observe an almost complete disappearance  of the lines with the highest excitation 
during the days of absolute minimum of \Ha{} on BJD\,=\,8021.4600, 8023.4639, and 
8027.4785 (Sect. \ref{_sminn}). Furthermore, the Ca{\small{II}} lines almost 
disappear, as if the star had lost its chromosphere for a while. This period of minimum
activity of the chromosphere may last a few days or possibly weeks. 
Even the strongest of all lines, \Heps{}, is not seen, although this line is entirely
formed in the flare regions, contrary to \Ha{} which is still present in the quiet state.
The evaporated chromospheric matter in flares becomes cooler during minimum activity, suggesting that the
condensed matter absorbs the blue parts of the Na{\small{I}} emission 
lines. The emission in Na{\small{I}} \DO \ and \DT \ lines  is still present, implying that they likely 
form in the remnants of the formerly strong chromosphere.  

During the periods of strongest activity of Proxima Cen, an extended red wing is present in
the \Ha{} line, most likely as a result of the hot matter penetrating into the cooler layers.
Strong flares affect the formation region of Ca{\small{II}} lines and show the anti-correlation 
with \Ha{} \citep{pavl17}.  The intensity of the Na{\small{I}}, \DO{}, and \DT{} lines is reduced 
during strong flares, but not as much as the Ca{\small{II}} lines. 
Interestingly, different flares yield emission lines of different widths, suggesting that 
stronger flares occur in the regions of larger turbulent velocity dispersion.
Even moderately strong 1$_\epsilon$ flares produce the notable \heiv{} 
and \hev{} lines (Fig.\ \ref{_2mr}). We also emphasise that absorption components are seen
in the cores of all lines during maximum activity.

We analysed extended sets of spectra obtained on specific dates where flares of different 
strengths took place. The study of temporal sequences of \Ha{} reveals variability on 
timescales of 10 min. We observe a few flares at good temporal resolution even though
flares may also occur on shorter timescales. All flares show similar structures with several
phases, including a slow initialization, a fast phase, a phase of fading, a secondary increase 
of intensity, followed by another fading phase. Unfortunately, our interpretation of the different phases 
lacks critical information: Firstly, we do not know the precise geometry and 
size of flare regions. Secondly, we do not spatially resolve the stellar disc, implying that the
spectrum of Proxima Cen includes the integrated flux of the whole stellar surface. Finally,
flares  may appear at different parts of the stellar disc and even at the limb, or partially beyond the limb. 
All these cases of potential  variability may produce different shapes of the observed lines. 

Fortunately, we have at least four full nights with good time coverage where we see different 
profiles due to activity. On BJD\,=\,6417 we have a flare with a strength of $3_\epsilon$ with well-defined temporal behaviour. A similar flare structure may be present on the following
day (BJD\,=\,6418), however, the temporal changes here are smoother in time. 
On the other hand, we see a flare on BJD\,=\,6418, at least seen in the \Heps{} behaviour. 
 These results agree with \cite{dave14} who found that the majority of flares occurring
in the atmosphere of the active M4 dwarf GJ\,1243 last longer than approximately 30 minutes
and show multiple peaks depicting the complex nature of flares.
 
Nevertheless, we see different behaviour of emission lines formed in the flare regions and the 
chromosphere. Our analyses suggest the presence of a flare region located over the chromosphere, with both 
regions being bound by a transition zone. Strong flares penetrate the chromospheric layers and 
partially destroy them reducing the strength of chromospheric lines. In other words, the structure 
of the flare region is more complicated than in the case of the Sun. Strong emission in the 
Na{\small{I}}, \DO{}, and \DT{} lines is present during phases of minimum activity, however 
this emission is affected by absorption of cooling plasma located in formerly hot regions.
We believe that these lines form in the lower chromosphere, as their widths 
show little variation with the level of activity.

We encourage observers to acquire more continuous HARPS observations over several 
days/weeks to assess the structure, geometry, and evolution of the flare regions. This 
strategy should also be extended to other bright M dwarfs in order to compare these patterns 
in a broader range of sources.

%
%
\begin{acknowledgements}
The authors kindly thank M.\ A.\ Bautista, S.\ N.\ Nahar, M.\ J.\ Seaton, and D.\ A.\ Verner 
who supplied data compiled in the NIST database.
This research has made use of the Simbad and Vizier databases, operated
at the Centre de Donn\'ees Astronomiques de Strasbourg (CDS), and
of NASA's Astrophysics Data System Bibliographic Services (ADS).
Based on observations collected at the European Organisation for Astronomical Research 
in the Southern Hemisphere under ESO programme(s) 087.D-0300(A).
This is research has made use of the services of the ESO Science Archive Facility.
This paper makes use of data from the MEarth Project, which is a collaboration between Harvard University and the Smithsonian Astrophysical Observatory. The MEarth Project acknowledges funding from the David and Lucile Packard Fellowship for Science and Engineering and the National Science Foundation under grants AST-0807690, AST-1109468, AST-1616624 and AST-1004488 (Alan T. Waterman Award), and a grant from the John Templeton Foundation.
YP thanks financial support from the Fundaci\'on Jes\'us Serra for a 2 month stay 
(Sept--Oct 2016) as a visiting professor at the Instituto de Astrof\'isica de Canarias (IAC) 
in Tenerife. NL and VJSB are supported by the AYA2015-69350-C3-2-P program from 
Spanish Ministry of Economy and Competitiveness (MINECO). 
JIGH acknowledges financial support from the Spanish MINECO under the 2013 
Ram\'on y Cajal program MINECO RYC-2013-14875, and ASM, JIGH, and 
RRL also acknowledge financial support from MINECO (program AYA2014-56359-P). 
 We thank the anonymous referee for a thorough review and we highly
appreciate the comments and suggestions, which significantly contributed
 to improving the quality of the publication. YP thanks Language Editor
 Joshua Neve for the work on the article language.
\end{acknowledgements}

%

\bibliographystyle{aa}


\begin{onecolumn}
\begin{appendix}
\section {A}

\begin{figure*}
  \centering
\includegraphics[width=0.9\linewidth, angle=0]{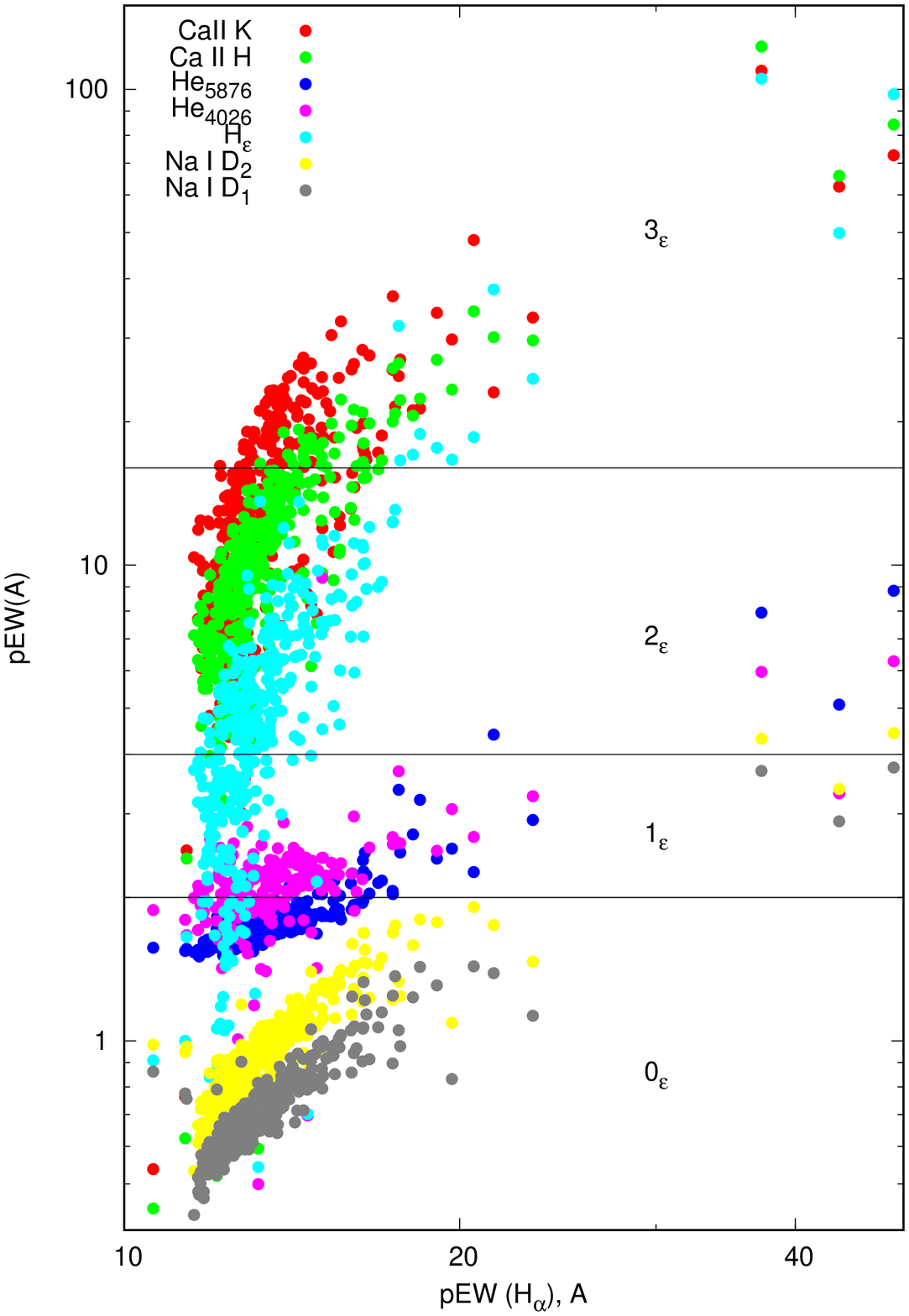}  \vskip 2cm
\caption{Correlation of the variations in the \pew{} of some emission lines with the \pew{} 
of \Ha{} for all dates. The horizontal lines split the figure with respect to our criteria of the 
strengths of \Heps{} (Section \ref{_allpew}).}
   \label{_2mr}
\end{figure*}
\begin{longtable}{cccc cccc c}
\hline\hline
\\
\caption{Changes of the \pew{} of several emission lines \label{_pew1}}
\\
\hline
\\
BJD     &  \Ha{} &Ca{\small{II}}(K)& Ca{\small{II}}(H)&  He$_{5876}$  & He$_{4026}$ &  \Heps{} & Na{\small{I}}(D$_2$) &   Na{\small{I}}(D$_1$) \\
\\
\hline
 3152.5999&  12.890&  19.066&  14.332&   1.746&   1.969&   9.508&   0.783&   0.616  \\
 3202.5847&  43.796&  62.410&  65.757&   5.088&   2.720&  49.893&   3.384&   2.894  \\
 3203.5371&  17.684&  27.040&  22.213&   2.487&   2.014&  16.606&   1.243&   0.975  \\
 3205.5317&  15.059&  11.952&   9.609&   1.829&   8.572&   4.514&   0.948&   0.794  \\
 3207.5051&  12.437&   5.008&   4.611&   1.623&   0.461&   1.937&   0.849&   0.646  \\
 3207.5110&  12.436&   5.141&   4.421&   1.636&   0.656&   2.414&   0.769&   0.635  \\
 3577.4956&  12.302&  13.018&   9.739&   1.696&   1.509&   5.877&   0.699&   0.560  \\
 3809.8936&  13.313&  18.012&  12.005&   1.591&   1.395&   5.211&   0.909&   0.703  \\
 3810.7834&  11.656&  11.859&   7.633&   1.513&   1.431&   3.862&   0.611&   0.482  \\
 3812.7822&  11.895&  12.353&   8.525&   1.571&   1.469&   4.745&   0.649&   0.524  \\
 3816.8096&  12.156&  12.984&   8.804&   1.590&   1.522&   4.935&   0.683&   0.544  \\
 4173.8184&  11.678&  10.228&   6.880&   1.505&   1.379&   3.748&   0.610&   0.475  \\
 4293.5864&  12.856&  16.315&  11.495&   1.715&   1.577&   6.255&   0.859&   0.661  \\
 4296.5933&  13.383&  14.243&  11.246&   1.801&   1.409&   7.216&   0.962&   0.752  \\
 4299.5977&  12.689&  12.003&   9.570&   1.626&   1.551&   4.785&   0.818&   0.629  \\
 4300.5576&  13.433&  17.932&  12.330&   1.694&   1.519&   6.094&   0.937&   0.737  \\
 4646.5693&  13.649&  19.344&  13.608&   1.822&   1.557&   7.357&   0.989&   0.760  \\
 4658.5645&  13.239&  14.329&  10.858&   1.687&   1.323&   5.207&   0.923&   0.691  \\
 4666.5435&  12.388&  14.018&   9.927&   1.675&   1.544&   5.435&   0.772&   0.607  \\
 4878.8638&  15.044&  24.824&  16.890&   2.035&   1.813&  11.295&   1.170&   0.874  \\
 4879.8301&  12.316&  13.903&   9.370&   1.612&   1.440&   4.533&   0.747&   0.574  \\
 4881.8735&  12.993&  17.230&  11.340&   1.722&   1.589&   6.705&   0.811&   0.617  \\
 4882.8535&  14.490&  24.215&  16.839&   1.751&   1.586&   7.478&   1.084&   0.809  \\
 4883.8618&  12.937&  17.681&  13.541&   1.652&   1.668&   8.912&   0.784&   0.591  \\
 4884.8354&  12.551&  13.546&   9.465&   1.620&   1.384&   4.976&   0.789&   0.605  \\
 4885.8750&  12.190&  16.156&  10.349&   1.577&   1.743&   5.487&   0.708&   0.550  \\
 4886.8438&  12.254&  11.473&   7.947&   1.602&   1.357&   4.385&   0.720&   0.561  \\
 4989.6523&  14.229&  21.399&  14.083&   1.707&   1.460&   6.406&   0.867&   0.674  \\
 4992.6504&  11.707&  12.407&   8.193&   1.548&   1.291&   4.038&   0.632&   0.502  \\
 5053.5562&  13.055&  19.046&  14.432&   1.677&   1.635&   6.294&   0.908&   0.692  \\
 5057.5386&  12.810&  17.949&  12.613&   1.715&   1.618&   6.653&   0.876&   0.683  \\
 5275.7715&  13.849&  21.512&  15.388&   1.770&   1.672&   7.411&   0.976&   0.741  \\
 5276.8716&  12.699&  14.710&  10.893&   1.631&   1.276&   4.489&   0.794&   0.616  \\
 5277.8062&  12.215&  15.401&  10.607&   1.559&   1.434&   4.604&   0.743&   0.578  \\
 5278.7905&  14.300&  20.514&  15.465&   1.954&   1.745&   9.308&   0.924&   0.713  \\
 5279.7920&  13.347&  11.439&   9.874&   1.707&   1.223&   4.173&   0.986&   0.724  \\
 5280.7939&  13.983&  22.077&  15.265&   1.743&   1.444&   7.126&   1.076&   0.820  \\
 5281.7598&  14.344&  26.349&  19.259&   1.933&   1.824&  13.570&   1.023&   0.782  \\
 5282.8140&  14.923&  22.895&  17.440&   2.036&   1.630&   9.737&   1.278&   0.980  \\
 5283.7788&  12.705&  13.590&   9.549&   1.621&   1.395&   4.515&   0.866&   0.663  \\
 5292.8027&  13.913&  16.956&  12.184&   1.762&   1.302&   6.342&   0.996&   0.754  \\
 5611.8071&  15.581&  24.732&  18.560&   1.890&   1.586&   8.670&   1.384&   1.012  \\
 5641.7515&  13.495&  17.931&  13.400&   1.830&   1.570&   6.470&   0.913&   0.709  \\
 5644.7910&  13.526&  17.867&  11.860&   1.667&   1.388&   5.601&   0.859&   0.657  \\
 5645.7593&  13.720&  16.356&  12.060&   1.727&   1.469&   6.379&   0.840&   0.640  \\
 5647.7300&  37.302& 109.430& 123.000&   7.946&   5.102& 105.250&   4.318&   3.691  \\
 5648.8286&  12.826&  15.777&  11.225&   1.665&   1.406&   5.034&   0.760&   0.591  \\
 5653.6494&  12.629&  15.355&  11.026&   1.693&   1.484&   6.154&   0.749&   0.589  \\
 5656.6514&  11.553&  10.381&   7.119&   1.535&   1.320&   3.708&   0.533&   0.431  \\
 5659.7778&  11.786&   9.915&   6.983&   1.536&   1.326&   3.157&   0.594&   0.467  \\
 5663.6284&  12.289&   9.918&   7.882&   1.605&   1.160&   3.499&   0.666&   0.536  \\
 5673.7227&  11.785&   9.952&   7.198&   1.616&   1.380&   4.766&   0.598&   0.482  \\
 5715.5605&  13.882&  16.343&  12.489&   1.729&   1.402&   5.511&   0.996&   0.766  \\
 6310.8408&  13.167&  13.643&  11.507&   1.669&   1.585&   4.713&   0.899&   0.684  \\
 6320.8667&  16.072&  26.470&  21.220&   1.930&   2.258&  11.110&   1.287&   0.942  \\
 6322.8281&  12.535&  14.110&  11.043&   1.596&   1.702&   5.558&   0.772&   0.599  \\
 6354.8208&  13.840&  15.120&  11.604&   1.684&   1.528&   4.944&   1.009&   0.757  \\
 6355.8647&  14.611&   1.183&   0.803&   1.778&   0.314&   0.702&   0.896&   0.797  \\
 6356.8047&  13.227&  14.417&  10.664&   1.606&   1.462&   4.596&   0.948&   0.714  \\
 6361.8389&  13.061&  14.120&  11.607&   1.601&   1.737&   5.260&   0.851&   0.635  \\
 6362.8809&  13.280&  16.030&  12.216&   1.613&   1.848&   5.268&   0.927&   0.706  \\
 6363.7783&  13.247&  15.810&  16.343&   1.984&   3.200&  13.605&   0.830&   0.647  \\
 6365.8184&  13.902&  23.503&  19.003&   1.938&   2.190&  11.971&   1.046&   0.791  \\
 6366.7949&  12.956&  12.878&  10.572&   1.651&   1.600&   5.694&   0.795&   0.599  \\
 6367.7729&  12.315&  11.695&   9.746&   1.594&   1.399&   4.983&   0.766&   0.589  \\
 6369.7544&  12.357&  11.586&   8.773&   1.577&   1.432&   5.029&   0.724&   0.568  \\
 6370.7476&  12.516&  15.564&  11.719&   1.629&   1.686&   6.599&   0.808&   0.614  \\
 6371.8052&  11.940&  12.188&   9.538&   1.549&   1.854&   4.436&   0.685&   0.545  \\
 6373.7373&  12.520&  15.054&  11.915&   1.605&   1.896&   5.916&   0.808&   0.621  \\
 6374.7646&  12.445&  13.615&  10.970&   1.582&   1.620&   4.681&   0.764&   0.596  \\
 6417.5044&  12.853&   6.855&   6.137&   1.701&   7.019&   3.242&   0.951&   0.717  \\
 6417.5117&  12.499&   6.492&   5.210&   1.617&   6.727&   1.474&   0.918&   0.678  \\
 6417.5190&  13.452&   9.492&   8.084&   1.874&   8.070&   5.289&   1.018&   0.786  \\
 6417.5264&  12.733&   6.340&   5.064&   1.712&   7.245&   2.227&   0.982&   0.724  \\
 6417.5337&  12.335&   5.744&   4.974&   1.633&   7.438&   1.479&   0.884&   0.666  \\
 6417.5410&  12.374&   5.840&   4.923&   1.676&   7.502&   1.688&   0.866&   0.658  \\
 6417.5483&  12.268&   5.877&   4.896&   1.511&   7.536&   1.237&   0.906&   0.642  \\
 6417.5557&  12.128&   4.543&   4.372&   1.608&   8.082&   0.892&   0.899&   0.638  \\
 6417.5630&  12.373&   5.897&   5.015&   1.730&   7.739&   1.513&   0.912&   0.652  \\
 6417.5703&  11.930&   4.819&   3.972&   1.604&   7.703&   0.839&   0.855&   0.613  \\
 6417.5786&  12.394&   5.098&   5.136&   1.668&   8.697&   1.630&   0.910&   0.683  \\
 6417.5859&  12.275&   4.520&   4.387&   1.619&   7.164&   1.582&   0.844&   0.650  \\
 6417.5933&  13.104&   5.298&   4.300&   1.602&   6.211&   1.258&   0.867&   0.656  \\
 6417.6006&  13.057&   5.500&   5.139&   1.542&   7.222&   2.418&   0.887&   0.654  \\
 6417.6079&  13.066&   6.437&   5.097&   1.611&   7.572&   2.222&   0.871&   0.655  \\
 6417.6152&  12.842&   5.664&   5.000&   1.588&   7.280&   1.831&   0.885&   0.646  \\
 6417.6226&  12.407&   4.356&   4.562&   1.659&   8.277&   1.077&   0.847&   0.638  \\
 6417.6299&  12.638&   5.648&   5.299&   1.609&   8.611&   2.201&   0.891&   0.629  \\
 6417.6372&  13.487&   6.714&   6.816&   1.707&   6.762&   3.669&   0.935&   0.684  \\
 6417.6445&  14.596&   8.690&   8.519&   1.967&   7.775&   7.300&   1.027&   0.789  \\
 6417.6543&  21.457&  23.078&  30.127&   4.402&   7.785&  37.944&   1.751&   1.387  \\
 6417.6621&  18.424&  21.342&  22.382&   3.210&   7.487&  18.869&   1.800&   1.430  \\
 6417.6694&  17.499&  21.526&  20.838&   2.626&   6.372&  13.070&   1.748&   1.367  \\
 6417.6768&  16.397&  19.851&  18.034&   2.269&   6.457&  10.032&   1.685&   1.329  \\
 6417.6841&  16.014&  15.614&  14.593&   2.248&   6.414&   8.208&   1.600&   1.239  \\
 6417.6914&  16.443&  17.121&  16.055&   2.486&   6.800&  10.095&   1.554&   1.217  \\
 6417.6987&  16.495&  16.682&  15.274&   2.229&   6.534&   8.905&   1.455&   1.136  \\
 6417.7056&  15.973&  13.144&  13.186&   2.142&   6.626&   7.071&   1.395&   1.051  \\
 6417.7134&  15.619&  12.114&  10.779&   2.020&   6.625&   6.004&   1.314&   1.018  \\
 6417.7207&  15.413&  10.649&   9.295&   1.967&   7.584&   5.058&   1.222&   0.932  \\
 6417.7280&  14.362&   9.803&   9.452&   1.820&   6.715&   4.288&   1.089&   0.853  \\
 6417.7354&  13.835&   7.442&   6.306&   1.773&   7.986&   2.992&   1.087&   0.844  \\
 6417.7427&  13.351&   6.348&   6.342&   1.678&   7.495&   2.702&   1.010&   0.735  \\
 6417.7505&  12.945&   5.992&   5.181&   1.622&   7.066&   1.960&   0.932&   0.701  \\
 6417.7573&  12.703&   6.050&   6.239&   1.627&   6.735&   1.909&   0.890&   0.658  \\
 6417.7642&  12.431&   5.184&   4.577&   1.595&   7.311&   1.827&   0.831&   0.630  \\
 6417.7720&  12.272&   6.054&   5.173&   1.645&   7.049&   3.017&   0.873&   0.620  \\
 6417.7793&  12.314&   6.740&   5.787&   1.653&   6.820&   3.430&   0.820&   0.612  \\
 6417.7866&  12.197&   5.113&   4.773&   1.639&   7.520&   1.087&   0.851&   0.637  \\
 6417.7939&  12.057&   4.415&   4.326&   1.650&   7.906&   1.667&   0.802&   0.614  \\
 6417.8037&  12.602&   6.596&   6.733&   1.705&   5.914&   3.311&   0.853&   0.678  \\
 6417.8105&  12.336&   5.492&   4.764&   1.589&   6.410&   1.438&   0.902&   0.672  \\
 6417.8179&  12.299&   6.333&   5.437&   1.615&   7.652&   1.738&   0.881&   0.689  \\
 6417.8257&  12.176&   5.365&   4.563&   1.583&   7.374&   0.835&   0.914&   0.663  \\
 6417.8325&  12.210&   4.888&   4.137&   1.591&   7.637&   1.181&   0.879&   0.662  \\
 6417.8403&  12.511&   5.888&   6.523&   1.579&   7.553&   1.472&   0.915&   0.679  \\
 6417.8477&  12.509&   7.564&   5.940&   1.612&   6.069&   1.655&   0.870&   0.650  \\
 6417.8545&  12.304&   5.414&   4.496&   1.558&   7.006&   1.052&   0.861&   0.655  \\
 6417.8623&  12.342&   7.500&   5.274&   1.617&   7.184&   1.745&   0.859&   0.655  \\
 6417.8691&  12.297&   4.855&   4.995&   1.561&   7.629&   1.867&   0.883&   0.667  \\
 6417.8774&  12.902&   7.207&   5.807&   1.799&   7.253&   4.022&   0.925&   0.719  \\
 6417.8848&  12.582&   7.012&   5.415&   1.678&   6.322&   2.917&   0.936&   0.711  \\
 6417.8921&  12.764&   6.141&   5.850&   1.762&   6.747&   2.748&   0.973&   0.700  \\
 6417.8994&  12.850&   4.932&   4.836&   1.818&   7.729&   2.112&   0.945&   0.717  \\
 6417.9072&  12.768&   6.155&   5.552&   1.723&   6.952&   2.760&   0.898&   0.675  \\
 6418.5054&  12.067&   8.944&   6.769&   1.588&   4.445&   3.284&   0.753&   0.572  \\
 6418.5122&  11.925&   6.859&   6.649&   1.553&   4.131&   2.709&   0.717&   0.557  \\
 6418.5181&  12.002&   7.799&   6.805&   1.565&   4.418&   2.904&   0.734&   0.563  \\
 6418.5244&  11.957&   7.290&   5.986&   1.587&   4.594&   2.600&   0.751&   0.583  \\
 6418.5308&  11.876&   7.188&   5.496&   1.560&   5.016&   2.431&   0.736&   0.577  \\
 6418.5366&  11.876&   6.767&   5.841&   1.540&   4.447&   2.177&   0.737&   0.559  \\
 6418.5430&  11.771&   6.698&   5.689&   1.563&   5.110&   1.961&   0.730&   0.553  \\
 6418.5493&  11.976&   9.011&   6.786&   1.576&   4.718&   3.407&   0.763&   0.558  \\
 6418.5552&  12.277&   8.807&   8.210&   1.695&   5.208&   5.551&   0.784&   0.598  \\
 6418.5615&  11.980&   8.088&   6.308&   1.591&   4.735&   2.818&   0.758&   0.577  \\
 6418.5684&  11.937&   8.199&   6.614&   1.619&   5.907&   3.845&   0.767&   0.579  \\
 6418.5742&  12.141&   7.595&   6.646&   1.633&   5.309&   3.488&   0.759&   0.588  \\
 6418.5806&  12.148&   6.684&   6.204&   1.595&   5.573&   2.321&   0.777&   0.589  \\
 6418.5864&  12.135&   7.366&   6.044&   1.600&   5.415&   3.123&   0.797&   0.600  \\
 6418.5928&  11.912&   7.670&   5.983&   1.550&   5.469&   2.448&   0.753&   0.559  \\
 6418.5991&  11.931&   6.395&   5.963&   1.553&   5.253&   2.498&   0.726&   0.558  \\
 6418.6050&  12.031&   8.181&   6.982&   1.634&   5.479&   3.231&   0.730&   0.589  \\
 6418.6113&  11.835&   7.779&   6.049&   1.548&   5.108&   2.958&   0.735&   0.554  \\
 6418.6177&  11.884&   7.224&   6.554&   1.588&   5.508&   1.963&   0.735&   0.563  \\
 6418.6235&  12.023&   8.266&   7.460&   1.648&   4.903&   3.131&   0.759&   0.588  \\
 6418.6304&  11.785&   5.727&   5.501&   1.578&   4.869&   2.354&   0.748&   0.558  \\
 6418.6367&  11.876&   6.623&   5.548&   1.631&   6.465&   2.213&   0.767&   0.599  \\
 6418.6426&  11.933&   7.250&   6.622&   1.592&   5.275&   2.534&   0.744&   0.570  \\
 6418.6489&  11.717&   7.685&   6.122&   1.531&   5.108&   3.569&   0.719&   0.531  \\
 6418.6548&  11.783&   8.633&   7.294&   1.556&   4.446&   3.094&   0.737&   0.553  \\
 6418.6611&  11.821&   8.038&   6.659&   1.581&   4.921&   3.103&   0.732&   0.560  \\
 6418.6675&  11.934&   7.770&   7.309&   1.569&   5.509&   4.237&   0.748&   0.572  \\
 6418.6733&  12.179&   9.773&   8.277&   1.621&   4.194&   4.917&   0.799&   0.595  \\
 6418.6797&  12.179&   9.419&   8.385&   1.582&   4.461&   3.895&   0.827&   0.626  \\
 6418.6860&  12.177&  10.281&   8.647&   1.617&   4.522&   4.084&   0.828&   0.631  \\
 6418.6929&  12.246&   9.867&   7.818&   1.619&   4.350&   3.663&   0.795&   0.601  \\
 6418.6992&  12.321&  10.394&   8.596&   1.636&   4.731&   5.253&   0.776&   0.593  \\
 6418.7051&  12.652&  12.186&   9.971&   1.721&   4.857&   6.292&   0.850&   0.630  \\
 6418.7114&  12.650&  11.822&   9.333&   1.674&   4.986&   4.413&   0.863&   0.644  \\
 6418.7178&  12.676&  11.919&  10.329&   1.611&   4.680&   4.794&   0.920&   0.702  \\
 6418.7236&  12.855&  12.650&   9.961&   1.636&   4.705&   4.536&   0.969&   0.728  \\
 6418.7300&  13.086&  12.344&  10.772&   1.699&   5.338&   4.818&   1.018&   0.766  \\
 6418.7363&  13.261&  12.998&  10.733&   1.740&   5.023&   5.336&   1.028&   0.787  \\
 6418.7422&  13.279&  12.699&  10.263&   1.706&   4.814&   4.305&   1.074&   0.803  \\
 6418.7485&  13.162&  12.099&  10.639&   1.652&   5.253&   4.528&   1.039&   0.761  \\
 6418.7554&  13.172&  12.160&  10.341&   1.602&   4.662&   4.575&   1.047&   0.778  \\
 6418.7617&  13.126&  10.839&   9.453&   1.599&   4.853&   3.349&   1.031&   0.774  \\
 6418.7676&  13.126&  12.320&   9.717&   1.681&   4.498&   4.098&   0.994&   0.757  \\
 6418.7739&  12.905&  11.456&   8.680&   1.594&   4.556&   4.054&   0.972&   0.739  \\
 6418.7798&  12.747&   9.631&   7.932&   1.589&   4.407&   3.174&   0.953&   0.720  \\
 6418.7861&  12.591&   9.287&   8.290&   1.557&   4.808&   3.289&   0.914&   0.698  \\
 6418.7920&  12.507&   9.545&   7.859&   1.550&   4.443&   3.388&   0.882&   0.678  \\
 6418.7983&  12.791&  10.716&   8.603&   1.580&   4.430&   4.255&   0.917&   0.709  \\
 6418.8047&  12.878&  11.202&   9.215&   1.630&   4.437&   3.846&   0.952&   0.719  \\
 6418.8105&  12.736&   8.995&   7.406&   1.591&   5.435&   3.415&   0.921&   0.704  \\
 6418.8169&  12.674&   9.014&   7.140&   1.582&   5.219&   3.373&   0.912&   0.689  \\
 6418.8232&  12.613&   8.791&   7.416&   1.576&   5.316&   2.100&   0.890&   0.679  \\
 6418.8291&  12.734&   9.851&   7.487&   1.590&   4.870&   2.985&   0.909&   0.678  \\
 6418.8354&  12.492&   8.331&   6.829&   1.547&   5.051&   2.279&   0.876&   0.662  \\
 6418.8418&  12.497&   7.842&   6.707&   1.545&   4.501&   1.962&   0.857&   0.649  \\
 6418.8477&  12.607&   8.741&   7.328&   1.612&   5.512&   3.593&   0.858&   0.643  \\
 6418.8540&  13.325&  11.225&  10.094&   1.876&   5.171&   7.026&   0.963&   0.731  \\
 6418.8599&  13.431&  11.912&   9.502&   1.827&   4.814&   5.744&   0.994&   0.747  \\
 6418.8662&  13.049&   9.691&   7.886&   1.735&   5.056&   4.449&   0.970&   0.710  \\
 6418.8721&  12.846&   8.328&   7.180&   1.668&   5.190&   2.984&   0.924&   0.685  \\
 6418.8799&  12.825&   8.414&   7.392&   1.661&   5.368&   3.496&   0.898&   0.682  \\
 6418.8862&  12.514&   7.183&   6.179&   1.610&   5.244&   2.995&   0.856&   0.635  \\
 6418.8921&  12.414&   7.691&   6.550&   1.551&   5.235&   3.067&   0.840&   0.626  \\
 6418.8984&  12.328&   6.648&   5.903&   1.584&   5.635&   2.281&   0.793&   0.600  \\
 6419.6675&  12.727&  10.426&   8.272&   1.635&   4.766&   3.976&   0.887&   0.660  \\
 6419.6738&  12.893&  11.290&   9.688&   1.674&   4.876&   5.829&   0.865&   0.642  \\
 6420.5552&  12.127&   8.202&   6.737&   1.584&   4.671&   2.927&   0.772&   0.577  \\
 6420.5615&  12.058&   7.492&   6.163&   1.598&   5.030&   2.372&   0.754&   0.571  \\
 6420.5688&  11.882&   6.849&   5.898&   1.539&   5.454&   2.507&   0.749&   0.559  \\
 6420.5747&  11.933&   7.067&   6.253&   1.587&   5.434&   2.970&   0.732&   0.562  \\
 6420.5811&  11.863&   6.915&   5.730&   1.547&   4.811&   2.431&   0.715&   0.549  \\
 6420.5874&  11.891&   7.584&   6.115&   1.570&   5.383&   3.013&   0.716&   0.563  \\
 6420.5933&  13.629&  13.173&  11.070&   1.986&   4.431&   7.729&   0.982&   0.774  \\
 6420.5996&  13.180&  11.443&  10.267&   1.841&   5.316&   4.943&   1.049&   0.816  \\
 6420.6055&  12.918&  11.598&   9.585&   1.721&   5.656&   4.327&   0.956&   0.739  \\
 6420.6118&  12.649&   9.849&   9.001&   1.770&   5.819&   4.955&   0.893&   0.687  \\
 6420.6182&  12.882&  11.196&   9.620&   1.760&   4.597&   5.397&   0.908&   0.689  \\
 6420.6240&  12.681&  10.793&   8.875&   1.679&   5.482&   4.007&   0.863&   0.656  \\
 6420.6309&  12.242&   8.165&   6.951&   1.570&   5.237&   2.890&   0.800&   0.619  \\
 6420.6372&  12.172&   8.884&   6.747&   1.610&   5.196&   3.517&   0.766&   0.576  \\
 6420.6436&  12.188&   8.936&   6.992&   1.586&   4.700&   3.367&   0.769&   0.594  \\
 6420.6494&  12.056&   9.223&   7.058&   1.573&   4.546&   2.697&   0.748&   0.581  \\
 6420.6558&  12.408&  11.212&   9.149&   1.681&   4.315&   6.759&   0.786&   0.603  \\
 6420.6616&  13.588&  15.807&  13.646&   1.842&   4.140&   9.089&   1.012&   0.767  \\
 6420.6680&  14.089&  17.955&  14.921&   1.804&   4.655&   7.080&   1.131&   0.864  \\
 6420.6738&  13.790&  16.518&  13.525&   1.712&   4.652&   6.342&   1.130&   0.883  \\
 6420.6802&  14.317&  17.979&  15.046&   1.745&   4.720&   6.448&   1.228&   0.940  \\
 6420.6865&  14.107&  16.544&  12.886&   1.691&   4.564&   5.793&   1.219&   0.920  \\
 6421.5015&  11.727&   6.065&   4.591&   1.558&   6.548&   1.842&   0.744&   0.575  \\
 6421.5078&  12.106&   7.072&   5.765&   1.719&   5.916&   3.677&   0.774&   0.584  \\
 6422.4927&  13.462&  10.689&   8.858&   1.726&   6.435&   3.206&   1.007&   0.745  \\
 6422.7227&  16.900&  17.240&  15.889&   2.148&   6.350&   8.999&   1.441&   1.067  \\
 6422.7305&  17.029&  18.719&  16.591&   2.172&   6.382&   9.224&   1.494&   1.148  \\
 6422.7383&  16.389&  17.486&  15.222&   1.974&   6.530&   7.077&   1.460&   1.066  \\
 6422.7466&  15.601&  12.640&  10.600&   1.839&   5.419&   4.620&   1.332&   0.968  \\
 6423.5220&  12.878&   8.877&   7.812&   1.701&   5.379&   3.295&   0.865&   0.640  \\
 6423.8257&  16.376&  17.757&  16.482&   2.396&   6.796&  10.876&   1.359&   1.061  \\
 6424.4878&  12.240&   7.157&   5.594&   1.648&   6.389&   3.396&   0.822&   0.629  \\
 6424.4956&  12.267&   6.994&   6.380&   1.620&   6.119&   2.428&   0.816&   0.610  \\
 6424.6519&  12.222&  10.645&   8.304&   1.657&   4.777&   4.423&   0.781&   0.589  \\
 6426.4980&  12.926&  12.753&   9.698&   1.600&   4.570&   3.798&   0.870&   0.644  \\
 6426.5205&  12.842&  11.242&   9.050&   1.636&   4.677&   3.806&   0.867&   0.650  \\
 6426.5425&  14.837&  18.459&  15.145&   1.846&   5.637&   6.409&   1.284&   0.938  \\
 6426.5645&  14.218&  14.295&  11.192&   1.735&   4.804&   4.366&   1.120&   0.832  \\
 6426.5869&  13.671&  13.191&  11.188&   1.768&   5.246&   5.193&   1.008&   0.752  \\
 6426.6089&  12.849&  11.913&   8.972&   1.611&   4.956&   3.935&   0.868&   0.660  \\
 6426.6309&  12.981&  11.530&   9.750&   1.681&   5.572&   4.836&   0.896&   0.665  \\
 6426.7393&  13.827&  11.956&   9.359&   1.811&   6.531&   5.636&   1.083&   0.819  \\
 6426.7632&  18.156&  21.164&  20.598&   2.715&   6.895&  17.063&   1.589&   1.234  \\
 6427.4824&  12.437&   7.664&   6.408&   1.566&   6.006&   2.521&   0.777&   0.587  \\
 6427.4990&  12.869&   8.074&   7.317&   1.817&   6.282&   4.623&   0.875&   0.642  \\
 6428.4985&  12.674&   6.867&   6.705&   1.689&   6.479&   2.195&   0.885&   0.674  \\
 6428.5283&  13.394&  11.346&   8.836&   1.661&   5.772&   3.657&   0.965&   0.705  \\
 6656.8022&  12.774&  14.599&  11.017&   1.606&   1.349&   4.527&   0.807&   0.608  \\
 6656.8154&  12.695&  10.841&   9.241&   1.633&   1.133&   4.030&   0.793&   0.594  \\
 6657.8018&  12.705&  16.110&  10.999&   1.573&   1.390&   4.765&   0.781&   0.595  \\
 6657.8169&  12.606&  14.394&  10.067&   1.574&   1.562&   4.294&   0.756&   0.585  \\
 6658.8120&  16.358&  28.313&  20.938&   1.951&   1.707&  11.035&   1.208&   0.905  \\
 6658.8271&  15.999&  25.759&  19.644&   1.888&   1.610&   9.561&   1.137&   0.858  \\
 6659.8232&  12.598&  14.926&  11.102&   1.621&   1.440&   5.130&   0.755&   0.583  \\
 6659.8379&  12.991&  16.808&  12.266&   1.642&   1.406&   5.959&   0.843&   0.641  \\
 6660.8032&  13.135&  12.094&   9.993&   1.631&   1.332&   4.715&   0.820&   0.646  \\
 6660.8574&  12.976&  15.067&  11.119&   1.605&   1.403&   4.712&   0.836&   0.644  \\
 6661.8062&  12.538&  12.263&   9.347&   1.565&   1.440&   3.529&   0.752&   0.574  \\
 6661.8511&  12.683&  14.837&  10.835&   1.589&   1.361&   4.678&   0.803&   0.624  \\
 6663.8560&  13.555&  18.862&  13.991&   1.855&   1.362&   7.636&   0.914&   0.695  \\
 6663.8706&  14.059&  18.137&  14.756&   2.108&   1.742&  11.117&   1.008&   0.792  \\
 6665.7827&  13.570&  16.283&  12.602&   1.749&   1.370&   6.128&   0.925&   0.704  \\
 6665.8130&  13.263&  11.411&   8.948&   1.594&   1.114&   4.221&   0.922&   0.685  \\
 6665.8672&  14.647&  19.055&  14.913&   1.842&   1.568&   7.089&   1.120&   0.831  \\
 6666.7861&  13.500&  14.258&  11.316&   1.630&   1.242&   4.886&   0.881&   0.645  \\
 6666.8164&  13.738&  17.160&  14.352&   1.834&   1.836&   8.392&   0.861&   0.667  \\
 6666.8623&  13.043&  16.767&  11.546&   1.607&   1.255&   4.806&   0.807&   0.612  \\
 6667.7803&  12.680&  14.746&  11.277&   1.606&   1.544&   5.246&   0.841&   0.636  \\
 6667.8291&  13.529&  19.423&  13.997&   1.659&   1.510&   5.955&   0.976&   0.739  \\
 6667.8628&  12.993&  17.311&  11.537&   1.581&   1.608&   5.218&   0.842&   0.644  \\
 7406.8706&  19.677&  29.783&  23.379&   2.535&   2.321&  16.663&   1.092&   0.831  \\
 7407.8618&  13.193&   0.021&   0.594&   1.771&   0.291&   0.543&   0.910&   0.693  \\
 7410.8677&  15.438&  18.531&  17.151&   2.173&   1.743&  11.524&   1.290&   0.979  \\
 7412.8672&  13.485&  19.634&  13.432&   1.674&   1.526&   6.076&   0.964&   0.723  \\
 7413.8696&  13.543&  18.554&  13.554&   1.707&   1.524&   6.199&   0.925&   0.688  \\
 7414.8706&  14.302&  20.918&  15.038&   1.798&   1.544&   7.129&   1.065&   0.797  \\
 7415.8789&  13.451&  20.635&  14.148&   1.692&   1.610&   6.324&   0.945&   0.707  \\
 7416.8721&  15.649&  32.510&  22.279&   1.800&   1.669&   8.581&   1.408&   1.029  \\
 7417.8789&  13.687&  19.669&  13.204&   1.704&   1.545&   6.018&   0.948&   0.699  \\
 7418.8750&  20.585&  48.212&  34.128&   2.263&   2.041&  18.567&   1.911&   1.434  \\
 7420.8760&  13.520&  23.220&  15.779&   1.770&   1.734&   7.424&   1.012&   0.759  \\
 7420.8906&  13.084&  18.458&  12.158&   1.655&   1.581&   5.477&   0.911&   0.688  \\
 7421.8799&  13.232&  21.094&  14.027&   1.690&   1.570&   6.300&   0.915&   0.684  \\
 7422.8677&  13.896&  22.625&  15.241&   1.751&   1.688&   7.241&   0.971&   0.727  \\
 7423.8765&  13.604&  23.280&  15.605&   1.747&   1.610&   7.227&   1.015&   0.750  \\
 7425.8828&  13.718&  21.646&  14.212&   1.674&   1.699&   6.185&   0.997&   0.744  \\
 7427.8833&  17.418&  36.709&  25.927&   2.033&   1.903&  12.332&   1.685&   1.245  \\
 7428.8804&  14.479&  27.294&  18.344&   1.816&   1.708&   8.753&   1.163&   0.849  \\
 7429.8818&  14.028&  24.742&  17.070&   1.865&   1.748&   9.581&   1.012&   0.760  \\
 7431.8872&  14.489&  25.869&  17.377&   1.769&   1.641&   7.762&   1.086&   0.799  \\
 7432.8833&  13.396&  21.882&  14.051&   1.676&   1.527&   6.303&   0.966&   0.715  \\
 7433.8911&  13.726&  22.090&  15.388&   1.727&   1.648&   7.140&   1.049&   0.795  \\
 7434.8799&  12.807&  17.495&  11.918&   1.645&   1.522&   5.703&   0.856&   0.649  \\
 7436.8589&  13.634&  23.268&  15.813&   1.721&   1.678&   7.531&   0.959&   0.726  \\
 7437.8618&  15.345&  30.427&  19.847&   1.825&   1.677&   8.480&   1.329&   0.972  \\
 7438.8579&  14.705&  26.546&  18.489&   1.840&   1.618&   8.542&   1.230&   0.907  \\
 7439.8579&  14.107&  24.974&  16.776&   1.774&   1.687&   8.197&   1.092&   0.818  \\
 7440.8628&  13.408&  18.581&  13.578&   1.810&   1.582&   8.531&   0.978&   0.732  \\
 7442.8687&  14.064&  16.390&  13.015&   1.892&   1.733&   8.505&   1.058&   0.795  \\
 7444.8872&  19.078&  33.895&  27.008&   2.415&   1.874&  17.630&   1.774&   1.308  \\
 7445.8706&  14.645&  20.241&  14.068&   1.779&   1.486&   6.832&   1.109&   0.815  \\
 7446.8716&  14.415&  22.201&  14.869&   1.717&   1.471&   6.486&   1.102&   0.805  \\
 7447.8765&  15.303&  21.047&  15.326&   1.826&   1.527&   7.638&   1.268&   0.927  \\
 7448.8804&  15.180&  21.885&  16.069&   1.830&   1.624&   7.598&   1.265&   0.917  \\
 7449.8760&  13.594&  17.315&  12.678&   1.799&   1.529&   6.562&   1.057&   0.783  \\
 7450.8975&  15.073&  23.171&  18.062&   2.120&   1.796&  11.023&   1.324&   0.999  \\
 7452.8789&  13.465&  22.219&  14.870&   1.722&   1.567&   6.889&   0.978&   0.731  \\
 7453.9116&  13.558&  18.365&  12.928&   1.653&   1.462&   5.412&   1.084&   0.798  \\
 7454.9087&  13.950&  19.692&  14.563&   1.735&   1.647&   6.815&   1.109&   0.817  \\
 7455.9092&  17.411&  25.719&  20.024&   2.066&   1.967&  12.293&   1.221&   0.897  \\
 7456.9082&  13.642&  18.608&  13.181&   1.677&   1.575&   6.049&   0.997&   0.734  \\
 7457.8818&  14.545&  21.970&  16.394&   1.876&   1.800&   9.142&   1.068&   0.792  \\
 7458.8984&  14.735&  23.535&  18.004&   1.842&   1.735&   8.482&   1.186&   0.874  \\
 7460.8833&  13.435&  19.209&  13.353&   1.694&   1.610&   6.402&   0.949&   0.705  \\
 7461.8843&  16.599&  27.590&  19.851&   2.041&   1.895&  12.016&   1.231&   0.915  \\
 7462.8882&  13.630&  20.506&  15.272&   1.783&   1.765&   7.526&   1.022&   0.760  \\
 7463.8901&  15.037&  19.436&  16.241&   1.855&   1.474&   7.596&   1.282&   0.970  \\
 7465.8823&  49.008&  72.685&  84.356&   8.833&   5.388&  97.632&   4.435&   3.754  \\
 7466.9131&  13.074&  16.625&  12.239&   1.704&   1.509&   6.172&   0.913&   0.702  \\
 7467.9097&  12.795&  15.592&  11.283&   1.619&   1.431&   5.319&   0.869&   0.654  \\
 7473.8726&  12.828&   3.548&   3.918&   1.708&   1.214&   1.686&   0.906&   0.738  \\
 7473.8926&  12.483&   8.239&   6.561&   1.606&   1.261&   3.193&   0.810&   0.621  \\
 7474.8945&  14.192&  21.417&  16.216&   1.868&   1.837&   8.964&   1.109&   0.831  \\
 7475.8921&  12.558&  13.939&  10.013&   1.636&   1.521&   5.193&   0.813&   0.622  \\
 7476.8892&  12.957&  15.540&  11.992&   1.763&   1.484&   7.566&   0.831&   0.635  \\
 7477.8979&  12.862&  15.174&  11.055&   1.657&   1.471&   5.576&   0.838&   0.647  \\
 7905.4707&  17.631&  24.980&  26.565&   3.370&   2.950&  31.806&   1.332&   1.053  \\
 7905.5288&  15.026&  15.060&  12.403&   1.861&   1.510&   7.160&   1.105&   0.836  \\
 7905.6182&  15.089&  18.555&  13.867&   1.841&   1.601&   7.367&   1.119&   0.845  \\
 7934.6421&  14.166&   9.700&   8.195&   1.699&   1.590&   3.892&   1.163&   0.883  \\
 7935.6562&  13.512&  11.040&   9.451&   1.653&   1.326&   4.540&   0.971&   0.721  \\
 7936.6221&  13.720&  11.755&   9.573&   1.709&   1.447&   4.931&   0.951&   0.725  \\
 7937.6421&  12.714&   7.830&   6.815&   1.627&   1.185&   3.579&   0.854&   0.649  \\
 7942.5752&  13.320&  13.798&  10.690&   1.668&   1.543&   5.151&   0.896&   0.694  \\
 7943.5610&  12.488&  12.904&   9.937&   1.655&   1.275&   4.689&   0.787&   0.619  \\
 7944.5454&  14.017&  17.246&  14.003&   1.799&   1.453&   7.837&   1.047&   0.795  \\
 7945.5918&  14.759&  23.442&  17.556&   1.808&   1.551&   8.028&   1.073&   0.804  \\
 7946.5405&  13.821&  20.436&  14.644&   1.742&   1.711&   7.403&   0.934&   0.721  \\
 7948.6543&  13.514&   9.600&   8.859&   1.749&   1.210&   4.431&   0.991&   0.753  \\
 7949.6699&  13.106&   6.627&   6.111&   1.755&   0.496&   3.067&   0.923&   0.711  \\
 7951.4634&  13.403&   8.135&   7.531&   1.805&   1.034&   4.383&   0.902&   0.674  \\
 7952.6406&  12.857&  11.536&   9.361&   1.615&   1.038&   4.127&   0.776&   0.602  \\
 7953.4731&  13.517&  18.296&  14.287&   1.787&   1.748&   7.715&   0.940&   0.722  \\
 7954.6279&  14.484&  16.052&  13.038&   1.786&   1.312&   7.713&   0.944&   0.714  \\
 7955.6372&  14.889&   7.890&   7.555&   1.680&   1.088&   2.161&   1.073&   0.830  \\
 7956.4727&  13.095&  19.019&  13.453&   1.707&   1.583&   6.245&   0.828&   0.646  \\
 7959.4731&  12.593&  14.209&  10.473&   1.711&   1.374&   6.010&   0.786&   0.603  \\
 7960.4629&  12.643&  13.662&  10.526&   1.607&   1.388&   4.897&   0.865&   0.677  \\
 7961.4854&  12.074&  10.035&   7.928&   1.624&   1.421&   3.889&   0.703&   0.549  \\
 7962.4712&  12.171&  12.274&   9.099&   1.675&   1.593&   5.346&   0.727&   0.556  \\
 7964.4727&  12.108&   7.860&   6.688&   1.592&   1.199&   3.406&   0.697&   0.523  \\
 7965.4761&  12.650&   4.817&   4.796&   1.672&   0.719&   2.215&   0.823&   0.648  \\
 7966.4746&  11.771&   9.732&   7.981&   1.552&   1.342&   3.425&   0.673&   0.530  \\
 7967.5088&  14.478&  10.267&   9.621&   1.807&   0.793&   4.776&   1.190&   0.916  \\
 7968.4785&  11.648&   7.700&   6.316&   1.554&   1.179&   3.173&   0.664&   0.517  \\
 7969.4712&  12.540&  11.756&   9.341&   1.651&   1.394&   4.562&   0.801&   0.611  \\
 7970.4761&  12.904&  15.024&  10.904&   1.680&   1.525&   6.111&   0.828&   0.633  \\
 7971.4668&  12.553&  14.052&  10.386&   1.671&   1.467&   5.330&   0.820&   0.646  \\
 7972.4653&  13.630&  15.064&  11.640&   1.751&   1.541&   6.320&   0.991&   0.754  \\
 7973.4653&  13.570&  10.449&   9.530&   1.685&   1.264&   4.259&   1.008&   0.782  \\
 7974.4624&  14.147&  17.958&  13.729&   1.874&   1.582&   8.179&   1.049&   0.808  \\
 7979.4590&  13.414&  10.904&   8.243&   1.632&   1.427&   3.882&   0.949&   0.735  \\
 7980.4609&  13.432&  15.123&  11.394&   1.801&   1.544&   6.347&   1.077&   0.828  \\
 7981.5054&  15.453&  15.274&  12.928&   1.875&   1.501&   7.663&   1.115&   0.851  \\
 7984.5039&  13.538&  13.041&  11.668&   1.719&   1.397&   4.723&   0.979&   0.749  \\
 7985.5034&  12.895&  10.103&   8.333&   1.633&   1.389&   3.652&   0.870&   0.660  \\
 7986.5156&  23.263&  33.132&  29.664&   2.912&   2.514&  24.623&   1.465&   1.129  \\
 7987.5249&  13.465&  14.628&  11.889&   1.727&   1.403&   5.739&   0.977&   0.747  \\
 7990.4902&  13.813&  14.746&  13.084&   1.907&   1.441&   9.011&   1.017&   0.788  \\
 7991.5078&  13.537&  13.213&  10.777&   1.676&   1.329&   4.907&   0.954&   0.710  \\
 7992.5029&  12.787&  12.982&  10.182&   1.632&   1.554&   4.451&   0.869&   0.671  \\
 7993.4897&  12.409&  12.295&   9.991&   1.644&   1.586&   5.471&   0.787&   0.619  \\
 7994.4746&  12.645&  10.458&   8.650&   1.653&   1.217&   4.652&   0.825&   0.638  \\
 7995.4780&  16.158&  19.434&  16.153&   1.936&   1.463&   8.375&   1.254&   0.964  \\
 7996.4805&  14.066&  14.697&  12.151&   1.737&   1.221&   5.677&   0.993&   0.753  \\
 7997.4678&  16.093&  14.582&  12.443&   1.883&   1.238&   5.941&   1.437&   1.072  \\
 7998.4819&  12.367&  10.043&   8.047&   1.580&   1.149&   3.560&   0.758&   0.582  \\
 7999.4756&  12.238&   3.638&   3.217&   1.769&   1.008&   2.520&   0.834&   0.642  \\
 8000.4780&  13.079&  10.159&   8.924&   1.667&   0.882&   3.857&   0.947&   0.711  \\
 8001.4912&  12.608&  13.270&   9.891&   1.633&   1.234&   4.665&   0.826&   0.622  \\
 8002.4766&  12.514&   9.687&   8.363&   1.726&   1.421&   4.683&   0.788&   0.615  \\
 8007.5283&  12.666&   6.580&   7.267&   1.686&   1.008&   3.024&   0.930&   0.708  \\
 8011.4893&  14.721&  16.474&  13.983&   1.777&   1.230&   5.549&   1.215&   0.930  \\
 8012.4766&  12.905&   7.497&   6.946&   1.768&   0.995&   4.465&   0.904&   0.687  \\
 8013.4751&  13.084&  10.954&   9.233&   1.766&   1.444&   4.668&   0.933&   0.716  \\
 8017.4683&  14.716&   8.229&   6.129&   1.833&   1.478&   3.972&   1.398&   1.058  \\
 8018.5127&  12.994&  12.005&  10.280&   1.676&   1.550&   5.589&   0.860&   0.662  \\
 8019.4790&  12.143&   9.785&   7.503&   1.580&   1.178&   3.529&   0.729&   0.565  \\
 8020.4834&  11.836&   8.020&   6.802&   1.584&   1.321&   3.174&   0.668&   0.533  \\
 8021.4600&  10.618&   0.537&   0.444&   1.569&   1.194&   0.909&   0.982&   0.862  \\
 8022.4614&  12.113&   0.563&   0.520&   1.818&   1.207&   1.061&   0.910&   0.790  \\
 8023.4639&  11.342&   0.764&   0.624&   1.546&   1.143&   1.000&   0.946&   0.775  \\
 8025.4741&  12.738&   3.012&   2.120&   1.676&   1.238&   1.706&   1.192&   0.903  \\
 8026.4785&  12.453&   8.081&   7.296&   1.633&   1.417&   3.708&   0.886&   0.688  \\
 8027.4785&  11.379&   2.514&   2.416&   1.560&   1.213&   1.655&   0.973&   0.755  \\
\hline
\end{longtable}

\clearpage

\end{appendix}
\end{onecolumn}
\end{document}